\documentclass{article}
\usepackage{PRIMEarxiv}
\usepackage[utf8]{inputenc} % allow utf-8 input
\usepackage[T1]{fontenc}    % use 8-bit T1 fonts
\usepackage{yfonts}
\usepackage{hyperref}       % hyperlinks
\usepackage{url}            % simple URL typesetting
\usepackage{booktabs}       % professional-quality tables
\usepackage{amsfonts}       % blackboard math symbols
\usepackage{nicefrac}       % compact symbols for 1/2, etc.
\usepackage{microtype}      % microtypography
\usepackage{lipsum}
\usepackage{fancyhdr}       % header
\usepackage{graphicx}       % graphics
\graphicspath{{./}}     % organize your images and other figures under media/ folder

\newcommand\BibTeX{{\rmfamily B\kern-.05em \textsc{i\kern-.025em b}\kern-.08em
T\kern-.1667em\lower.7ex\hbox{E}\kern-.125emX}}

\usepackage{moreverb}
\usepackage{multirow}
\usepackage{bm}
\usepackage{amsmath}
\usepackage[table]{xcolor}

\newcommand{\Normal} {{\mathcal N}}
\newcommand{\bbeta}  {\mathbf{\beta}}
\newcommand{\kr}     {k_{r}}
\newcommand{\thetar} {\theta_{r}}
\newcommand{\tzr}    {t_{0,r}}
\newcommand{\Nr}     {N_{r}}
\newcommand{\Nir}    {N_{i, r}}
\newcommand{\Nimr}  {N_{i-1, r}}
\newcommand{\nir}    {n_{i,r}}
\newcommand{\Nreg}   {R}
\newcommand{\lphi}   {\lambda_{\phi}}
\newcommand{\tphi}   {\tau_{\phi}}
\newcommand{\gr}     {\gamma_r}
\newcommand{\bgamma} {\mathbf{\gamma}}
\newcommand{\bsigma} {\mathbf{\sigma}}
\newcommand{\bpsi} {\mathbf{\psi}}
\newcommand{\bp} {\mathbf{p}}
\newcommand{\Mcal} {\mathcal{M}}
\newcommand{\bveps} {\mathbf{\varepsilon}}
\newcommand{\dcor} {$d_{\textrm{cor}}$}

\newcommand{\etal}{\textit{et al}. }

\title{Detecting Outbreaks Using a Latent Field: Part I - Spatial Modeling
}

\author{
Cosmin Safta \\
Sandia National Laboratories \\
Livermore, CA\\
\texttt{csafta@sandia.gov} \\
\And
Jaideep Ray \\
Sandia National Laboratories \\
Livermore, CA \\
\texttt{jairay@sandia.gov} \\
\AND
Wyatt Bridgman \\
Sandia National Laboratories \\
Livermore, CA \\
\texttt{whbridg@sandia.gov} \\
}

\begin{document}
\maketitle

\begin{abstract}
	In this paper, we develop a method to estimate the infection-rate of a disease, over a region, as a \emph{field} that varies in space and time. To do so, we use time-series of case-counts of symptomatic patients as observed in the areal units that comprise the region. We also extend an epidemiological model, initially developed to represent the temporal dynamics in a single areal unit, to encompass multiple areal units. This is done using a (parameterized) Gaussian random field, whose structure is modeled using the dynamics in the case-counts, and which serves as a spatial prior, in the estimation process. The estimation is performed using an adaptive Markov chain Monte Carlo method, using COVID-19 case-count data collected from three adjacent counties in New Mexico, USA. We find that we can estimate both the temporal and spatial variation of the infection with sufficient accuracy to be useful in forecasting. Further, the ability to ``borrow'' information from neighboring areal units allows us to regularize the estimation in areal units with high variance (``poor quality'') data. The ability to forecast allows us to check whether the estimated infection-rate can be used to detect a change in the epidemiological dynamics e.g., the arrival of a new wave of infection, such as the fall wave of 2020 which arrived in New Mexico in mid-September 2020. We fashion a simple anomaly detector, conditioned on the estimated infection-rate and find that it performs better than a conventional surveillance algorithm that uses case-counts (and not the infection-rate) to detect the arrival of the same wave.
\end{abstract}


\maketitle

\section{Introduction}
\label{sec:intro}
The infection-rate of a disease, especially a (human-to-human) communicable one, is perhaps the most concise distillation of the epidemiological dynamics of an outbreak. It waxes and wanes as a  population's mixing patterns change with the seasons or when a new variant arrives. It varies in space, modulated by risk factors \emph{viz.}, socioeconomic conditions, population density and demographic profile. It could potentially be a very informative quantity to monitor as part of disease surveillance, but is rarely ever done. This is because the infection-rate of an outbreak cannot be directly observed; instead, it has to be estimated, most commonly using a time-series of case-counts of patients (i.e., infected people who have tested positive). Depending on the quality of case-count data, which could have large reporting errors and display a considerable amount of variability if obtained from a small population where case-counts are low, the estimation of the infection-rate can be a difficult task.

Regardless of these difficulties, there have been many studies that estimate the infection-rate, particularly for the COVID-19 pandemic~\cite{DazaTorres:2022,WangZ:2020,ChenP:2021}. Our own work~\cite{Blonigan:2021,Lin:2021,Safta:2021} parameterized a temporally-varying infection-rate and convolved it with the incubation period of COVID-19 to construct a disease model; when fitted to COVID-19 case-count data using Bayesian inference, it yielded parameters of the infection-rate model. This model could be used to provide 2-week-ahead forecasts of the behavior of the outbreak; when the observed data disagreed with the forecasts consistently, it indicated a change in epidemiological dynamics (e.g., the effect of lockdowns in California~\cite{Safta:2021} or the start of the fall wave of COVID-19 in New Mexico~\cite{Blonigan:2021}). All these studies aggregate case-counts over large populations (usually above 250,000) to reduce the  variability in the observed case-counts and thus ease the estimation problem for the infection-rate. However, this aggregation can be problematic if performed over a large, sparsely populated region (e.g., the state of New Mexico, USA). The infection-rate estimated is necessarily an average over the regional population and may bear little resemblance to the local population if the population displays large spatial heterogeneity; this is certainly the case with New Mexico due to the presence of urban areas as well as remote, sparsely-populated desert counties.  Since public health measures are often decided at the county-level, these regionally-averaged estimates of infection-rate are only used as a rough guide by public health professionals.

In this paper, we develop a method to estimate the infection-rate as a spatiotemporal field, described over areal units that comprise a region. Each areal unit supplies a time-series of case-counts for the estimation of the infection-rate field. For the purposes of this paper, we will use the COVID-19 outbreak in New Mexico (NM) and its counties as the test case, using data collected between June 1, 2020 and September 15, 2020; after September 15, the case-counts in NM steadily rose into the winter, an event we will refer to colloquially as the ``Fall 2020'' wave. Our approach is based on two key hypotheses. Our first premise is that the parameterized model for the time-varying infection-rate, as developed by Safta~\etal\cite{Safta:2021}, can be used to model the temporal evolution of the outbreak in each areal unit. This will lead to an inverse/estimation problem that will scale with the number of areal units and could quickly become intractable. Our second premise is that the spatial correlations in the epidemiological dynamics, as observed in the case-count data, can be fashioned into a random field model to regularize the high-dimensional field inversion and render it tractable. As part of this investigation, our method will be exposed to observational data of variable ``quality'', from relatively low-variability observations from populous counties, such as Bernalillo, to high-variability low case-count data from smaller counties around it.

The development of the this method will require us to address the following research questions:
\begin{itemize}
\item How does one fashion a random field model, from observational data of case-counts, to regularize the estimation problem for the infection-rate field?
\item How does one include the random field model into the estimation of the infection-rate field? Does its inclusion improve the quality of the estimated infection-rate vis-\`a-vis an estimation performed using data from a areal unit independently? In particular, for counties/areal units with poor quality data, does the inclusion of the random field model (i.e., incorporate the ability to ``borrow'' information from neighbors) improve the estimation of the infection-rate?
\item Can we use the estimated infection-rate to detect the arrival of the Fall 2020 wave in the counties of NM? How does it compare to a conventional outbreak-detector (specifically H{\"o}hle and Paul, 2008~\cite{Hohle:2008})? In addition, in the absence of the Fall 2020 wave, does the use of the infection-rate lead to a false positive?
\end{itemize}

We will address the questions using data from three adjoining NM counties \emph{viz.} Bernalillo, Santa Fe and Valencia. The inverse problem is sufficiently low-dimensional to be solved exactly using an adaptive Markov chain Monte Carlo (AMCMC; see Haario~\etal\cite{Haario:2001}). A companion paper (see Ray~\etal\cite{23rs5a} for the technical report version) extends the method to all 33 counties (areal units) of NM, using mean-field Variational Inference to solve the inverse problem for the infection-rate approximately, as the problem becomes too high-dimensional for AMCMC.

The main contribution of the paper is in illustrating the use of random field models in inverse problems to yield local epidemiological information, using the spatial correlation extant in epidemiological dynamics (caused by population mixing) to compensate for high-variability in the case-count time-series observational data.  A second contribution of the paper is to demonstrate that the information so obtained (in the form of a local infection-rate) contains actionable public health information; we will do so by detecting the arrival of the Fall 2020 wave. Note that we do not attempt to make a proper outbreak detector in this paper; that is left to future work. Also note that the use of random field models in disease mapping is well-established~\cite{Best:2005,Waller:2010}; however, these methods seek to only \emph{smooth} observed case-count data rather than estimate the underlying infection-rate.

The paper is structured as follows. In \S~\ref{sec:litrev} we review existing literature on infection-rate estimation, the empirical construction and parameterization of random field models, especially in disease mapping, and how outbreak-detectors function. In \S~\ref{sec:form}, we parameterize a Gaussian random field (GRF) model to represent spatial correlations in epidemiological dynamics and  formulate a general inverse problem for the infection-rate. In \S~\ref{sec:res}, we present the results of the infection-rate estimation, jointly for the three counties, and compare them with the results obtained from independent estimation. We also discuss how the estimated infection-rate performs in detecting the Fall 2020 wave, compared to conventional techniques (\S~\ref{sec:disc}). We conclude in \S~\ref{sec:concl}.

\section{Literature Review}
\label{sec:litrev}
{\bf Covariates and spatial autocorrelation in COVID-19 dynamics: }Huang~\etal~\cite{Huang:2021} analyzed the spatial relationship between the main environmental and meteorological factors and COVID-19 cases in Hubei province of China using a geographically weighted regression (GWR) model. Results suggest that the impacts of environmental and meteorological factors on the development of COVID-19 were not significant, something we also found in NM (see \S~\ref{sec:expl}). Their findings indicate that measures such as social distancing and isolation played the primary role in controlling the development of the COVID-19 epidemic. Geng~\etal~\cite{Geng:2021} analyzed spatio-temporal patterns of COVID-19 infections at scales spanning from county to continental. They found that spatial evolution of COVID-19 cases in the United States followed multifractal scaling. A rapid increase in the spatial correlation was identified early in the outbreak (March to April 2020) followed by an increase at a slower rate until approaching the spatial correlation of human population. For this study, the multiphase COVID-19 epidemics were modeled by a kernel-modulated susceptible–infectious–recovered (SIR) algorithm. Schuler~\etal~\cite{Schuler:2021} employed a compartmental model for all 412 districts of Germany coupled with non-pharmaceutical intervention (NPI) models. They identify disease spread dynamics that corresponds to different spatial correlation levels, obtained via variogram estimation, between adjacent districts. McMahon~\etal~\cite{McMahon:2022} analyzed the spatial correlations of new active cases in the USA at the county level and showed that various stages of the epidemic are distinguished by significant differences in the correlation length. Their results indicate that the correlation length may be large even during periods when the number of cases declines and that correlations between urban centers were more significant than between rural areas. Rendana~\etal~\cite{Rendana:2021} analyzed the spatial distribution of COVID-19 cases, epidemic infection-rate, spatial pattern during the first and second waves in the South Sumatra Province of Indonesia. The study found little to no correlation between different regions. Air temperature, wind speed, and precipitation have contributed to the high epidemic infection-rate in the second wave. Indika~\etal~\cite{Sathish:2023} inspect the daily count data related to the total cases of COVID-19 in 93 counties in the state of Virginia using a Bayesian conditional autoregressive (CAR) modeling framework. The authors find that Moran statistic values at specific time points are impacted by, and linked to, the executive orders at the state level. In summary, there is some evidence that modeling of COVID-19 over small areal units might need to accommodate spatial auto-correlation, and might also require the inclusion of other covariates.

{\bf Random fields and disease maps:} There is little literature on the use of a random field to estimate the infection-rate of a disease. However, the estimation of a latent field called \emph{relative risk} $r(\bm x)$ is central to disease mapping.~\cite{Lawson:2017,MacNab:2022} A disease map is a 2D plot of the risk of contracting a disease, computed from case-counts collected over areal units e.g., counties, that comprise a region e.g., a province. First, one obtains an ``expected'' value $e_i$ for the observed case counts $y^{(obs)}_i$ for areal unit $i$, usually from a region-wide average of disease incidences and demographics.  It is then locally adjusted (in space) using the relative risk field to bring is closer to observations i.e., $y^{obs}_i \sim {\rm Poisson}(r_i e_i)$.  The risk $r_i$ is then modeled as $\log(r_i) = \boldsymbol{z}_i \cdot \boldsymbol{\beta} + \phi_i$, where $\boldsymbol{z}_i$ are co-variate risk factors for areal unit $i$, $\boldsymbol{\beta}$ are regression weights and $\phi_i$ captures auto-correlated random effects in space using a random field model. The simplest random field model is iCAR (intrinsic Conditional AutoRegressive~\cite{Lawson:2017}), a specific type of Gaussian Markov Random Field (GMRF). Thus
\[
\phi = \{\phi_i\} \sim \Normal\left(0, \{\tau^2 Q\}^{-1} \right), \mbox{\hspace{1cm}} Q = {\rm diag}(W\boldsymbol{1}) - W,
\]
where $W$ is the adjacency matrix of the areal units (i.e., $w_{ij} = 1$ if areal units $i$ and $j$ share a boundary). The object of estimation from data is $\tau^2$. The precision matrix $Q$ tends to be sparse. This formulation leads to an improper jont distribution for $\phi$. The Besag-York-Mollie (BYM) model~\cite{Besag:1991} overcomes this issue by extending iCAR as $\phi = \phi^1 + \phi^2, \phi^1 \sim \Normal(0, \{\tau^2 Q\}^{-1})$ and $\phi^2 \sim \Normal(0, \sigma^2I)$. We will use a variation of BYM in our work. The objects of estimation from case-count data are $(\tau^2, \sigma^2)$. A second variation, called pCAR (proper CAR~\cite{Stern:1999,Cressie:2015}), modifies the precision matrix $Q = {\rm diag}(W\boldsymbol{1}) - \rho W$, where the objects of estimation are $(\tau^2, \rho, \sigma^2)$. The idea of a random field being used to smooth areal units in \emph{feature}-space (as opposed to geometrical space) has also been developed using GMRF~\cite{Baptista:2016}. Such a method is useful for diseases like alcohol abuse where similarity of socioeconomic and health factors in areal units, rather than the geometric distance between them, are more relevant for smoothing. The difference lies in how $Q$ is modeled using a similarity $S$ matrix~\cite{Best:1999}.

{\bf Outbreak detectors: } Outbreak detection functions primarily as anomaly detection in space and time~\cite{12uf5a}. The case-count at time $t$, $y_t$, is often modeled as a normal random variate $y_t \sim \Normal(\mu_t, \sigma_t^2)$; an alarm is raised if $y_t - \mu_t > \kappa \sigma_t$, where $\kappa$ is a threshold value adjusted to trade-off specificity and sensitivity of the detection. This approach can be considered as an expansion of Shewhart charts~\cite{Shewhart:1930} and is sometimes referred to as ``statistical process control'' (SPC) methods. Methods differ on how $(\mu_t, \sigma_t)$ are computed. Serfling~\cite{Serfling:1963} fitted historical data of case-counts from influenza outbreaks with a linear trend and trigonometric functions (to account for their seasonality) to obtain estimates (and forecasts of) $(\mu_t, \sigma_t)$. A zero-mean Gaussian was assumed as a model for the fitting errors. The method is widely used and  over time the linear and periodic components have been adapted for local conditions and specific diseases~\cite{Pelat:2007}. For outbreaks with low counts, this approach has been modified to use Poisson error models, where the log-mean is modeled as a function of time, much like Serfling's method~\cite{Parker:1989,Jackson:2007}. Farrington's widely used method~\cite{Farrington:1996} parallels Serfling's approach, with linear and periodic trends, but the quasi-Poisson model accommodates the over-dispersion  observed in epidemiological surveillance data as ${\rm var}(y_t) = \phi \mu_t$, where $\phi$ is estimated from the data.
$(\mu_t, \sigma_t)$ have also been modeled and forecast using time-series model~\cite{Reis:2003} such as AutoRegressive Integrated Moving Average (ARIMA) but the surveillance time-series has to be first rendered stationary by subtracting out any trends and seasonality (which incurs errors). A comparison of ARIMA and SPC methods for detecting outbreaks showed that ARIMA methods were unremarkable in their ability to model surveillance data~\cite{Williamson:1999}, due to non-stationarity and sparsity. Outbreaks detection can also be modeled as state-transition events and thus based on Hidden Markov Models~\cite{LeStrat:1999} and Markov switching models~\cite{MartinezBeneito:2008,Conesa:2010,LuHM:2011}. Outbreak detection can also be formulated as a two-component model consisting of an endemic phase (modeled using a Poisson distribution) and an epidemic one (modeled using an autoregressive parameter). Both components are fitted to the data in a time-window around $t$ and a likelihood ratio test is used to evaluate which model fits better~\cite{06hh4a,Hohle:2008}. This can be used to detect when an epidemic starts. We will use such a  model~\cite{Hohle:2008} as a baseline in \S~\ref{sec:disc}.

Perhaps the investigations that are closest to ours, in modeling philosophy, are those by Lawson and collaborators\cite{10ls2a,23la1a,23kl6a}. Fundamentally, our approach consists of ``stitching together'' models meant for individual areal units\cite{Safta:2021, Blonigan:2021} via CAR models (specifically, the BYM model). Lawson and co-workers model case-counts directly, whereas we use a parametric model of a temporally-variable (and, in this paper, also spatially-variable) infection-rate field that is related to the case-counts via the incubation period distribution. The use of the incubation-period model (see \S~\ref{sec:form}) makes our model computationally more expensive than the ones used by Lawson and collaborators. Case-counts, in Lawson's formulation, are modeled using a Susceptible-Infected-Removed compartmental formalism with a one-lagged-in-time auto-correlation and a BYM CAR model to couple with adjoining areal units; the clearest description of the model is in Lawson and Song, 2010\cite{10ls2a}, which was applied to four counties in South Carolina. The same model was adapted to COVID-19 data from all counties of South Carolina\cite{21lk2a} and the UK\cite{21sl3a}. In an allied work, Lawson investigates, and selects between, various formulations of their basic model, as applied to COVID-19 data, with 1-day-ahead forecasting accuracy in mind; he finds no clear benefits between using a space-time versus a purely temporal model\cite{23la1a}. The group has also investigated, much like us, whether departures from forecasts could be used to detect anomalies within the context of epidemiological surveillance\cite{22lk2a,23kl6a}.They devised metrics such as the Surveillance Kullback-Liebler\cite{16rl2a} (SKL) and Surveillance Conditional Predictive ordinate\cite{11cl2a} (SCPO) to monitor and detect outlier epidemiological behavior. Lawson and Kim\cite{22lk2a} found that one needed to include a leading indicator/syndrome of epidemiological activity e.g., absenteeism, as a modeling covariate to detect epidemiological changes in a timely manner. A more methodologically-oriented paper\cite{23kl6a} investigated whether Poisson or Negative Binomial  (NB) distributions should be use to link the observed case-counts to the modeled values in a likelihood function. They found that the NB distribution provided better goodness-of-fits (perhaps because the two-parameter distribution is more flexible than Poisson) but for small datasets, Poisson provided more predictive forecasts. To summarize, one can use cases-counts directly for (spatio-temporal) model-based syndromic surveillance and there is some uncertainty over whether one should use Poisson or NB distributions to capture the stochasticity in the observation. However, the possibility of using a latent variable that might be better behaved, e.g., infection-rate, has not been investigated.

\section{Exploratory Data Analysis}
\label{sec:expl}
In this section we perform an exploratory data analysis on the COVID-19 data from New Mexico (NM), in order to design the spatial problem.

\subsection{The COVID-19 Dataset}

The COVID-19 dataset covers the duration from 2020-01-22 to 2022-05-13, and consists of daily (new) case-counts of COVID-19 from each of the 33 counties of NM; the data is available online.~\cite{nytcovid,jhucovid} The 73 covariates (i.e., risk factors) of COVID-19 span demographics, socioeconomic information (income, business and home ownership etc.) and infrastructure. These were obtained from another group in Sandia National Laboratories and is described in their publication~\cite{20sf5a}; we provide a summary below. Demographic data on age distribution, gender, racial orgins, housing, family units and living arrangements, education, health etc. were obtained from US Census Bureau's QuickFacts for New Mexico~\cite{QF:NM}, representing 5-year estimates between 2014-2018 and the 2013-2017 American Community Survey estimates. Geographical information e.g., area of counties,population densities etc. were also obtained from the Census dataset. Infrastructure represents the resources needed by a county to operate, such as number of COVID testing sites, nursing homes and K-12 schools.~\cite{Infra:NM,Infra:Screen} Geospatial data was also extracted from University of New Mexico Earth Data Analysis Center which develops the Resource Geographic Information System~\cite{Infra:RGIS}. In total, data was compiled from 40 sources, manually down-selected to 73 features and adjusted (when needed) to each county's population.

\subsection{Data Analysis}
\label{sec:datanl}

Let $Y_t = \{y_{t,1}, y_{t,2}, \ldots y_{t, \Nreg}\}$ be the vector of case-counts reported on day $t$ in each of the $\Nreg$ areal units (i.e., counties of NM). Let $Y^{\ast}_t = \{y^{\ast}_{t, 1}/p_1, y^{\ast}_{t, 2}/p_2, \ldots y^{\ast}_{t, \Nreg}/p_{\Nreg}\}$ be the vector of normalized cumulative case-counts over the duration $\left(t - 90, t\right]$ i.e., $y^{\ast}_{t, r}$ is the cumulative number of case-counts over the 90-day period $\left(t - 90, t\right]$ for areal unit $r$ and $p_r$ is the areal unit's population. The 90-day window is adopted to average out the effect of reporting errors, as well as to reduce the effect of low case-counts in some of the very sparsely populated desert counties of NM.  We assume that the case-counts can be modeled as a linear function of risk factors i.e., $Y^{\ast}_t \approx {v}_{0,t} + \left[{\mathbf Z}\right] {\mathbf v}_t$ where the $k^{th}$ column of ${\mathbf Z}$ contains the value of the $k^{th}$ risk factor for all $\Nreg$ areal units and ${\mathbf v}_t = \{v_{k,t}\}, k = 1 \ldots K$ are their relative weights in time-window $t$. The risk factors ${\mathbf Z}$ are constant in time but vary between areal units. In disease mapping terms, the model ${\mathbf v}_{0,t} + \left[{\mathbf Z}\right] {\mathbf v}_t$ provides the expected value of $Y^{\ast}_t$ and any deviations would be deemed ``random'', to be modeled statistically.

Some of the risk factors are very correlated and thus carry little independent information, and consequently we simplify the model via sparse Principal Component Analysis~\cite{20ez6a} (PCA) to a set of principal component $\left[{\mathbf \Phi}\right] = \{\phi_k\}$ to remove unnecessary risk factors i.e., $Y^{\ast}_t \approx {v}_{0,t} + \left[{\mathbf Z}\right] {\mathbf v}_t \approx {w}_{0,t} + \left[{\mathbf \Phi}\right] {\mathbf w}_t$. Note that the principal components $\phi_k$ from sparse PCA do \emph{not} form an orthogonal basis set. We see from the scree plot in Fig.~\ref{fig:scree} (in the Appendix) that $K = 10$ is sufficient to explain 95\% of the variation in $Y^{\ast}_t$. Further, sparse PCA constructs $\phi_k$  using the most important risk factors. The main components of the sparse PCA modes are percent elderly, affluence, medical institutions per capita, size of population, percent native American and percent male.

\begin{figure}[!ht]
\centerline{
\includegraphics[width=0.45\textwidth]{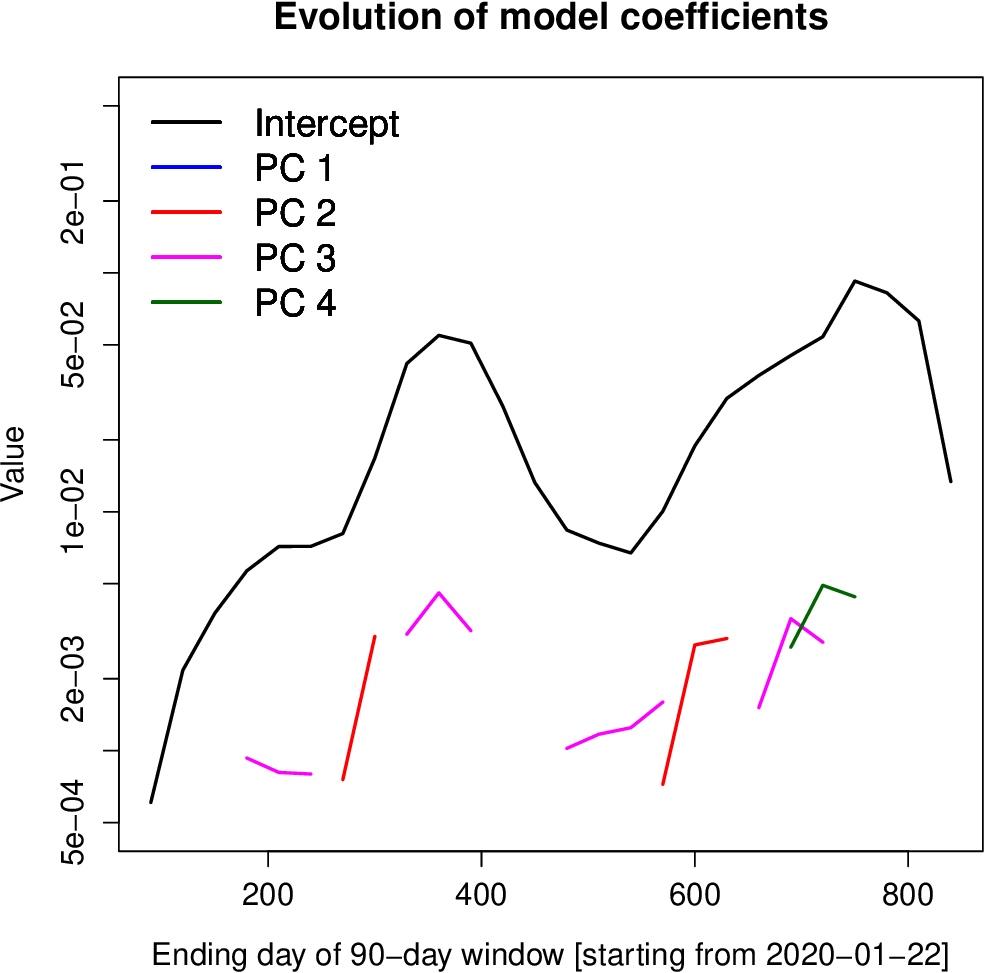}
\includegraphics[width=0.45\textwidth]{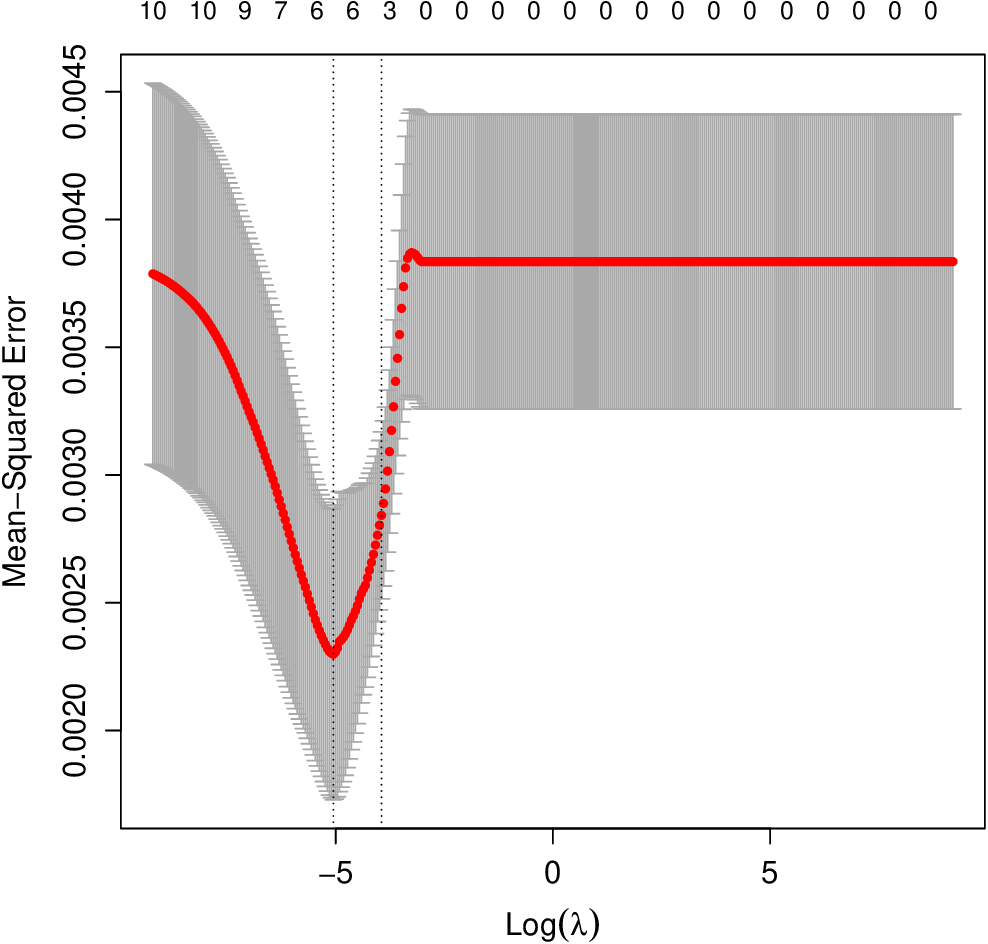}
}
\centerline{
\includegraphics[width=0.45\textwidth]{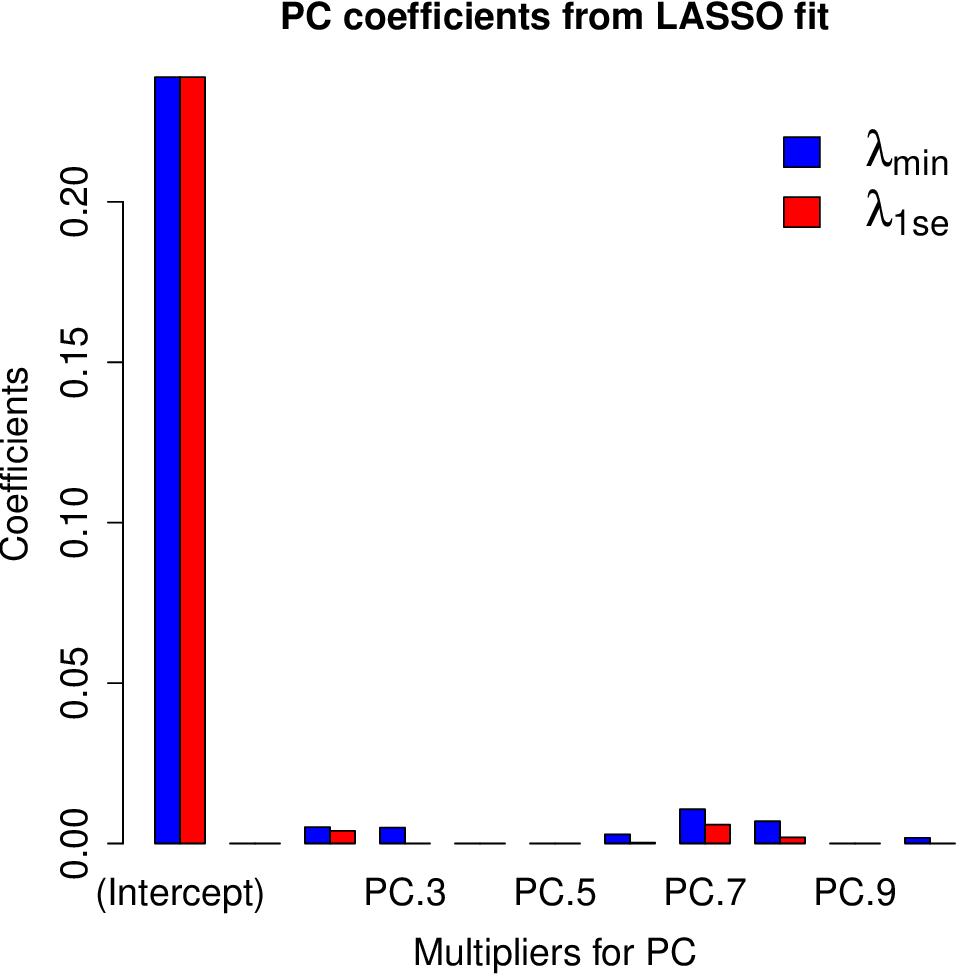}
\includegraphics[width=0.45\textwidth]{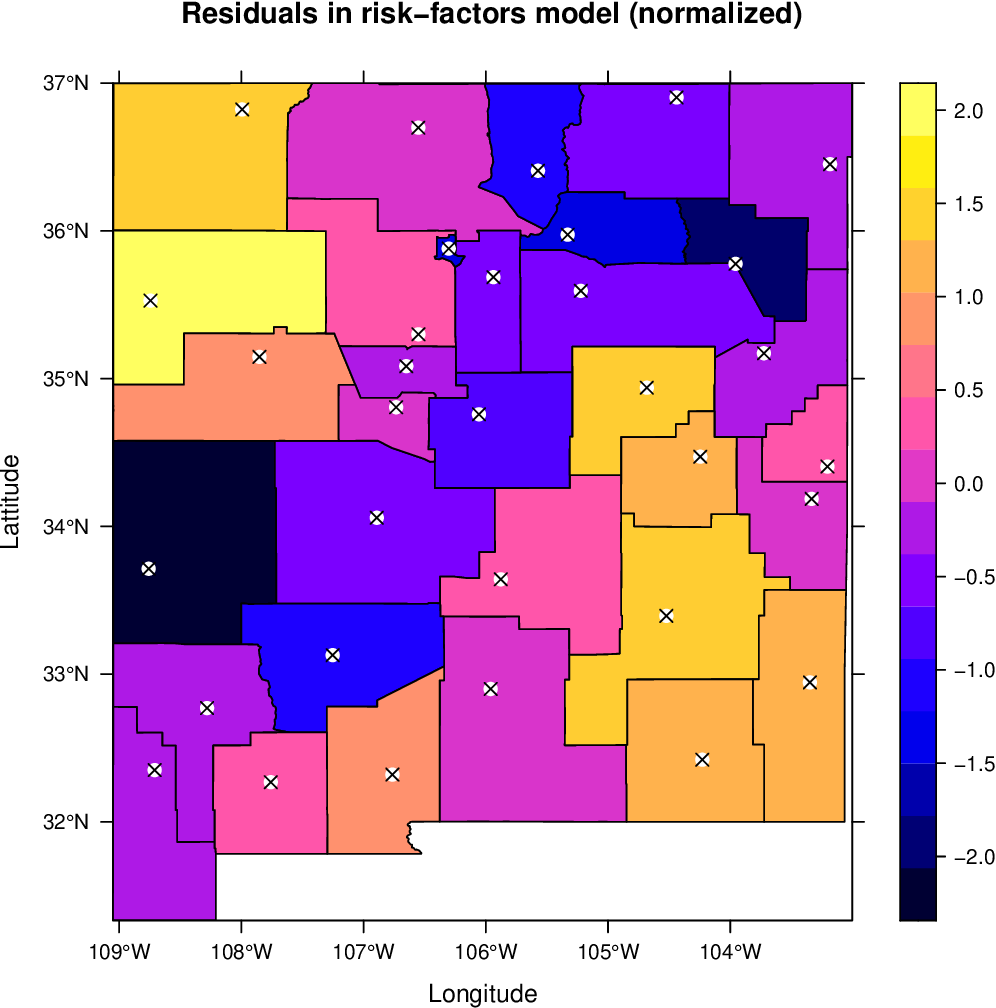}
}
\caption{Top left: Evolution of coefficients $w_{k,t}$ over time as the risk-factor model is fitted to cumulative case-counts $y_{t,r}$ normalized by county populations. Results are plotted for the intercept and four principal components (PC). Only the intercept survives and is far larger that the weights associated with the principal components. Top right: Plot of the prediction error from a 7-fold cross-validation performed with the risk-factor model and LASSO, on case-count data accumulated over the entire two-and-a-half-year duration (and normalized by county populations). The figures on the upper horizontal axis denotes the number of principal components retained in the fitted model. $\lambda_{min}$ and $\lambda_{1se}$ are clearly marked. Bottom left: Distribution of coefficients, corresponding to penalties $\lambda_{min}$ and $\lambda_{1se}$; the intercept dominates. Bottom right: The residuals from the risk-factors model i.e., the component not explained by the risk-factors model. The spatial correlations are clear.}
\label{fig:eda}
\end{figure}

We fit a regression model $Y^{\ast}_t =  {w}_{0,t} + \left[{\mathbf \Phi}\right] {\mathbf w}_t + {\mathbf \eta}, {\mathbf \eta} = \{\eta_r\}, r = 1 \ldots \Nreg, \eta_r \sim \Normal(0, \sigma^2)$ and simplify it with backward-forward stepwise elimination for each time window. New time-windows are obtained by advancing the previous one by 30 days. Fig.~\ref{fig:eda} (top left) plots the variation of the absolute values of the coefficients ${\mathbf w}$ over time. We see that the intercept $w_{0}$ dominates and persists over the entire duration, whereas the others are present only episodically, suggesting that the model might be fitting to noise. To investigate whether the risk factors play any part in the regression model, we take the cumulative sum of the case-counts over the entire duration of the dataset $Y^{\ast\ast}_t$ and fit $Y^{\ast\ast}_t =  {u}_{0,t} + \left[{\mathbf \Phi}\right] {\mathbf u}_t + \epsilon$ via LASSO. Fig.~\ref{fig:eda} (top right) shows the MSE as a function of the sparsity penalty $\lambda$ in LASSO; the digits along the upper horizontal axis plots the PCA modes retained as $\log(\lambda)$ is increased. The ``error bars'' show the variation in MSE as we undergo 7-fold cross-validation. We use the value of $\lambda_{1se}$ in our regression model (the second vertical dotted line in Fig.~\ref{fig:eda} (top right), where the mean MSE corresponds to 1 standard deviation away from the minimum MSE observed for $\lambda_{min}$). The coefficients ${\mathbf u}$ obtained from these two values of $\lambda$ are plotted in Fig.~\ref{fig:eda} (bottom left). It is clear that the intercept $w_0$ dominates i.e., the case-counts for COVID-19 are not very dependent on $\left[{\mathbf \Phi}\right]$ and $Y^{\ast\ast}_t \approx  {u}_{0,t} + {\mathbf \epsilon}$. The implication is that over the time-period of interest, the spatial patterns observed in $Y^{\ast\ast}_t$ were not explained by the spatially-variable risk factors. Fig.~\ref{fig:eda} (bottom right) plots the $z-$score of ${\mathbf \epsilon}$ and the spatial correlation of the epidemiological dynamics not modeled by risk factors is clear. There is a ``blue'' diagonal of NM counties running Northeast to Southwest, where as the Northwest and Southeast corners are yellow. In between are ``magenta'' counties. Note that much of the blue diagonal is along the Rio Grande valley, and the population density falls as we travel away from it, into the desert. Clearly, a neighborhood matrix $W$ for a GMRF model could be made from this data, and we address this next. Note that this spatial variation is \emph{not} explained by risk factors, but perhaps is due to mixing of populations in the counties.

Moran's $I-$statistic test~\cite{18bw2a} is used to detect spatial autocorrelation in a variable defined over areal units. It requires an adjacency matrix $W$ between areal units as input. We consider three different definitions of $W$ \emph{viz.} ``binary'' where $w_{ij} = 1$ when areal units $i$ and $j$ share a border (i.e., they are immediate neighbors), ``binary-modified`` where $w_{ij}$ is weighed by the reciprocal of the distance between adjacent counties' county seat and ``row-standardised`` where $w_{ij}$ is weighed by the number of neighbors that areal unit $i$ has. Moran's $I-$statistic is computed with the ${\mathbf \epsilon}$ that is provided to the test (``observed $I-$statistic'') versus the null case where the elements of ${\mathbf \epsilon}$ are IID. The figure of merit is the standard deviate of the observed $I-$statistic. The standard deviate of the ${\mathbf \epsilon}$ shown in Fig.~\ref{fig:eda} (bottom right) is in Table~\ref{tab:Moran}, top row; clearly it is far from being IID random. Thereafter, we perform the same Moran's $I-$statistic test for the 90-day windows (Fig.~\ref{fig:eda} (top left)) and tabulate the mean and standard deviation of the the $I-$statistic in Table~\ref{tab:Moran}, bottom row; again, the $I-$statistic indicates significant spatial auto-correlation. We see that the ``binary'' and ``row-standardised'' versions of the adjacency matrix give similar results and they are both far superior to the ``binary-modified`` form of $W$. The computation was repeated with an adjacency matrix with a 2-hop neighborhood (where the immediate neighbors of an areal unit, and their immediate neighbors, were included in the adjacency matrix) and the $I-$statistic was indistinguishable from random ${\mathbf \epsilon}$. Henceforth, we will adopt the row-standardised form of $W$ as our spatial prior as we estimate the infection-rate field over multiple areal units, as it provides the largest  standard deviate of Moran's $I-$statistic.

\begin{table}
\begin{center}
\caption{Standard deviate of the $I-$statistic of the observed data with different adjacency matrices. In the second row, we tabulate the mean standard deviate over all windows; the number in parenthesis is the standard deviation.}
\label{tab:Moran}
\begin{tabular}{lccc}
\toprule
\textbf{Test case}                    & Binary $W$   & Binary-modified $W$ & Row-standardised $W$ \\
\midrule
Cumulative cases for the full dataset & 3.44      & 2.76            & 3.57 \\
90-day windows                        & 2.5 (1.1) & 2.08 (0.8)      & 2.7 (1.35)  \\
\bottomrule
\end{tabular}
\end{center}
\end{table}

\section{Formulation}
\label{sec:form}
Next, we propose an epidemiological model to forecast infection rates across adjacent geographical regions. The model is an extension of previous work by Safta~\etal\cite{Safta:2021} and Blonigan~\etal\cite{Blonigan:2021} for epidemic forecasts over a single region to multiple regions. In this section we will briefly describe the single region model and then present statistical approaches to estimate the model parameters over adjacent geographical regions.

\subsection{Epidemiological Model}
\label{sec:epimod}
The epidemiological model combines an infection-rate model and an incubation rate model. In a given areal unit $r$, the infection rate is assumed to follow a Gamma distribution (in time) with a probability density function (pdf) given by
\begin{equation}
f_{inf}(t;\kr,\thetar)=\thetar^{-\kr}t^{\kr-1}\exp(-t/\thetar)\big/\Gamma(\kr).
\label{eq:infrate}
\end{equation}
The infection-rate in Eq.~\eqref{eq:infrate} is controlled by two parameters, $\kr$ (shape) and $\thetar$ (scale), and is sufficiently flexible to capture a range of outbreaks. The third parameter, $\tzr$, represents the start of the outbreak and will be inferred jointly with the infection rate parameters. For incubation we employ a model calibrated against early COVID-19 data~\cite{Lauer:2020}. This model follows a lognormal distribution with a cumulative distribution function (CDF) given by
\begin{equation}
F_{inc}(t;\mu,\sigma)=\frac{1}{2}\mathrm{erfc}\left(-\frac{\log t-\mu}{\sigma\sqrt{2}}\right)
\label{eq:incmod}
\end{equation}
Note that $\mu$ and $\sigma$ are \emph{not} constants, but are random variables themselves.
The mean $\mu$ is approximated as a Student's $t-$distribution and $\sigma$ is assumed to have a chi-square distribution. These choices result in 95\% confidence intervals of $\left[1.48,1.76\right]$ and $\left[0.320,0.515\right]$ for $\mu$ and $\sigma$, respectively, as described in Safta~\etal~\cite{Safta:2021}. We will refer to this model as the \emph{stochastic incubation model}.

The cumulative number of people that have turned symptomatic between time $\tzr$ (the start of the current epidemic wave) and time $t_i$ is computed as a convolution between the infection rate and the CDF of the incubation model
\begin{equation}
\Nir = \Nr \int_{\tzr}^{t_i} f_{inf}(\tau-t_0; \kr,\thetar) F_{inc}(t_i-\tau;\mu,\sigma)d\tau,
\label{eq:symptN}
\end{equation}
where $\Nr$ is the total number of people that will get infected (and counted) during the entire epidemic wave in areal unit $r$. This model assumes that a person shows symptoms once the virus incubation has completed. Furthermore, once symptoms are evident, it is also assumed that individuals have prompt access to medical services or otherwise self-report the COVID-19 infection, getting counted without delay. These assumptions will be relaxed in future versions of this effort where the model above will be endowed with latent variables that account for uncertainties due to reporting delays and unreported positive counts.

The number of people that turn symptomatic over the time interval $[t_{i-1},t_i]$, in areal unit $r$, is estimated as
\begin{eqnarray}
\nir = \Nir - \Nimr  & = & \Nr \int_{\tzr}^{t_i} f_{inf}(\tau - \tzr; \kr,\thetar)
\left( F_{inc}(t_i-\tau;\mu,\sigma) - F_{inc}(t_{i-1}-\tau;\mu,\sigma) \right)d\,\tau \label{eq:sympt} \\
&\approx& \Nr(t_i - t_{i-1}) \int_{\tzr}^{t_i} f_{inf}(\tau-t_0; \kr, \thetar)
f_{inc}(t_i-\tau;\mu,\sigma)d\tau
\label{eq:symptApprox}
\end{eqnarray}
where $f_{inc}$ is the pdf of the incubation model. In transitioning from Eq.~\eqref{eq:sympt} to Eq.~\eqref{eq:symptApprox} we made use of the approximation
\[
f_{inc}(t_i-\tau;\mu,\sigma)\approx\frac{F_{inc}(t_i-\tau;\mu,\sigma)-F_{inc}(t_{i-1}-\tau;\mu,\sigma)}{t_i-t_{i-1}}
\]
which amounts to approximating the incubation model PDF with a histogram with bin of size $(t_i-t_{i-1})$. Thus the four parameters that describe the epidemiological dynamics in an areal unit $r$ are $\gr = \{\kr, \thetar, \tzr, \Nr\}$ and $\bgamma = \{\gr\}$ is the accumulation of parameters over all $\Nreg$ areal units. We will refer to them colloquially as the ``epidemiological'' parameters. In this paper we focus on outbreak detection and for this purpose a model that follows a single wave, as above, is sufficient for the task. Given the assumptions above, these outbreak forecasts represent a lower bound on the actual number of people that are infected with COVID-19. A fraction of the population infected with a novel disease might also exhibit minor or no symptoms at all and might not seek medical advice, further contributing to lowering the predicted counts compared to the actual size of the epidemic.

\subsection{Model Calibration}
\label{sec:binf}
Given data in the form of time-series of daily counts, labeled generically as $\bm{Y}$, as shown in
\S\ref{sec:datanl}, and the model predictions $\bm{n}$ for the number of new symptomatic counts daily, presented in \S\ref{sec:epimod}, we will employ a Bayesian framework to calibrate the epidemiological model parameters. The discrepancy between the data and the model is written as
\begin{equation}
{\bm Y} = {\bm n}(\bm{p})+\epsilon(\bm{p})
\end{equation}
where ${\bm p}$ are the parameters that describe both the epidemiological models and the statistical discrepancy $\epsilon$ between the data and the epidemiological model. These parameters will be detailed in the following sub-sections. The probabilistic error model
encapsulates both errors in the observations, e.g. availability of testing capabilities and test accuracy, as well as errors due to empirical modeling choices.

The multivariate distribution for the vector of parameters $\bm{p}$ can be estimated in a Bayesian framework as
\begin{equation}
P(\bm{p}\vert {\bm Y})\propto P({\bm Y}\vert\bm{p}) P(\bm{p})
\label{eq:bayes}
\end{equation}
where $P(\bm{p}\vert {\bm Y})$ is the posterior distribution we are seeking
after observing the data ${\bm Y}$, $P({\bm Y}\vert\bm{p})$ is the likelihood
of observing the data ${\bm Y}$ given a specific choice for parameters $\bm{p}$, and
$P(\bm{p})$ contains the prior information about the models parameters. The subsections below provide a detailed description about the setup of the likelihood and prior distributions.

\subsubsection{Likelihood Construction with Spatial Correlations}
\label{sec:lik}

We now derive a likelihood expression $\mathcal{L}_{\mathcal{D}}$ which accounts for the discrepancies between the number of people reported symptomatic daily and the number of new cases predicted by the model, via Eq.~\eqref{eq:symptApprox}. We denote the reported daily count $Y_i^{(o)}=\{y_{i,1},y_{i,2},\ldots,y_{i,\Nreg}\}$ for day $i$, and the daily predicted count $Y_i^{(p)} = \{n_{i,1},n_{i,2},\ldots,n_{i,\Nreg}\} = \Mcal(t_i; \bgamma)$, where $\Mcal(t_i; \bgamma)$ is the epidemiological model described in Eq.~\ref{eq:symptApprox}, with $\bgamma$ constituting the epidemiological parameters over $\Nreg$ regions, some of which might be adjacent. $\bgamma$ are the parameters that will be jointly inferred given the available data.

For a given data $i$, we state
\begin{equation}
Y_i^{(o)} = Y_i^{(p)} + \bveps_i = \Mcal(t_i; \bgamma) + \bveps_i, \bveps_i \sim \Normal \left(0, \Sigma_i\right),
\label{eqn:obs}
\end{equation}
i.e., we assume that the data -- model mismatch is a multivariate Gaussian distribution with a block covariance matrix. We will assume that the discrepancies are independent over the temporal axis and correlated in space, i.e.
\begin{equation}
\mathcal{L}_{\mathcal{D}}=\prod_{i=1}^{N_d}\frac{1}{(2\pi)^{N_r/2}
\mathrm{det}(\Sigma_i^{1/2})}\exp\left(-\frac{1}{2}(Y^{(o)}_i-Y^{(p)}_i)
\Sigma_i^{-1}(Y^{(o)}_i-Y^{(p)}_i)^T\right)
\label{eq:lik}
\end{equation}
Here $\Sigma_i$ is the block in the large covariance matrix (that spans over $N_d$ days of observations) that corresponds to the predictions for day $i$. Per the BYM model, we will model the discrepancy $Y^{(o)}_i-Y^{(p)}_i = \bveps_i$ with two components i.e., $\bveps_i = \bveps_{i,1} + \bveps_{i,2}$. Per Fig.~\ref{fig:eda} (bottom right), $\bveps_{i,1}$ will be modeled with a pCAR to capture spatial auto-correlation. In contrast $\bveps_{i,2}$ models random, temporally independent, reporting errors and any model shortcomings. Consequently the $Y^{(o)}_i-Y^{(p)}_i = \bveps_i$ discrepancy is modeled as the product of two independent, zero-mean multivariate Gaussian components~\cite{Cressie:2008}, with a resulting in a joint covariance matrix given by
\begin{equation}
\Sigma_i = P^{-1}+\mathrm{diag}\left(\sigma_a+\sigma_m Y^{(p)}_i\right)^2,
\label{eqn:covmat}
\end{equation}
where $P$ is the precision matrix associated with the Gaussian Markov Random Field (GMRF) model  assumed to account for the spatial correlations between adjacent regions (a proper Conditional Auto-Regressive (pCAR) model\cite{MacNab:2022}). We will refer to the parameters $\bsigma = \{\sigma_a, \sigma_m\}$ as the ``error model'' (or ErrM). The precision matrix $P$ is defined as
\begin{equation}
P=\frac{1}{\tphi^2}\left(\mathrm{diag}\{g_1,g_2,\ldots,g_{N_r}\}-\lphi W\right)
\end{equation}
Here, $g_j$ is the number of regions adjacent to region $j$, and $W$ is a matrix that encodes the relative topology of the regions considered in the joint inference, with entries defined as
\begin{equation}
w_{jj}=0\,\textrm{and}\,
w_{jk}=\begin{cases}
1 & \textrm{if regions {\it j} and {\it k} are adjacent,}\\
0 & \textrm{otherwise.}
\end{cases}
\end{equation}
Thus $P$ defines a pCAR spatial model with row-standardisation and is a function of the ``spatial coefficients'' (or SpC) $\bpsi = \{\tphi^2, \lphi\}$, which will also have to be estimated from the data. The inclusion of $\bpsi$ implies that the epidemiological parameters $\bgamma$ will display spatial correlation. The magnitude of the correlation is unknown \emph{a priori}, and will be estimated from the case-count data.

To summarize, the accuracy of the spatiotemporal model for epidemiological dynamics is controlled by the parameters $\bp = \{\bgamma, \bsigma, \bpsi\}$, which will be the object of inference from data from $\Nreg$ NM counties. The dimensionality of the inverse problem scales with $\Nreg$ and is limited by the scalability of the inversion method. We will use $\Nreg = 3$ and consider inferences using the following setups:
\begin{itemize}
\item independent inferences (i.e., $\Nreg = 1$), county by county, for the counties of Bernalillo, Santa Fe, and Valencia.
\item two adjacent counties (i.e., $\Nreg = 2$), i.e. Bernalillo \& Santa Fe and Bernalillo \& Valencia. For these cases the covariance matrix $P^{-1}$ corresponding to the GMRF model is given by
\begin{equation}
P^{-1} = \frac{\tphi^2}{1-\lphi^2}
\begin{bmatrix}
1 & \lphi  \\
\lphi  & 1 \\
\end{bmatrix}
\end{equation}
\item three counties (i.e., $\Nreg = 3$), Bernalillo, Santa Fe, and Valencia, jointly. Bernalillo is adjacent to the other two counties but Santa Fe and Valencia do not share a border. The GMRF covariance matrix $P^{-1}$ is given by
\begin{equation}
P^{-1} = \frac{\tphi^2}{2\left(1-\lphi^2\right)}
\begin{bmatrix}
1 & \lphi & \lphi \\
\lphi & 2-\lphi^2 & \lphi^2 \\
\lphi & \lphi^2 & 2-\lphi^2 \\
\end{bmatrix}
\end{equation}
\end{itemize}

\subsubsection{Prior Distributions}
\label{sec:prior}

We employ uninformative priors for the shape and scale parameters, $k_r$ and $\theta_r$, of the infection rate models, in Eq.~\eqref{eq:infrate}. We also employ an uninformative prior for the total count of infected people during the pandemic $N_r$. From our previous work~\cite{Safta:2021,Blonigan:2021} we observed that the convolution model in Eqs.~\eqref{eq:symptN}-\eqref{eq:symptApprox} exhibit sharp transitions when the inferred start time $t_0$ is not well constrained by the data, e.g. in situations where the daily counts are noisy in the low single digits. For this purpose for $t_0$ we selected a Gaussian distribution with a wide enough standard deviation, e.g. 10 days, to allow the data to easily overcome this prior when the number of counts increases beyond the low single digits count.

Further, to ensure the discrepancy model parameters, $\sigma_a$ and $\sigma_m$, are automatically positive, we work with their natural logarithm in the Bayesian framework. Consequently, the equivalent uninformative prior for the logarithm of standard deviations, $\log\sigma_a$ and $\log\sigma_m$, is the uniform distributions. For both these parameters, we bound the natural logarithms' values to $[-30,10]$, a range sufficiently wide to account for the discrepancies between model predictions and observations, while preventing numerical underflow/overflow errors during MCMC sampling.

For the parameters controlling the pCAR model, we employ a Gamma distribution with shape $10$ and scale $2$, $\Gamma (10,2)$, for $\tphi$ and a uniform distribution $U(0,0.9)$ for $\lphi$ following Shand \etal~\cite{20sf5a}

\subsubsection{Sampling the Posterior Distribution}
\label{sec:mcmcspl}

As in our previous work on epidemiological models~\cite{Safta:2021,Blonigan:2021}, we employ a Markov Chain Monte Carlo (MCMC) algorithm is used to sample from the posterior density $p(\bm{p}\vert\bm{Y})$, specifically the adaptive Metropolis (AMCMC) algorithm~\cite{Haario:2001}. To accommodate the stochastic incubation model (Eq.~\eqref{eq:incmod}), we employ an unbiased estimate of the likelihood presented in Eq.~\eqref{eq:lik}. For each MCMC step we select a random set of $(\mu,\sigma)$ for the incubation model according to their prescribed distributions, then run the epidemiological model to generate $\bm{Y}^{(p)}$ and estimate the likelihood. This approach is similar to the pseudo-marginal MCMC algorithm~\cite{Andrieu:2009} guaranteeing that the resulting samples correspond to the unbiased posterior distribution model. We use the Effective Sample Size (ESS)~\cite{Kass:1998} estimate to gauge the number of samples sufficient to describe the posterior distribution given the data available. For the results presented in this paper, we found that $1$ to $2$ million MCMC samples were needed to extract $5$K-$10$K effective samples required to estimate summary statistics and marginal distributions for the epidemiological models' parameters.

\subsubsection{Diagnostics}
\label{sec:diag}
The sampling process described in \S~\ref{sec:mcmcspl} yields ${\rm O(10^6)}$ samples of $\bp = \{\tzr, \kr, \Nr, \thetar, \tphi^2, \lphi\}$ from the posterior probability density function (PDF) and the question arises regarding how we assess the accuracy/predictive skill of the PDF. Primarily, we will use \emph{posterior predictive tests}, whereby we will select 100 samples from the posterior PDFs and use Eq.~\ref{eqn:obs} to predict case-counts. These forecasts will be limited to 14 days, beyond which, as described in our previous papers\cite{Safta:2021}, the model is not expected to be predictive. Fundamentally, observations up to time $t$ contains information about epidemiological dynamics up to time $t - \Delta$, $\Delta$ being a measure of the incubation period; after that, an increasing fraction of the infected people have yet to show symptoms and appear in the case-counts. Using the mean incubation period plus twice the standard deviation as an estimate for $\Delta$ (Eq.~\ref{eq:incmod}), we get $\Delta \approx \exp(1.76 + 2\times 0.515) = 16.3$ days, and so we curtail forecasting at a 2-week horizon. These forecasts are compared with the observed case-counts, and in case of a mismatch, the epidemiological dynamics are assumed to have changed after time $t$. Apart from forecasting, the correlation structure in the $\bp$ samples can be informative. For each of the areal units of interest, we plot 2D marginal plots (in the Appendix) and, in \S~\ref{sec:sda}, perform \emph{grouped} statistical dependence analysis to uncover how parameters for each areal unit vary with those from other areal units, or with global parameters such as $\{\lphi, \tphi^2. \sigma_a, \sigma_m\}$.

\section{Results}
\label{sec:res}
\begin{figure}[!htb]
\centering
\includegraphics[width=0.45\textwidth]{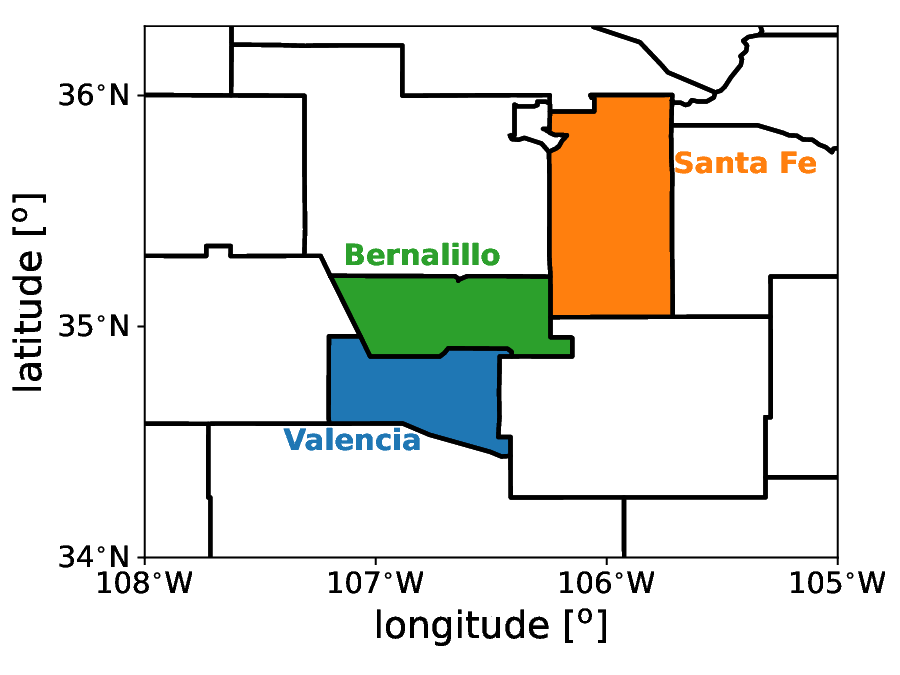}
\caption{The geographical extent of three adjacent New Mexico counties considered in this paper: Bernalillo (in green), Santa Fe (in orange) and Valencia (in blue).}
\label{fig:geoBSFV}
\end{figure}

Our use of AMCMC\cite{Haario:2001} (which is not very scalable when coupled with a moderately computationally expensive model) limits us to 10-15 dimensional posterior distributions. For this reason, we limit our study to three regions, $\Nreg = 3$, for a total of 16 parameters, i.e. $4$ parameters for each region, and $4$ parameters to describe the error model and correlations between regions. We selected three NM counties, Bernalillo, Santa Fe and Valencia, shown in Fig.~\ref{fig:geoBSFV}, as this allows to understand whether the adjacency between counties plays a role in the model calibration. Bernalillo is sandwiched between the other two counties and thus shares boundaries with the other two; while Santa Fe and Valencia do not share boundaries.

\subsection{Markov Chain Monte Carlo results}
\begin{figure}[t]
\centerline{
\includegraphics[width=0.3\textwidth]{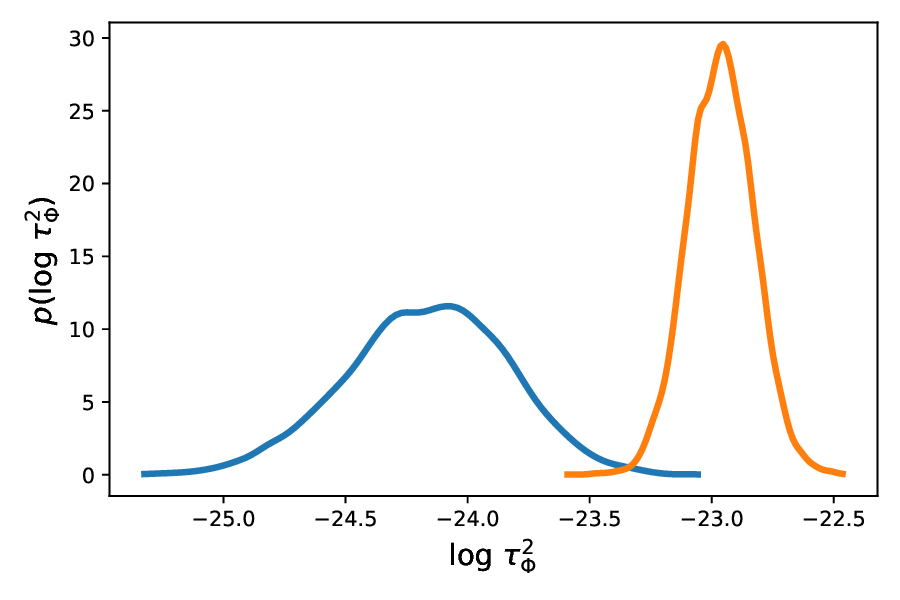}
\includegraphics[width=0.3\textwidth]{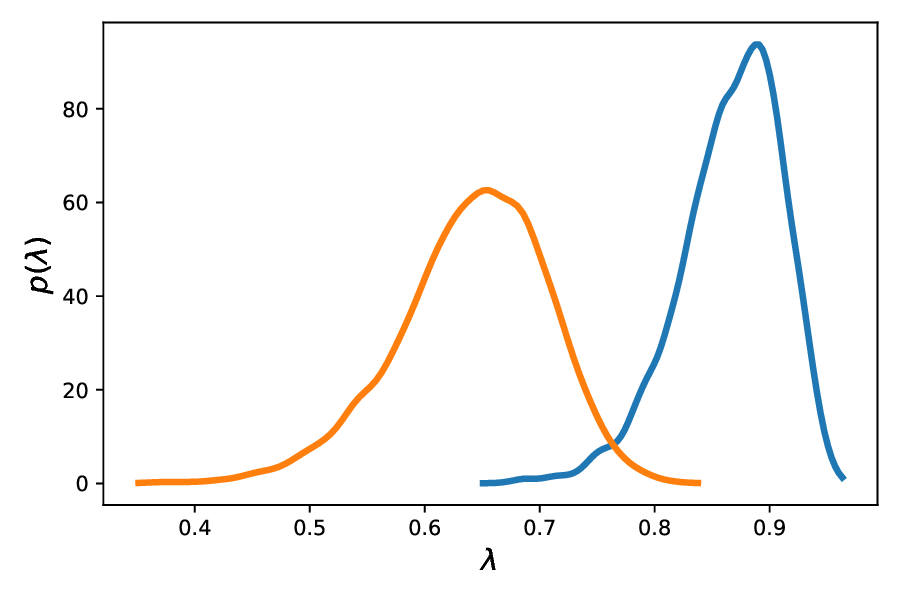}
}
\centerline{
\includegraphics[width=0.3\textwidth]{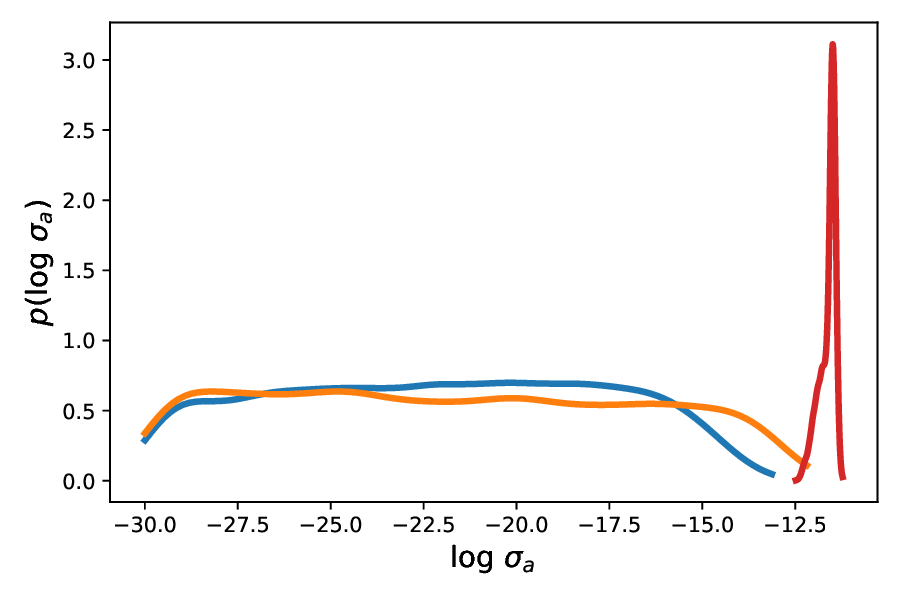}
\includegraphics[width=0.3\textwidth]{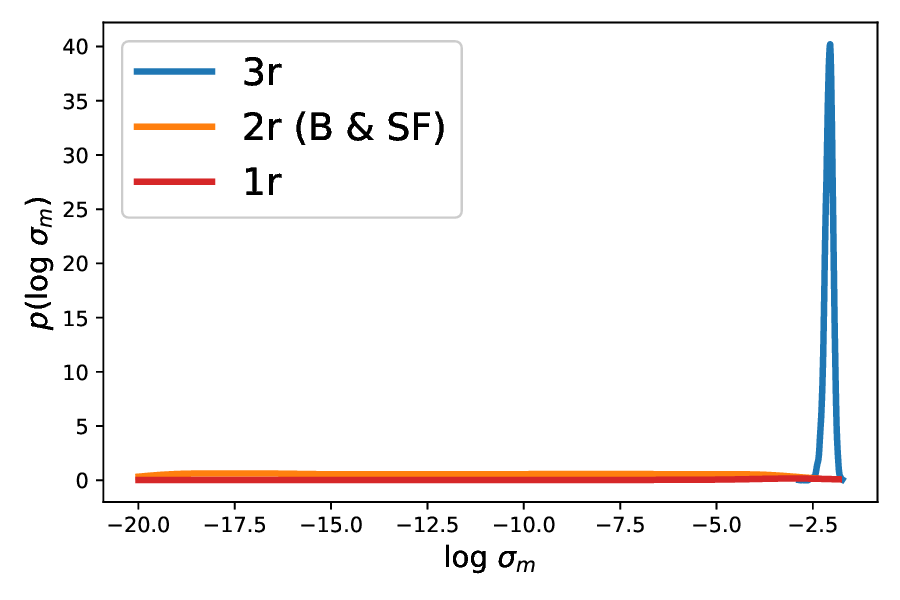}
}
\caption{Marginal posterior distributions GMRF parameters $(\tphi^2, \lphi)$ (top row) and noise parameters $(\sigma_a, \sigma_m)$ (bottom row), estimated via $2r$ and $3r$ joint estimations with data for Santa Fe.} \label{fig:SF_marg1D_NoiseSpace}
\end{figure}

In this section we will discuss summaries given samples from the posterior distributions sampled via MCMC. We first compare posterior results obtained for 1-, 2-, and 3-region statistical inference runs and the examine their impact on quality of model predictions vs the available observations.
\begin{figure}[!htb]
\centerline{
{\bf Bernalillo}\hfill
{\bf Santa Fe}\hfill
{\bf Valencia}
}
\centerline{
\includegraphics[width=0.33\textwidth]{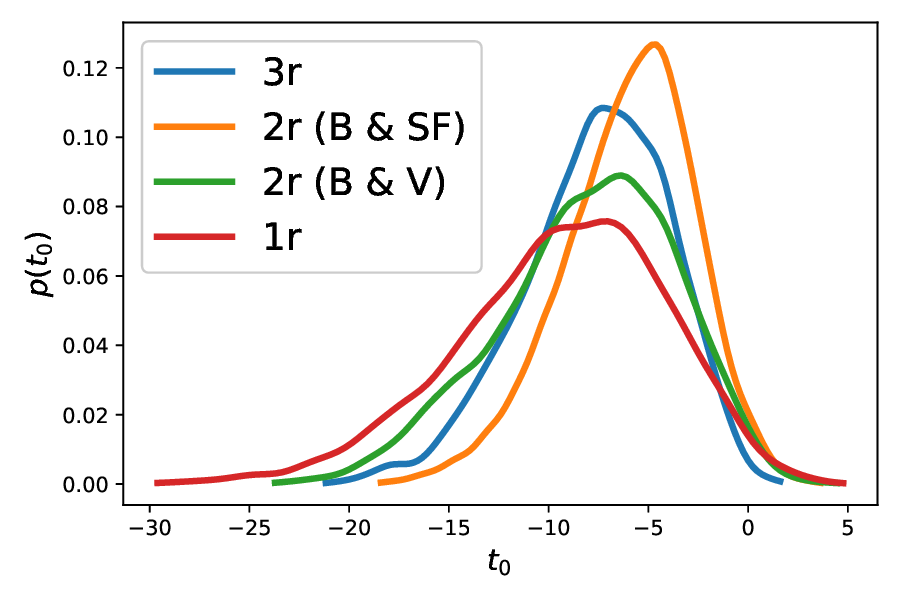}
\includegraphics[width=0.33\textwidth]{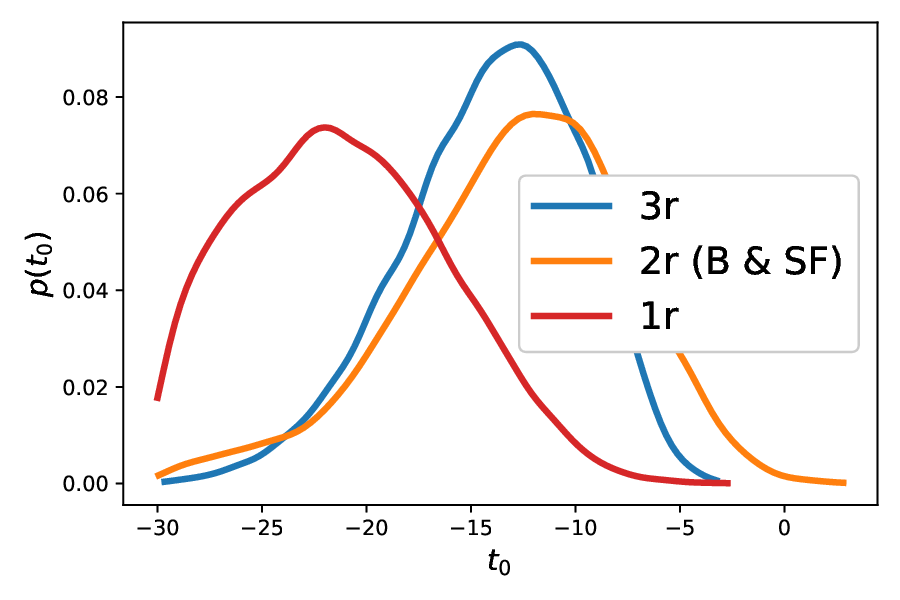}
\includegraphics[width=0.33\textwidth]{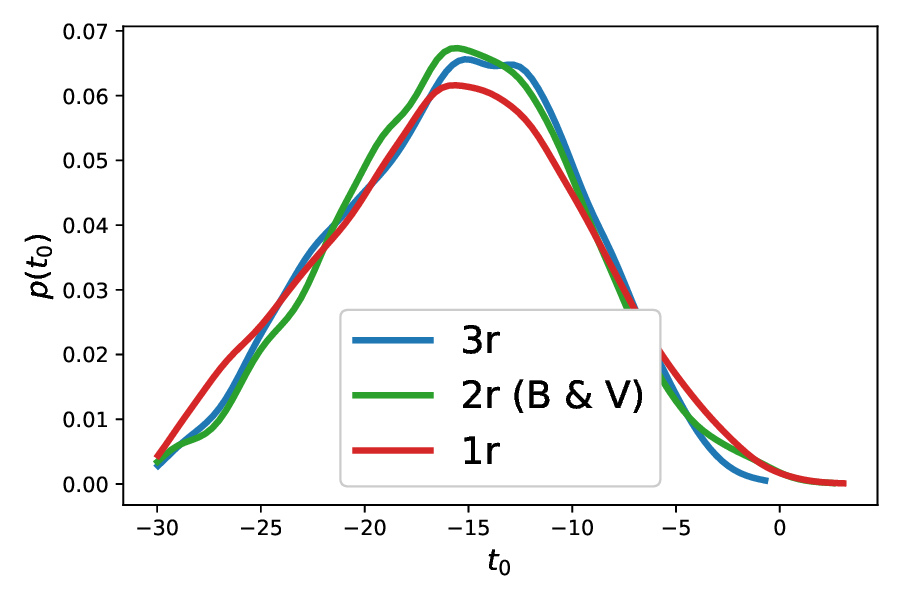}
}
\centerline{
\includegraphics[width=0.33\textwidth]{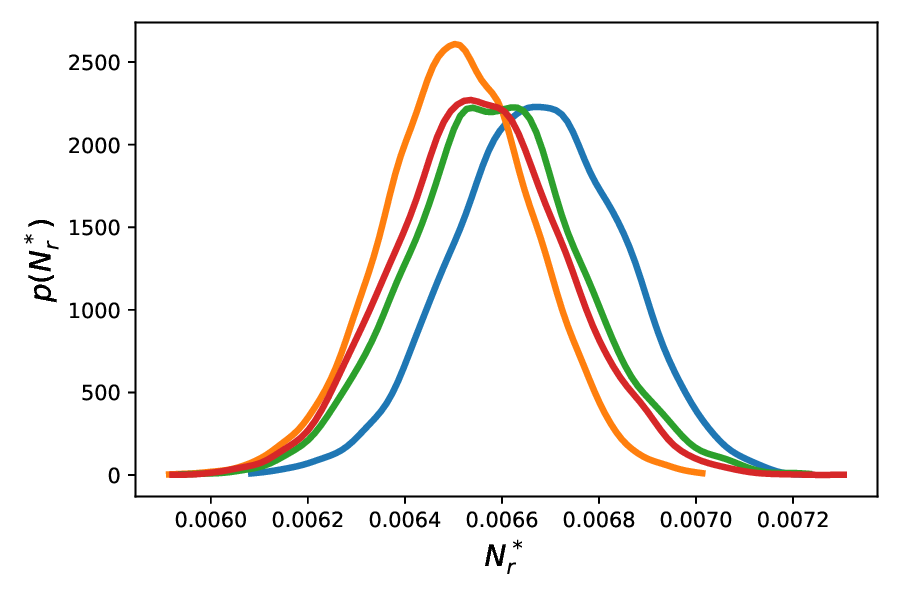}
\includegraphics[width=0.33\textwidth]{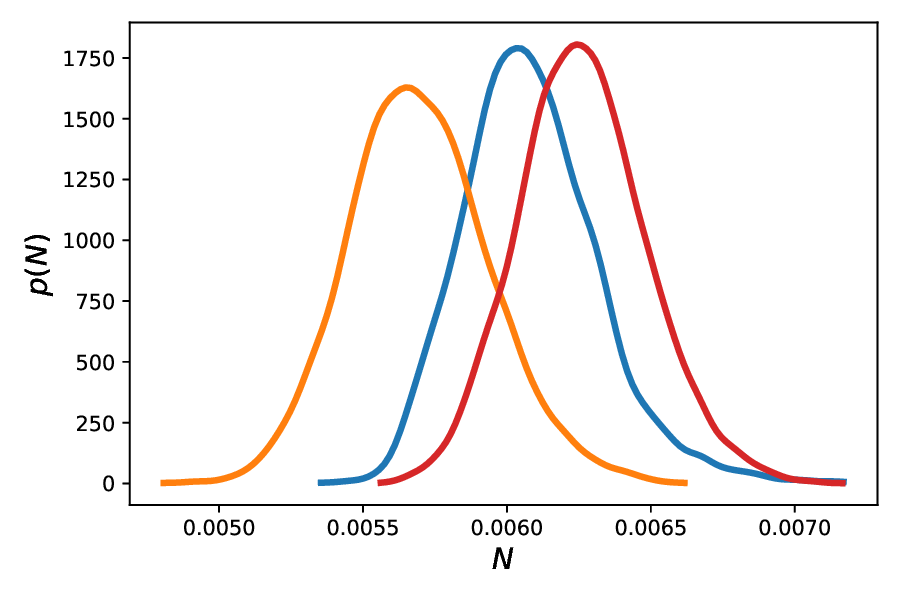}
\includegraphics[width=0.33\textwidth]{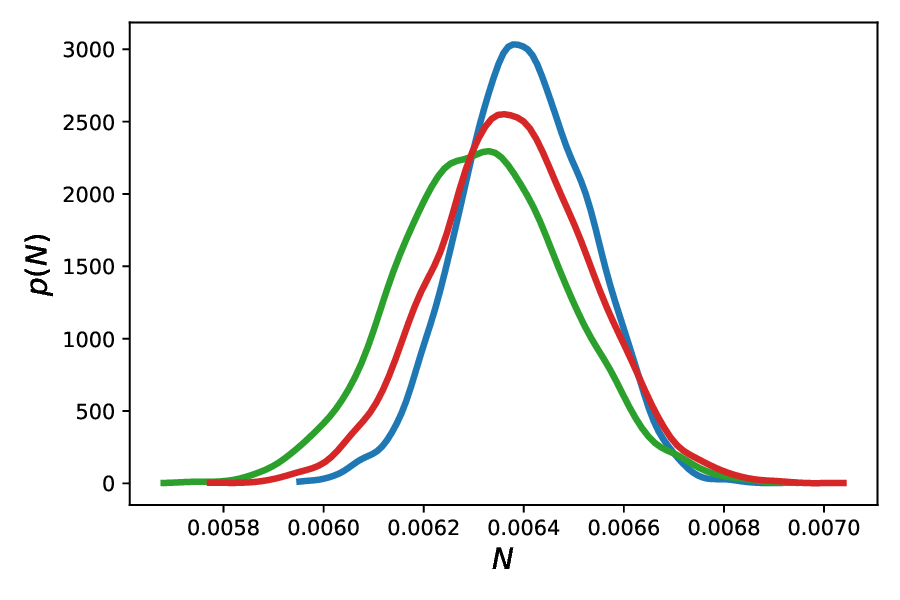}
}
\centerline{
\includegraphics[width=0.33\textwidth]{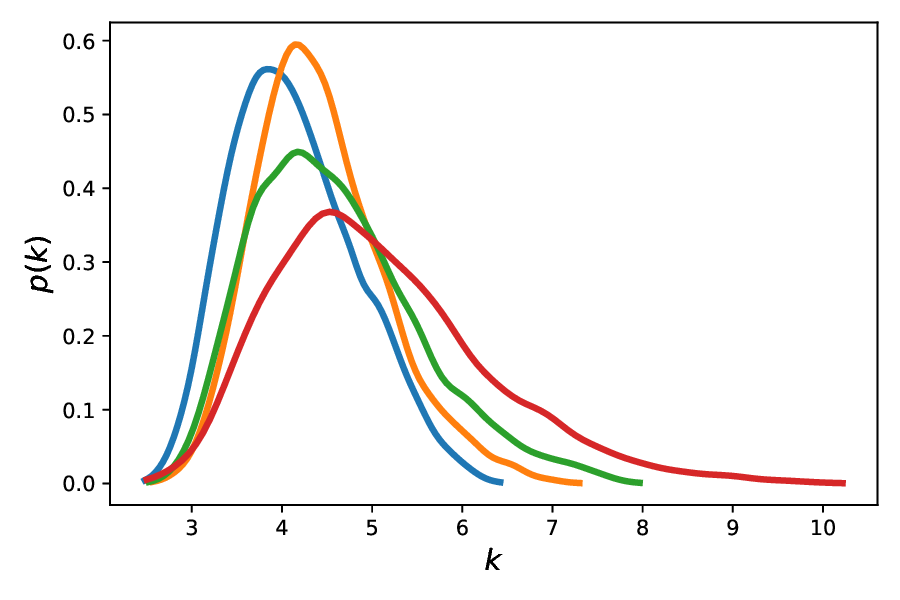}
\includegraphics[width=0.33\textwidth]{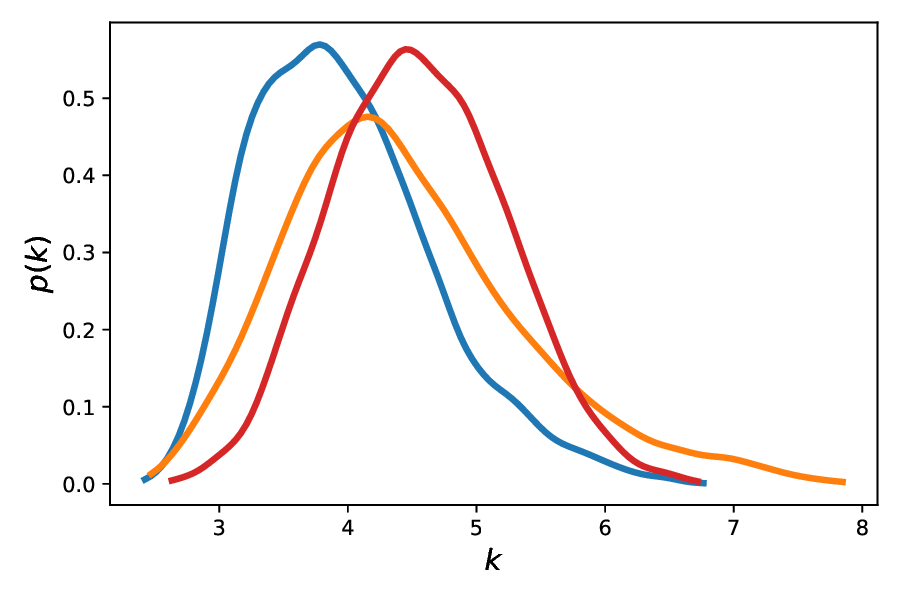}
\includegraphics[width=0.33\textwidth]{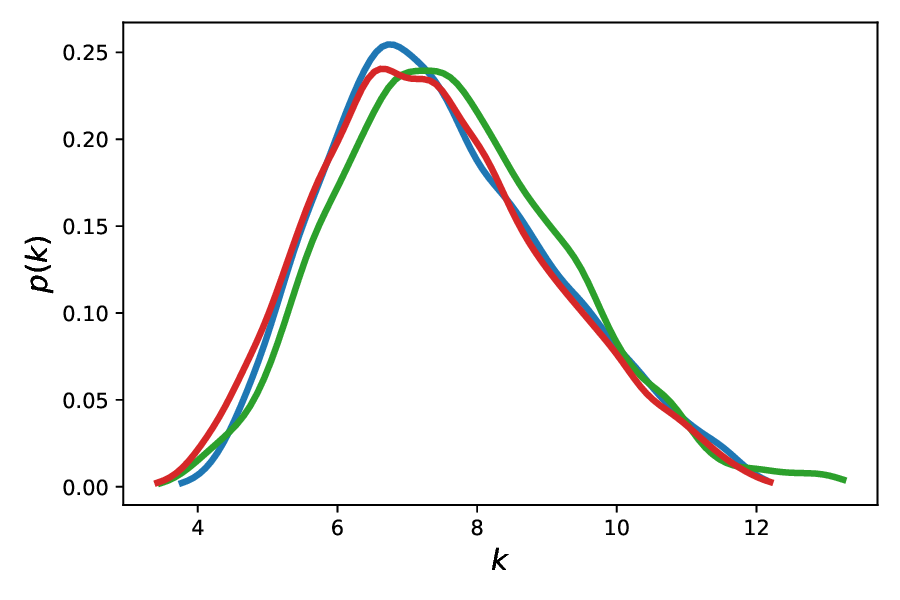}
}
\centerline{
\includegraphics[width=0.33\textwidth]{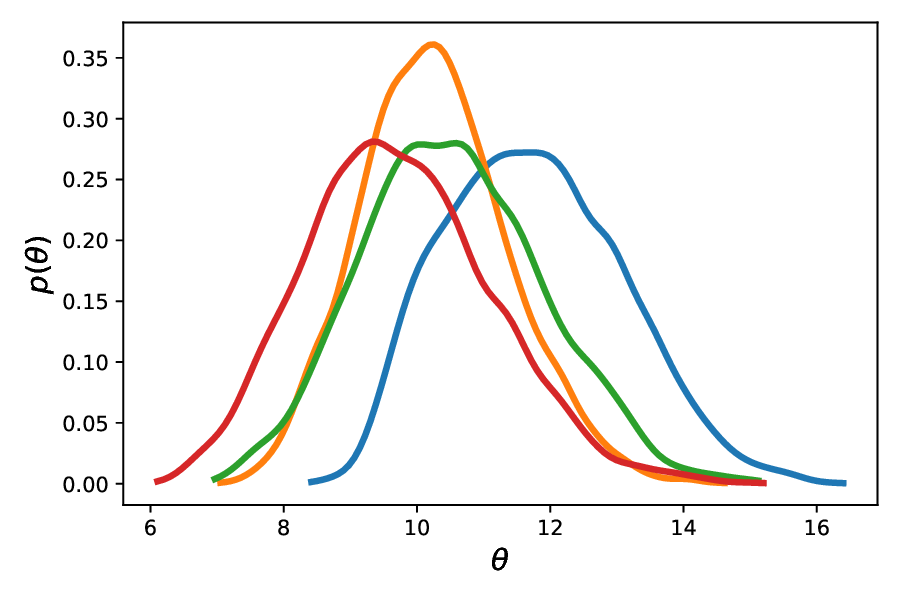}
\includegraphics[width=0.33\textwidth]{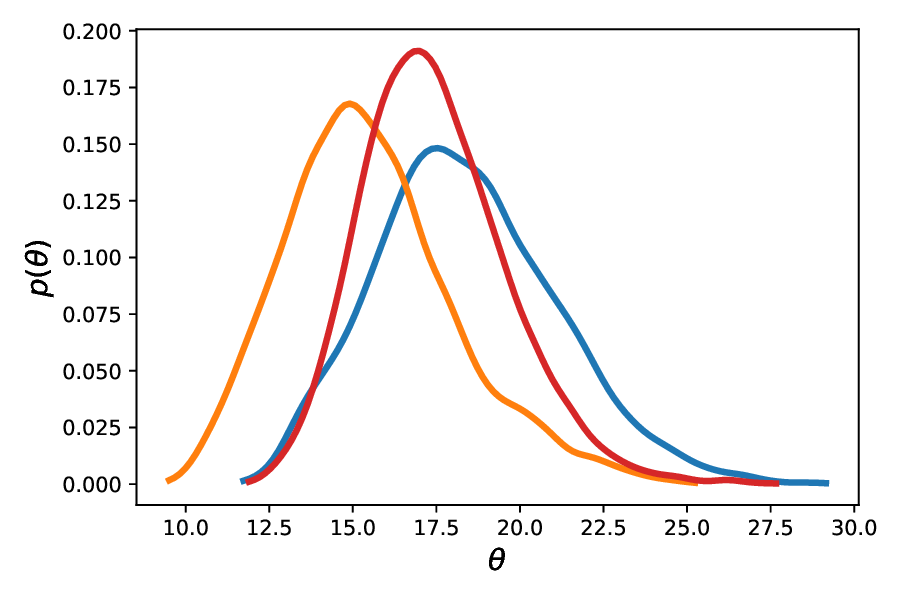}
\includegraphics[width=0.33\textwidth]{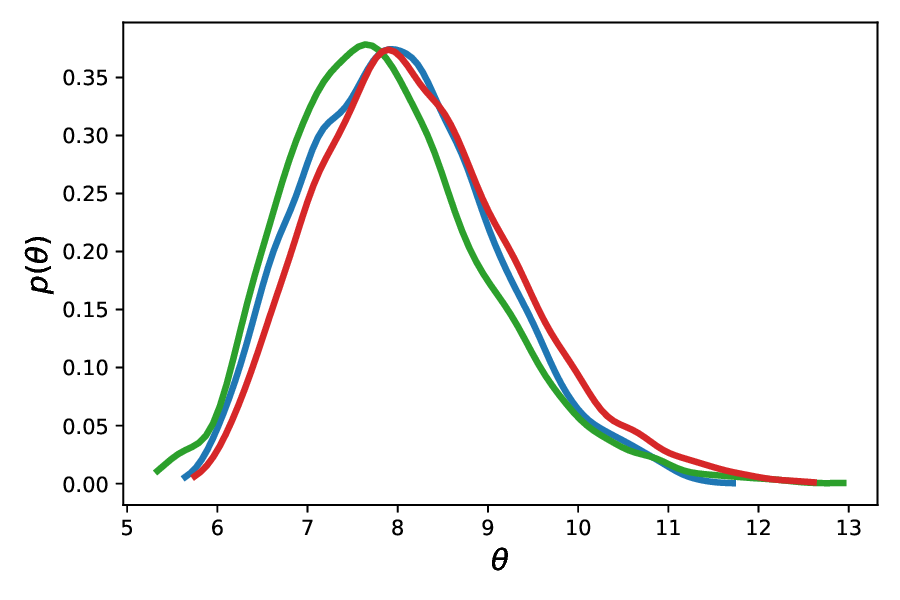}
}
\caption{1-D marginal posterior distributions to Bernalillo (left column), Santa Fe (middle column), and Valencia (right column). Top row: PDFs for $\tzr$. Second row: PDFs for $\Nr$. Third row: PDFs for $k$. Bottom row: PDFs for $\thetar$.} \label{fig:BSFV_marg1D}
\end{figure}
In Fig.~\ref{fig:BSFV_marg1D} we plot the 1D marginalized posterior PDFs of the epidemiological parameters i.e. $(\tzr, \kr, \Nr, \thetar)$ for all three counties. The 2D marginals are in the Appendix in Figs.~\ref{fig:B_marg}, Fig.~\ref{fig:V_marg} and Fig.~\ref{fig:SF_marg}. The 1D PDFs were computed using data from all three counties jointly (denoted ``3r'' in the legend), jointly using data from 2 counties at a time (denoted as ``2r'') and independently (denoted as ``1r'' inversions). We see that joint estimation does not noticeably sharpen the PDFs for any of the objects of interest (OOI), but does shift the PDFs for Santa Fe. This robustness to population size is because the likelihood for the inverse problem  is constructed with normalized counts, implying that the larger case-counts observed in Bernalillo (about 6 times larger than  Santa Fe or Valencia) do not bias the results against the smaller counties. We note that the PDFs for Valencia do not change much in the three estimations. In Fig.~\ref{fig:SF_marg1D_NoiseSpace}, top row, we plot the parameters of the GMRF $(\tphi^2, \lphi)$. It is clear that these spatial parameters can be estimated from the $2r$ and $3r$ inversions, with $\lphi$ becoming easier to estimate with specificity as we add more regions, at the expense of $\log(\tphi^2)$.

In Fig.~\ref{fig:SF_marg1D_NoiseSpace}, bottom row, we plot the noise parameters $(\sigma_a, \sigma_m)$, for Santa Fe, obtained from the same set of inversions. We see that the noise parameters are small and can be estimated, though it becomes progressively more difficult to estimate $\sigma_a$ with much specificity with joint estimation, while $\sigma_m$ becomes easier. This is because $\sigma_a$ estimates the magnitude of the epidemiological processes unexplained by our model and the genesis of these processes is likely to be different in the three counties, leading to the difficulty in estimation. This can be explained using Eq.~\ref{eqn:covmat}. Here $\tphi^2$ and $\sigma_a$ appear additively, and the uncertainties in one could be exchanged for the other, as can be seen in Fig.~\ref{fig:SF_marg1D_NoiseSpace} top left and bottom right.

\begin{figure}
\centerline{
{\bf Bernalillo}\hfill
{\bf Santa Fe}\hfill
{\bf Valencia}
}
\centerline{
\includegraphics[width=0.33\textwidth]{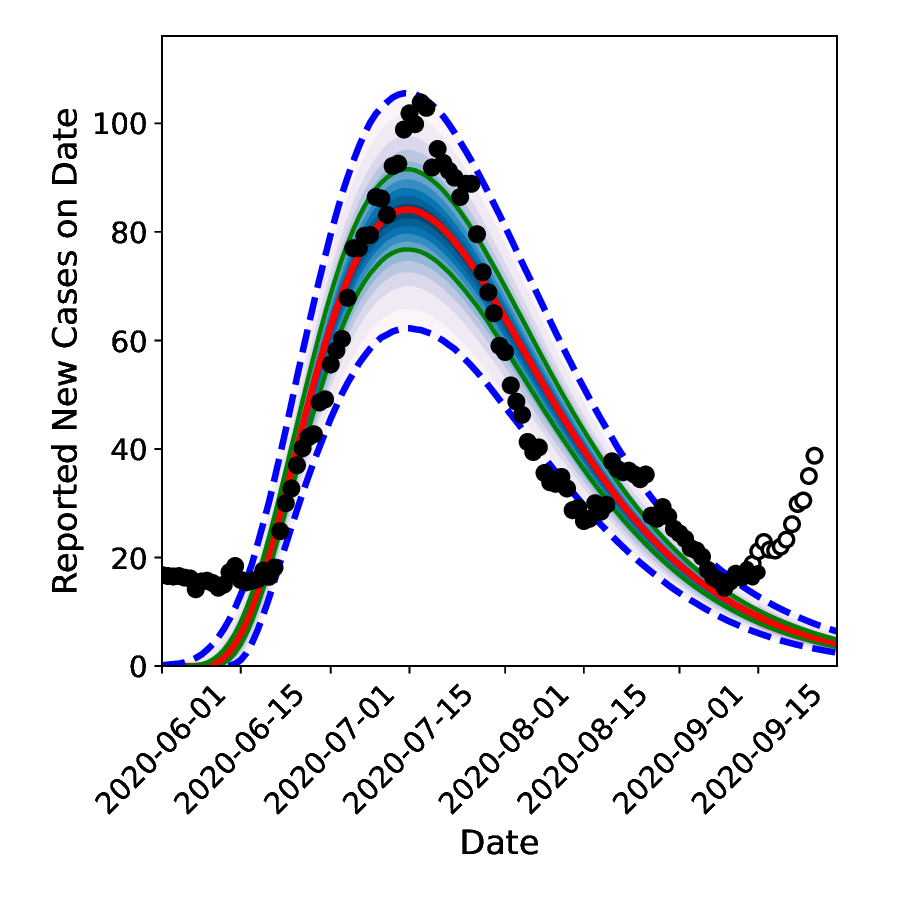}
\includegraphics[width=0.33\textwidth]{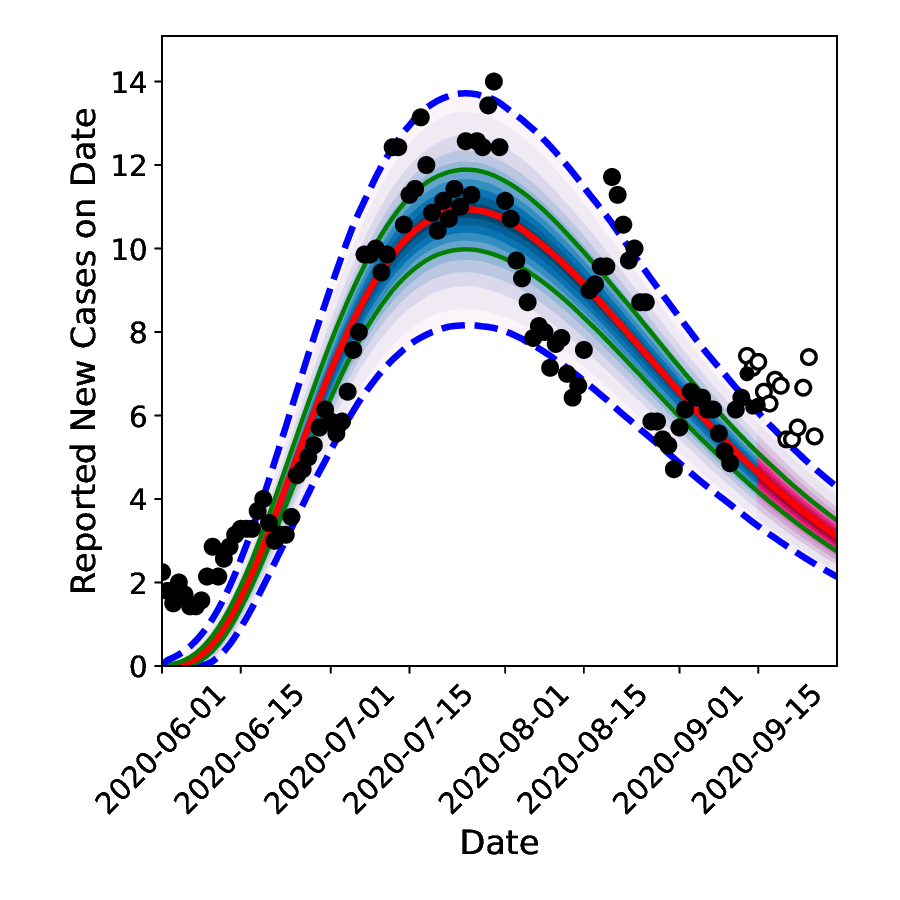}
\includegraphics[width=0.33\textwidth]{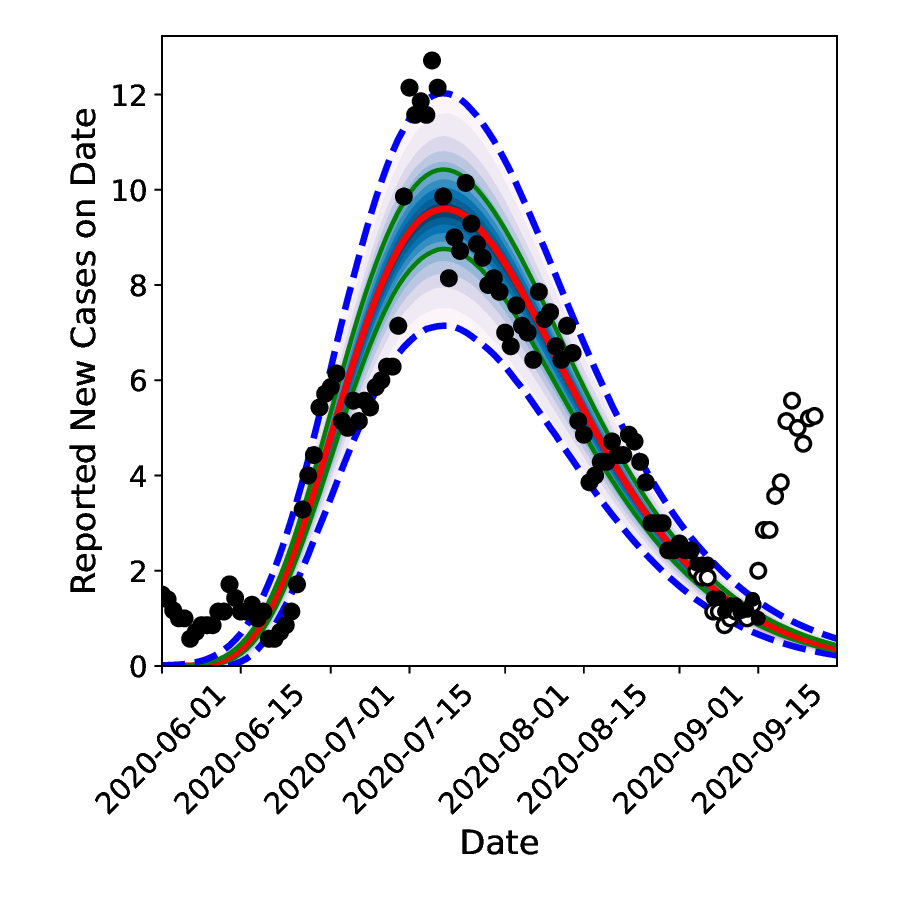}
}
\centerline{
\includegraphics[width=0.33\textwidth]{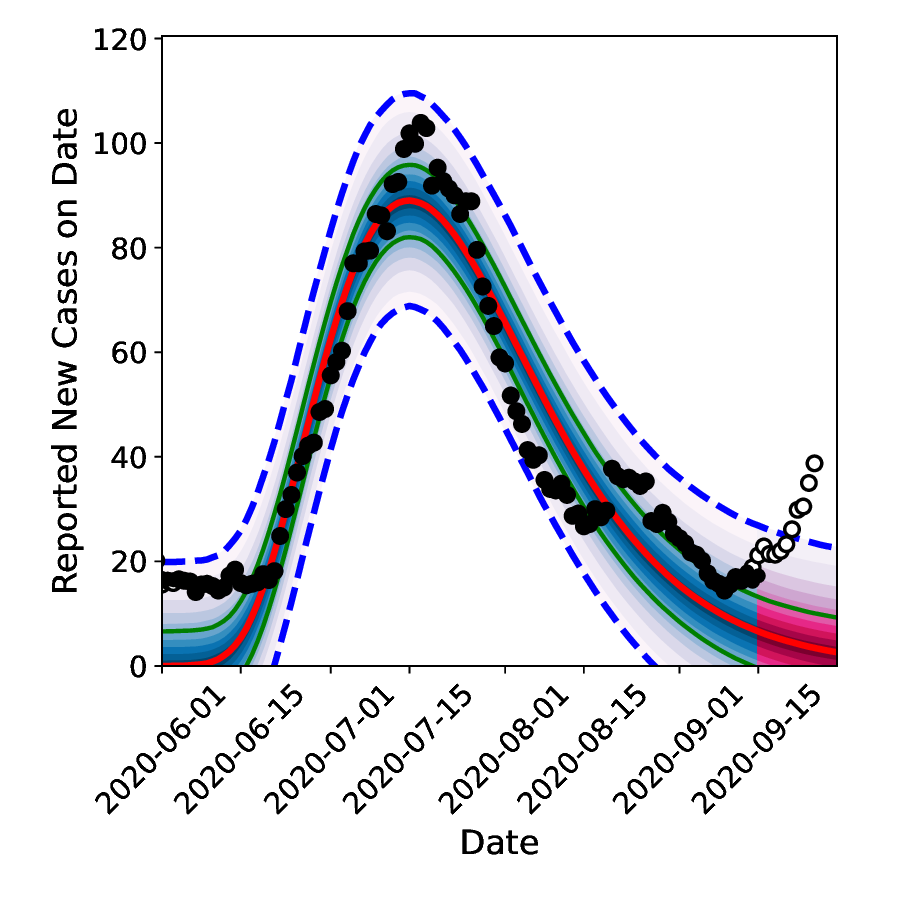}
\includegraphics[width=0.33\textwidth]{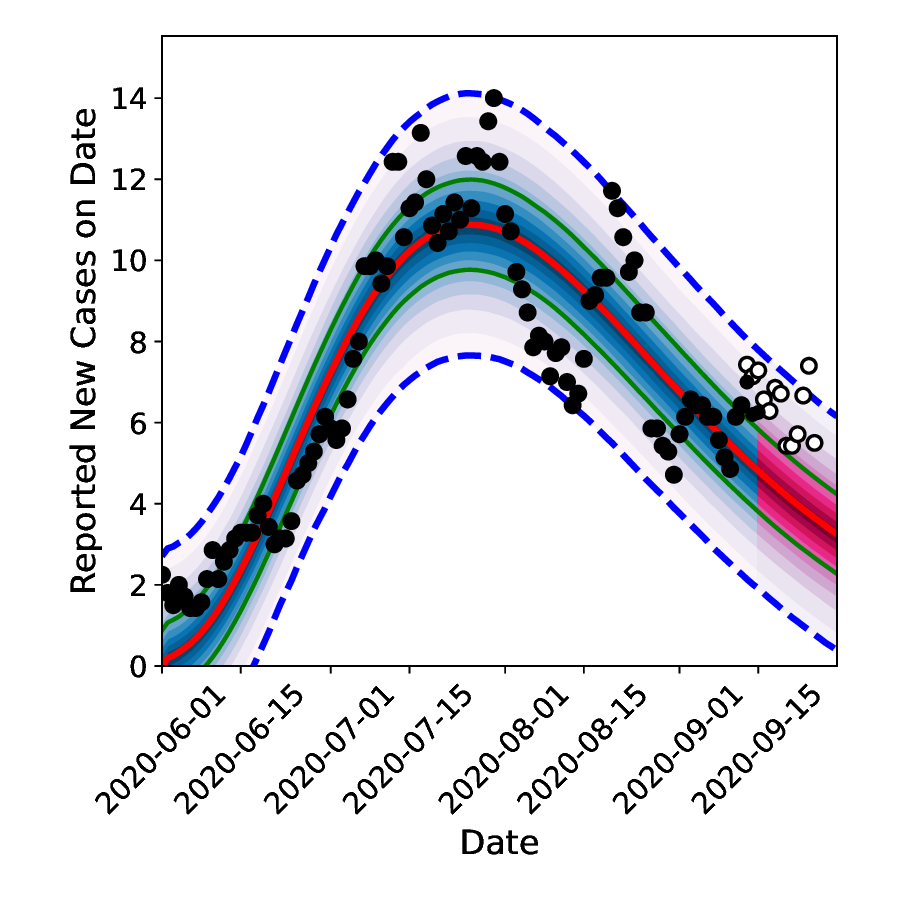}
\includegraphics[width=0.33\textwidth]{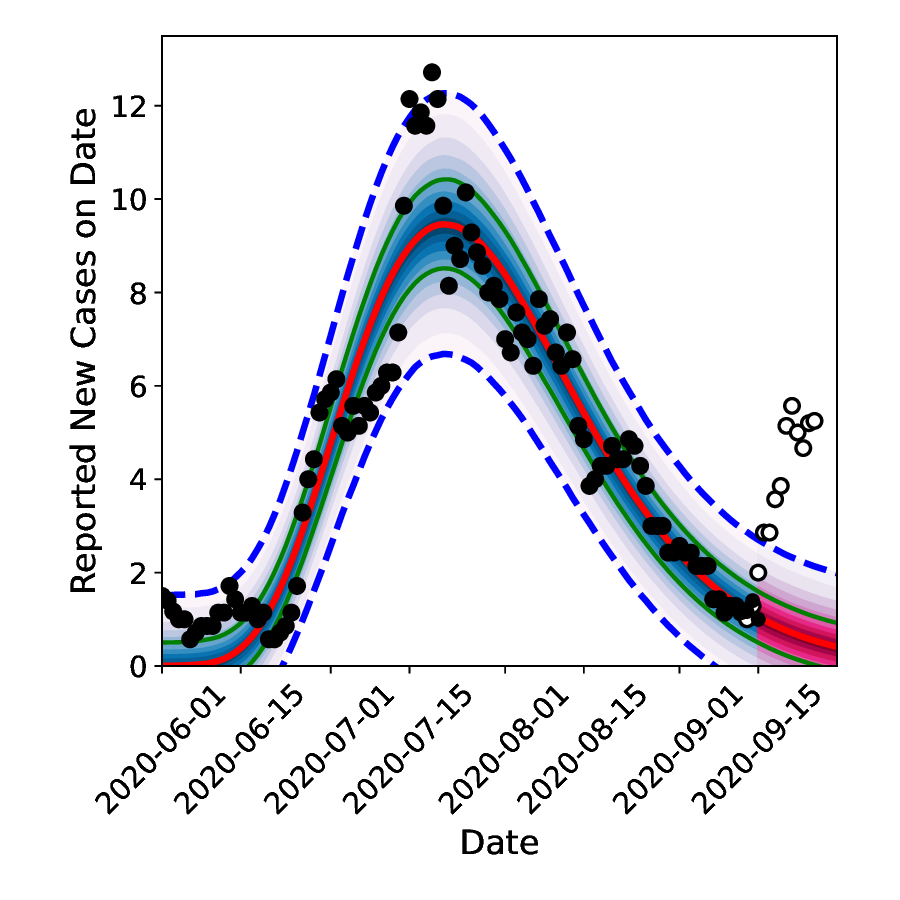}
}
\caption{Comparison of posterior predictive distribution results obtained via joint inference (using the GMRF model) for Bernalillo (left), Santa Fe (middle), and Valencia (right) shown on top row with equivalent results from independent inferences for each county separately, on the bottom row; data up to September ${\rm 15^{th}}$, 2020 is used and case-count data was smoothed with a 7-day running average. The red line is the median prediction, the shaded teal region is the inter-quartile range and the dashed lines are ${\rm 5^{th}}$ and ${\rm 95^{th}}$ percentiles.}
\label{fig:3_county_pp}
\end{figure}
In Fig.~\ref{fig:3_county_pp}, we plot the fit of the model to data till September 15, 2020 (the arrival of the Fall 2020 wave) and the two-week forecasts done after that. These predictions are performed by randomly sampling 100 $(\tzr, \kr, \Nr, \thetar, \tphi^2, \lphi)$ from the posterior distribution  (Fig.~\ref{fig:BSFV_marg1D}) and running the model forward from the start of our calibration period to the end of September 2020 (note that the calibration data stops at September 15, 2020, and the rest is a forecast). The data for the two-week period is also plotted and it not supposed to agree with the forecast, as the calibrated model does not contain information about the Fall 2020 wave. We see quite clearly that the uncertainies in forecast (the dashed blue line denoting the ${\rm 5^{th}}$ and ${\rm 95^{th}}$ percentiles are tighter for the 3-region joint inversion (top row) for all three counties. This tightness implies that it becomes easier for us to detect the discrepancy between the forecast and the data, the marker for the arrival of the Fall 2020 wave. This is particularly true for Santa Fe. The agreement between the predictions (up to September 15, 2020) and the reported case-counts are quantified using the (Continuous Ranked Probability Score~\cite{14gk2a}) and tabulated in Table~\ref{tab:crps}. We see that the most accurate forecasts do arise from independent estimations, but the $3r$ inversions are close behind.
\begin{figure}[t]
\centerline{
{\bf Bernalillo}\hfill
{\bf Santa Fe}\hfill
{\bf Valencia}
}
\centerline{
\includegraphics[width=0.33\textwidth]{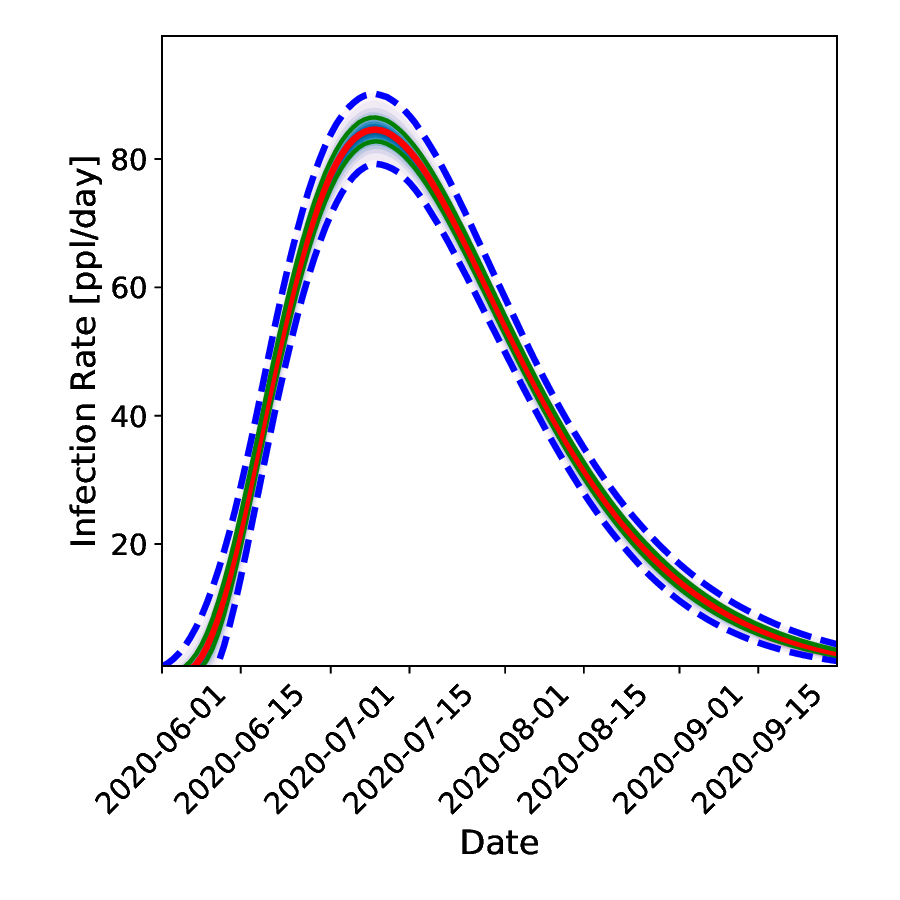}
\includegraphics[width=0.33\textwidth]{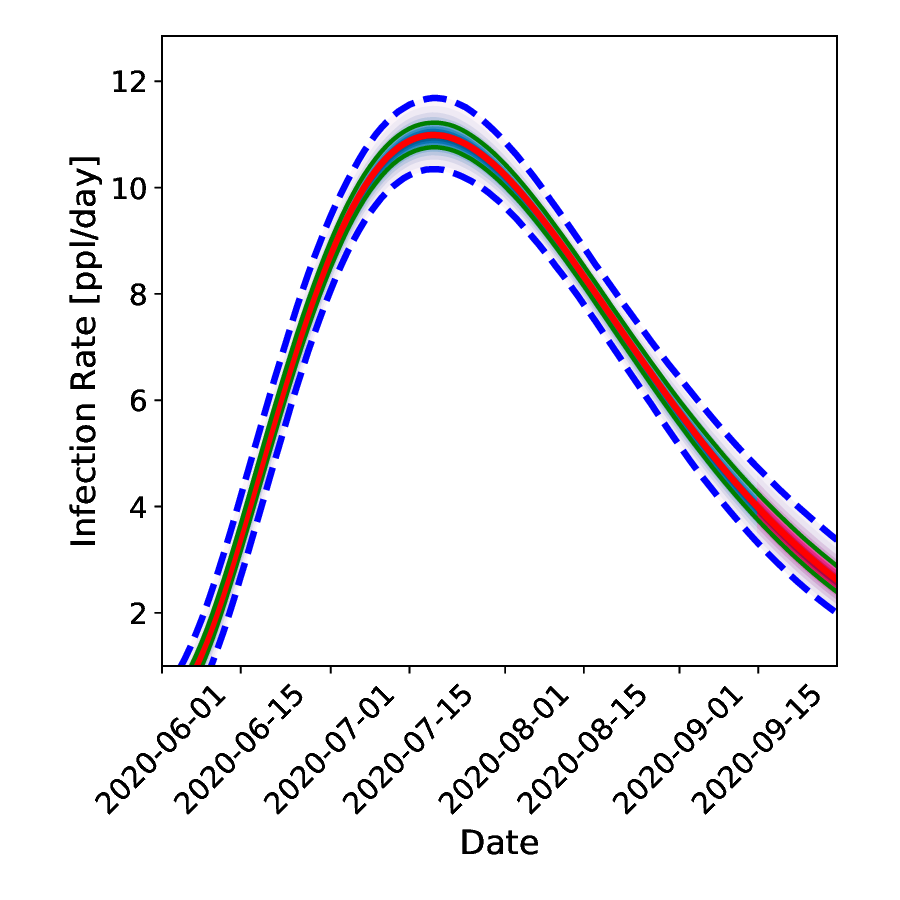}
\includegraphics[width=0.33\textwidth]{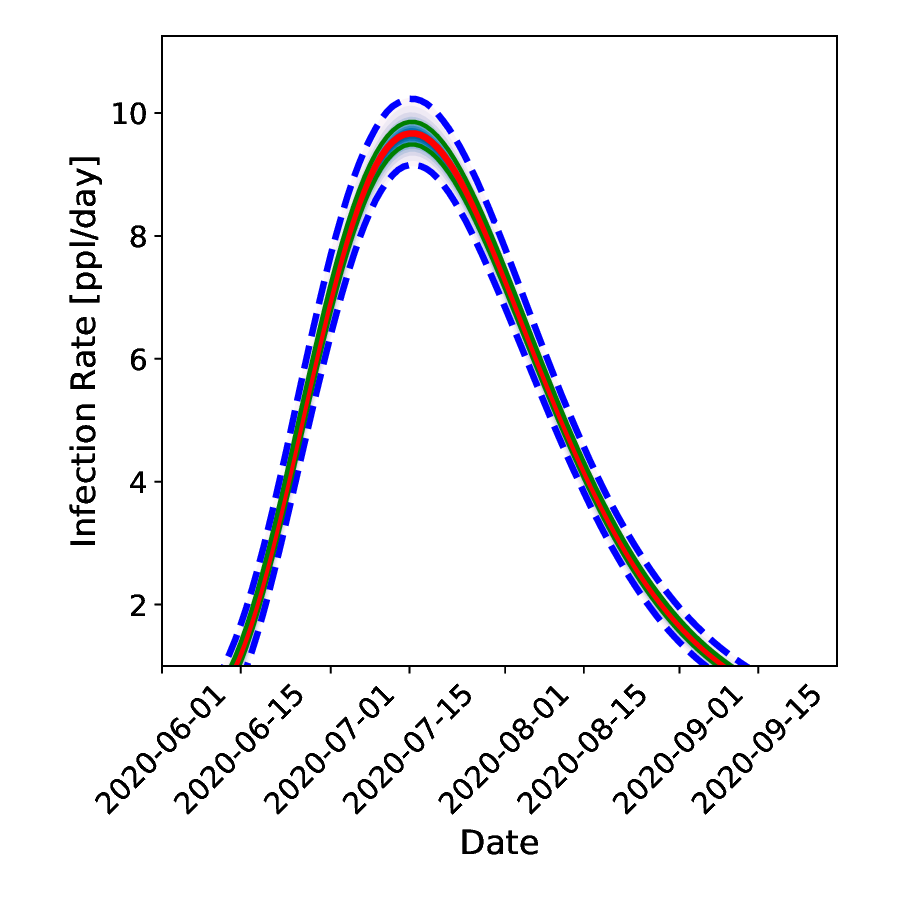}
}
\centerline{
\includegraphics[width=0.33\textwidth]{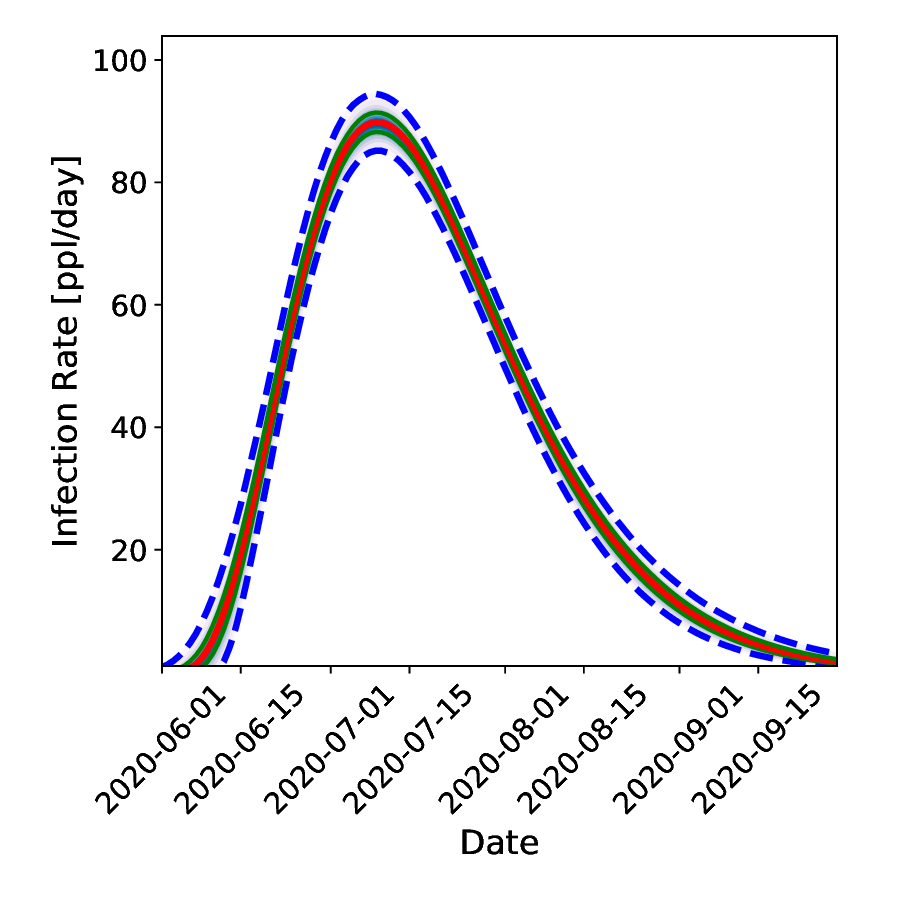}
\includegraphics[width=0.33\textwidth]{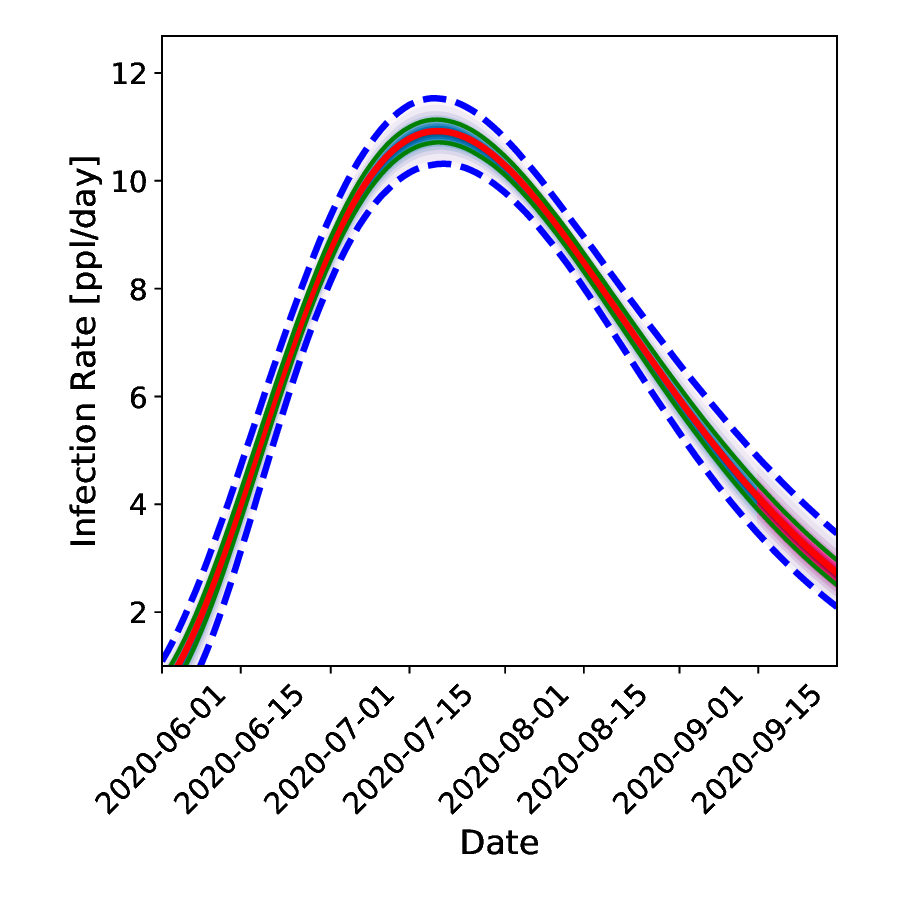}
\includegraphics[width=0.33\textwidth]{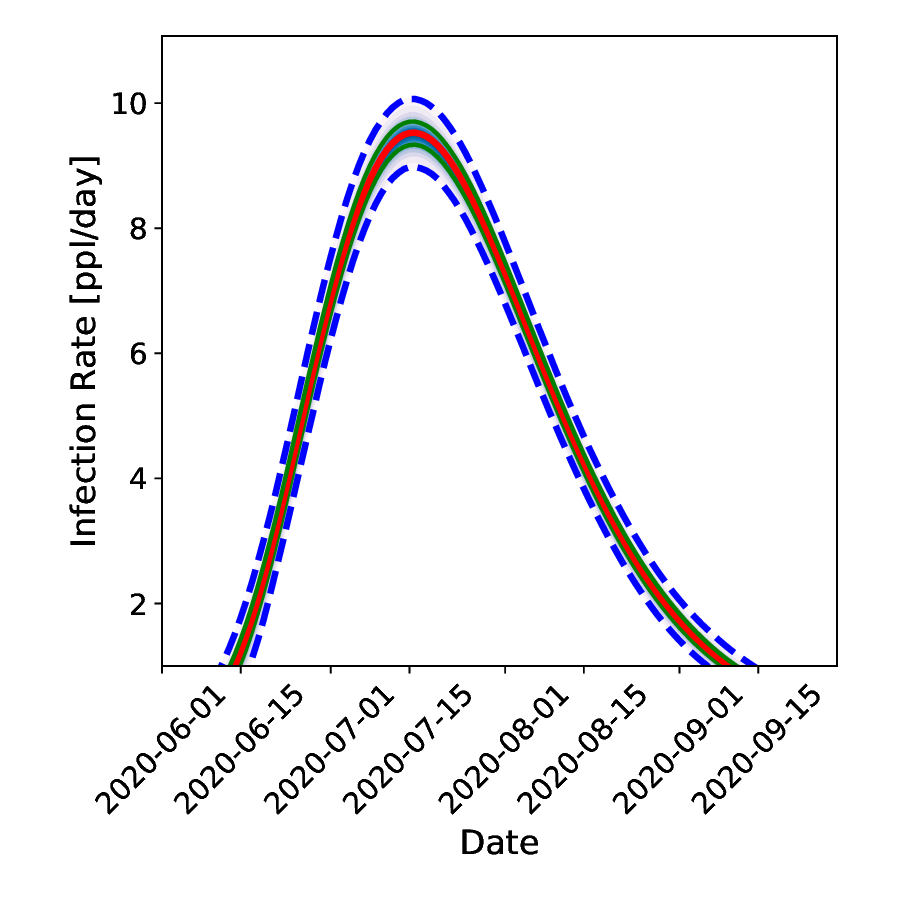}
}
\caption{Comparison of reconstructed infection-rate profiles that underly the predictions in Fig.~\ref{fig:3_county_pp}. The top row contains results obtained via joint inference (using the GMRF model) for Bernalillo (left), Santa Fe (middle), and Valencia (right).
Results from independent inferences for each county separately, are shown in the bottom row. The calibration data spans up to September ${\rm 15^{th}}$, 2020 and the case-count data was smoothed with a 7-day running average. The red line is the median prediction, the shaded teal region is the inter-quartile range and the dashed lines are ${\rm 5^{th}}$ and ${\rm 95^{th}}$ percentiles.}
\label{fig:3_county_inf}
\end{figure}
In Fig.~\ref{fig:3_county_inf}, we plot the corresponding infection rates for all three counties. Differences in the estimated infection rates, $3r$ joint estimation (top row) versus independent (bottom row), are difficult to discern. This is because the infection rate is only affected by $(\tzr, \kr, \Nr, \thetar)$ and, as is clear from Fig.~\ref{fig:BSFV_marg1D}, there is not much difference in their posterior PDFs. Instead, it is the noise and spatial parameters whose estimates differ as we add more regions to the joint estimation (see Fig.~\ref{fig:SF_marg1D_NoiseSpace}).
\begin{table}
\centering
\begin{tabular}{l|c|c|c|c|} \cline{2-5}
& 3 counties & 2 counties (B \& SF) & 2 counties (B \& V)  & 1 county \\ \hline
Bernalillo & 11.30 & 12.34 & 11.75 & \bf{10.20} \\ \hline
Santa Fe   & 2.65  & 2.87  & -     & \bf{2.48}  \\ \hline
Valencia   & 1.76  & -     & 1.82  & \bf{1.61} \\ \hline
\end{tabular}
\caption{Average CRPS values computed based on the discrepancy between the posterior predictive values corresponding to several model inference settings and the case counts recorded up to Sept. 15, 2020. The best forecasts arise when parameters are estimated for each county independently (last column), but the 3-county joint inversion is close behind (second column). }
\label{tab:crps}
\end{table}

\subsection{Statistical dependence analysis}
\label{sec:sda}

\newcommand{\co}{\cellcolor{pink!15}}
\newcommand{\cb}{\cellcolor{blue!15}}
\newcommand{\cg}{\cellcolor{green!15}}
\newcommand{\crd}{\cellcolor{red!15}}

In this section we use distance correlation~\cite{Szekely:2007} to ascertain the degree of dependence in the posterior distributions for individual parameters and between collections of parameters, e.g. parameters that define the model for individual counties. Distance correlation values, denoted \dcor, reveal the relationships between model parameters inside each region and between regions when the parameters are inferred jointly. This information can be used to aid in model construction and gauge the degree of which the parameters controlling the dynamics of the epidemics are connected across region boundaries and therefore can benefit a joint inference approach.

Numerically, we estimate the distance correlation using the algorithm presented in definition 3 in Sz\'{e}kely~\etal\cite{Szekely:2007}. This algorithm employs samples generated by the MCMC exploration of the joint posterior distribution of the model parameters and estimates the degree of dependency between individual parameters conditioned on the count data available. We also employ this approach to estimate the degree of dependence between parameter subsets, grouped by regions.

Table~\ref{tab:1regB_SF} shows \dcor values for the Bernalillo (left table) and Santa Fe (right) table. The entries in this table can be viewed as quantitative assessments of the shapes observed for the 2D marginal PDFs presented in the right frames of Figs.~\ref{fig:B_marg} and~\ref{fig:SF_marg} included in the Appendix. For both counties we observe strong dependencies between $k$ and $\theta$, the shape and scale parameters of the Gamma distribution used to model the infection rate, and $t_0$. These strong dependencies, explained by the corresponding narrow 2D marginal PDFs (in Figs.~\ref{fig:B_marg} and ~\ref{fig:SF_marg} in the Appendix) are induced by the strong constraints imposed by the available case-count data and the infection rate dynamics. The error model parameters, $\sigma_a$ and $\sigma_m$, exhibit little dependency among themselves and with other model parameters for Bernalillo county which is driven by larger case-counts values. However, for Santa Fe, which exhibits lower case-counts and hence changes in case count values are more relevant, the model discrepancy parameters show non-negligible dependencies with respect to each other and other model parameters. Similar trends are also observed for Valencia county (results not shown) for which the observed case counts are comparable in magnitude to Santa Fe.

Table~\ref{tab:3reg} shows \dcor values computed with MCMC samples corresponding to a joint inversion for the three counties simultaneously. The sections in this table were colored to highlight the different types of parameter dependencies. The \dcor values corresponding to Bernalillo and Santa Fe counties (colored in orange) are similar to the corresponding values when the epidemiological models are calibrated region by region. This is due to infection rate models being defined on a per region basis and hence it is expected to observe that similar trends for the corresponding parameters affected by regional case counts. Given the large discrepancy between the magnitude of the case counts in adjacent regions, the additive component $\sigma_a$ of the error model is now less impactful compared to the multiplicative component. The spatial correlation model parameters and the multiplicative error model component show non-negligible \dcor (with joint PDFs displaying negative correlations - results not shown). We also show, in Table~\ref{tab:3regGrouped}, the corresponding \dcor values between model parameters grouped by model components, i.e. by region, then spatial correlation and error models, respectively. These results are essentially summaries of the corresponding values aggregated in similarly colored regions in Table~\ref{tab:3reg}.

\begin{table}[!htb]
\centering
\begin{tabular}{l|cccc|c}
& $t_0$ & $N$ & $k$ & $\theta$ & $\sigma_a$ \\ \hline
$N$ & 0 &  &  &  &  \\
$k$ & 0.9 & 0.1 &  &  &  \\
$\theta$ & 0.8 & 0.3 & 0.9 &  &  \\ \hline
$\sigma_a$ & 0 & 0.1 & 0 & 0 &  \\
$\sigma_m$ & 0   & 0 & 0 & 0 & 0.1 \\
\end{tabular}
\hspace{1cm}
\begin{tabular}{l|cccc|c}
& $t_0$ & $N$ & $k$ & $\theta$ & $\sigma_a$ \\ \hline
$N$ & 0 &  &  &  &  \\
$k$ & 0.9 & 0.4 &  &  &  \\
$\theta$ & 0.6 & 0.6 & 0.9 &  &  \\ \hline
$\sigma_a$ & 0.4 & 0.3 & 0.2 & 0 &  \\
$\sigma_m$ & 0.3 & 0.2 & 0.2 & 0 & 0.7 \\
\end{tabular}
\caption{Distance correlation values between parameters corresponding to the Bernalillo county (left) and Santa Fe county (right) using samples resulted from model calibrations using data for one county at a time.}
\label{tab:1regB_SF}
\end{table}

\begin{table}[!htb]
\centering
\begin{tabular}{c|l|cccc||cccc||cccc||cc||cc}
\multicolumn{1} {c}{} & & \multicolumn{4} {c||}{Bernalillo} &
\multicolumn{4} {c||}{Santa Fe} &
\multicolumn{4} {c||}{Valencia} &
\multicolumn{2} {c||}{SpC} &
\multicolumn{1} {c}{ErrM} \\ \cline{3-17}
\multicolumn{1} {c}{} & & $t_0$ & $N$ & $k$ & $\theta$ & $t_0$ & $N$ & $k$ & $\theta$ &
$t_0$ & $N$ & $k$ & $\theta$ & $\tau_{\Phi}^2$ & $\lambda_{\Phi}$ & $\sigma_a$ \\ \hline
\multirow{4}{*}{\rotatebox[origin=c]{90}{Bernalillo}}  & $t_0$ &  & & & &  &  &  &  &  &  &  &  &  &  &  \\
& $N$ & \co 0.1 &  &  &  &  &  &  &  &  &  &  &  &  &  &  \\
& $k$ & \co 0.9 & \co 0.2 &  &  &  &  &  &  &  &  &  &  &  &  &  \\
& $\theta$ & \co 0.8 & \co 0.3 & \co 0.9 &  &  &  &  &   &  &  &  &  &  &  &  \\ \hline
\multirow{4}{*}{\rotatebox[origin=c]{90}{Santa Fe}} & $t_0$ & \cb 0.2 & \cb 0 & \cb 0.2 & \cb 0.2 &  &  &  &  &  &  &  &  &  &  &  \\
& $N$ & \cb 0 & \cb 0.3 & \cb 0.1 & \cb 0.2 & \co 0.3 &  &  &  &  &  &  &  &  &  &  \\
& $k$ & \cb 0.2 & \cb 0.1 & \cb 0.2 & \cb 0.2 & \co 0.9 & \co 0.5 &  &  &  &  &  &  &  &  &  \\
& $\theta$ & \cb 0.1 & \cb 0.1 & \cb 0.2 & \cb 0.2 & \co 0.8 & \co 0.6 & \co 0.9 &  &  &  &  &  &  &  &  \\  \hline
\multirow{4}{*}{\rotatebox[origin=c]{90}{Valencia}} & $t_0$ & \cb 0.2 & \cb 0 & \cb 0.2 & \cb 0.2 & \cb 0.2 & \cb 0.1 & \cb 0.2 & \cb 0.2 &  &  &  &  &  &  &  \\
& $N$ & \cb 0 & \cb 0.3 & \cb 0.1 & \cb 0.1 & \cb 0.1 & \cb 0.3 & \cb 0.1 & \cb 0.1 & \co 0.1 &  &  &  &  &  &  \\
& $k$ & \cb 0.2 & \cb 0 & \cb 0.2 & \cb 0.2 & \cb 0.2 & \cb 0 & \cb 0.2 & \cb 0.2 & \co 1.0 & \co 0.1 &  &  &  &  &  \\
& $\theta$ & \cb 0.2 & \cb 0.1 & \cb 0.2 & \cb 0.2 & \cb 0.1 & \cb 0.2 & \cb 0.1 & \cb 0.2 & \co 0.9 & \co 0.2 & \co 1.0 &  &  &  &  \\  \hline
\multirow{2}{*}{\rotatebox[origin=c]{90}{SpC}} & $\tau_{\Phi}^2$ & \cg 0 & \cg 0.1 & \cg 0.1 & \cg 0.2 & \cg 0 & \cg 0.1 & \cg 0.1 & \cg 0.1 & \cg 0.1 & \cg 0 & \cg 0.1 & \cg 0 &  &  &  \\
& $\lambda_{\Phi}$ & \cg 0.1 & \cg 0.1 & \cg 0.1 & \cg 0.1 & \cg 0.1 & \cg 0.1 & \cg 0.1 & \cg 0.1 & \cg 0.1 & \cg 0 & \cg 0.1 & \cg 0.1 & \co 0.9 &  &  \\ \hline
\multirow{2}{*}{\rotatebox[origin=c]{90}{ErrM}} & $\sigma_a$ & \crd 0.1 & \crd 0 & \crd 0 & \crd 0 & \crd  0.1 & \crd 0.1 & \crd 0.1 & \crd 0.1 & \crd 0 & \crd 0 & \crd 0 & \crd 0 & 0 & 0 &  \\
& $\sigma_m$ & \crd 0.1 & \crd 0.1 & \crd 0.1 & \crd 0.1 & \crd 0.1 & \crd 0.1 & \crd 0.1 & \crd 0.1 & \crd 0.1 & \crd 0.2 & \crd 0.1 & \crd 0.1 & 0.5 & 0.4 & \co 0.1 \\ \hline
\end{tabular}
\caption{Distance correlation values between model parameters corresponding to three adjacent counties, the spatial correlation model (SpC), and to the error model (ErrM). The light orange color corresponds to dependencies between model parameters corresponding to the same region, blue to values between pairs of parameters in difference regions, green denotes dcor values between the SpC and the region parameters and light red to dcor values that pertain between ErrM and the regional model parameters. }
\label{tab:3reg}
\end{table}

\begin{table}[!htb]
\centering
\begin{tabular}{l|c||c||c||c}
& Bernalillo & Santa Fe & Valencia & SpC\\ \hline
Santa Fe & \cb  0.2 &          &          &      \\ \hline
Valencia & \cb  0.2 & \cb  0.2 &          &      \\ \hline
SpC      & \cg  0.1 & \cg  0.1 & \cg  0.1 &      \\ \hline
ErrM     & \crd 0.1 & \crd 0.1 & \crd 0.05   & 0.05   \\ \hline
\end{tabular}
\caption{Distance correlation values between groups of parameters corresponding to three adjacent counties, the spatial correlation model (SpC), and the parameters of the error model (ErrM). The color scheme is similar to the one presented in Table~\ref{tab:3reg}.}
\label{tab:3regGrouped}
\end{table}

\section{Discussion}
\label{sec:disc}
The results in \S~\ref{sec:res} show that we can estimate the infection-rate with a sufficient degree of accuracy so as to be able to provide short-term (2-week-ahead) forecasts of the evolution. Given that the inversion is, in effect, a smoothing operation (i.e., the observations inform infection processes that happened in the past), any discrepancy between forecasts and observations could be caused by a sudden change in the infection-rate. Thus it may be feasible to detect the arrival of a new wave of infection using the (latent) infection-rates estimated in Fig.~\ref{fig:3_county_inf}.

\begin{figure}
\centerline{\includegraphics[width=0.45\textwidth]{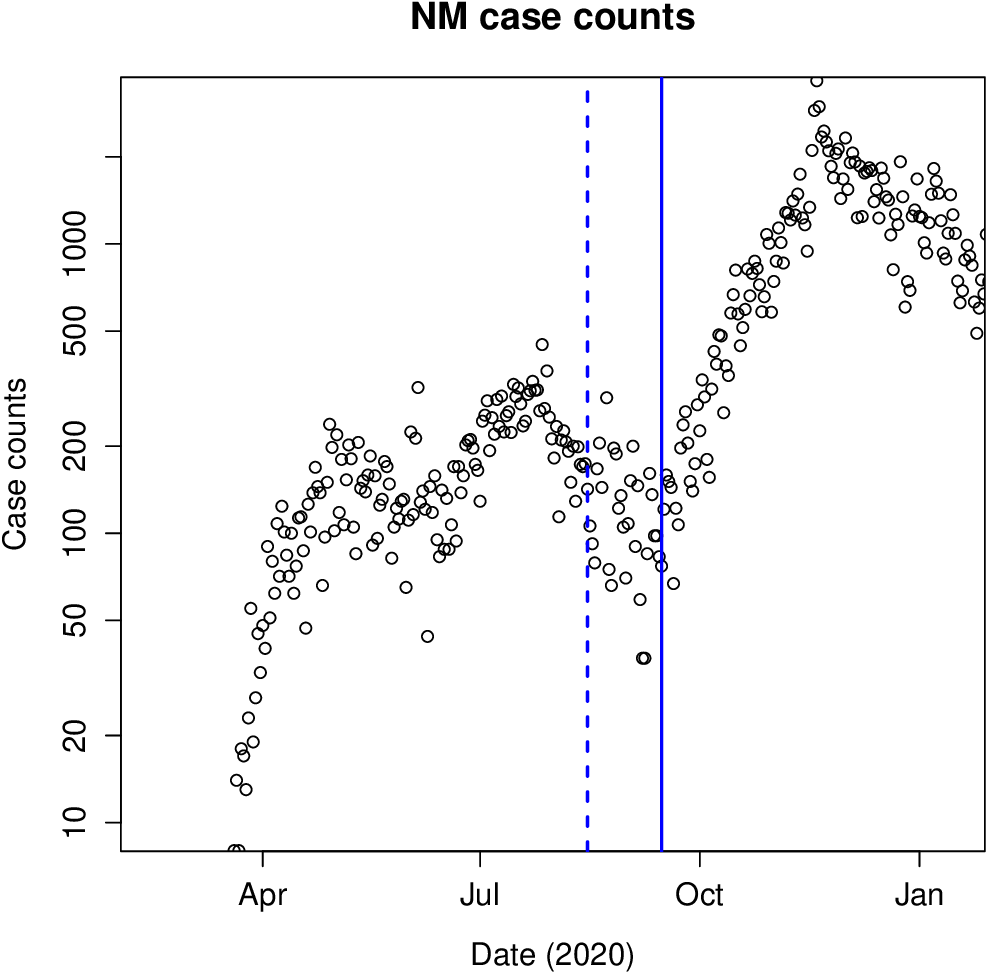}}
\caption{The waves of COVID-19 infection in New Mexico in 2020. The Fall 2020 wave started
around September 15, and is marked with a solid vertical line. The dashed line is August 15, where we will also check our detector.}
\label{fig:2020wave}
\end{figure}

The state of NM experienced three waves of COVID-19 infections in 2020; the state-wide totals of case-counts are shown in Fig.~\ref{fig:2020wave}. The second wave, that was felt between June ${\rm 1^{st}}$ and September ${\rm 15^{th}}$, provides us with ample data to infer an infection-rate, and forecast the outbreak till the end of September. As is clear from Fig.~\ref{fig:2020wave}, these forecasts will deviate from the data due to the arrival of the third wave (henceforth the ``Fall 2020" wave). Our aim is to use the estimated infection-rate to detect the Fall 2020 wave, and compare our performance versus a conventional method. We will also conduct such a test using data collected till August ${\rm 15^{th}}$ (before the Fall 2020 wave) and check whether our infection-rate method detects a (false) positive.

\begin{figure}
\centerline{
\includegraphics[height=0.3\textheight]{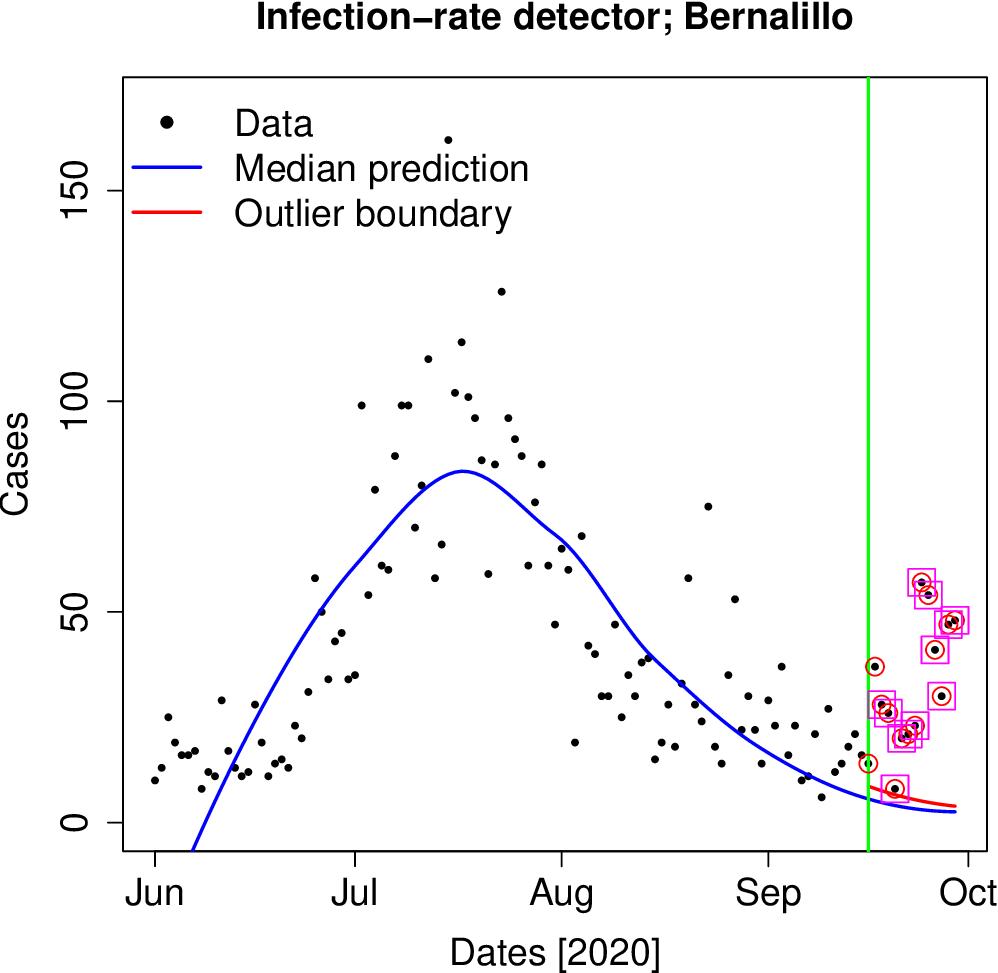}
\includegraphics[height=0.3\textheight]{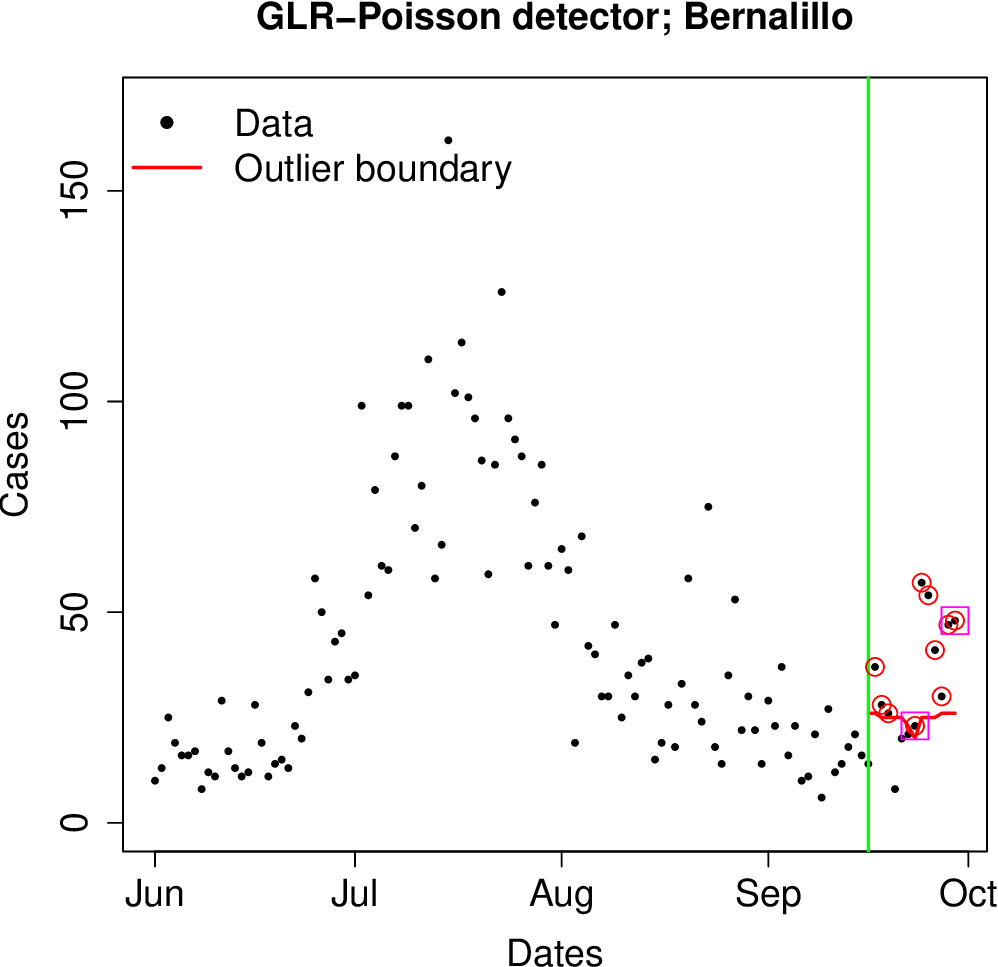}
}
\centerline{
\includegraphics[height=0.3\textheight]{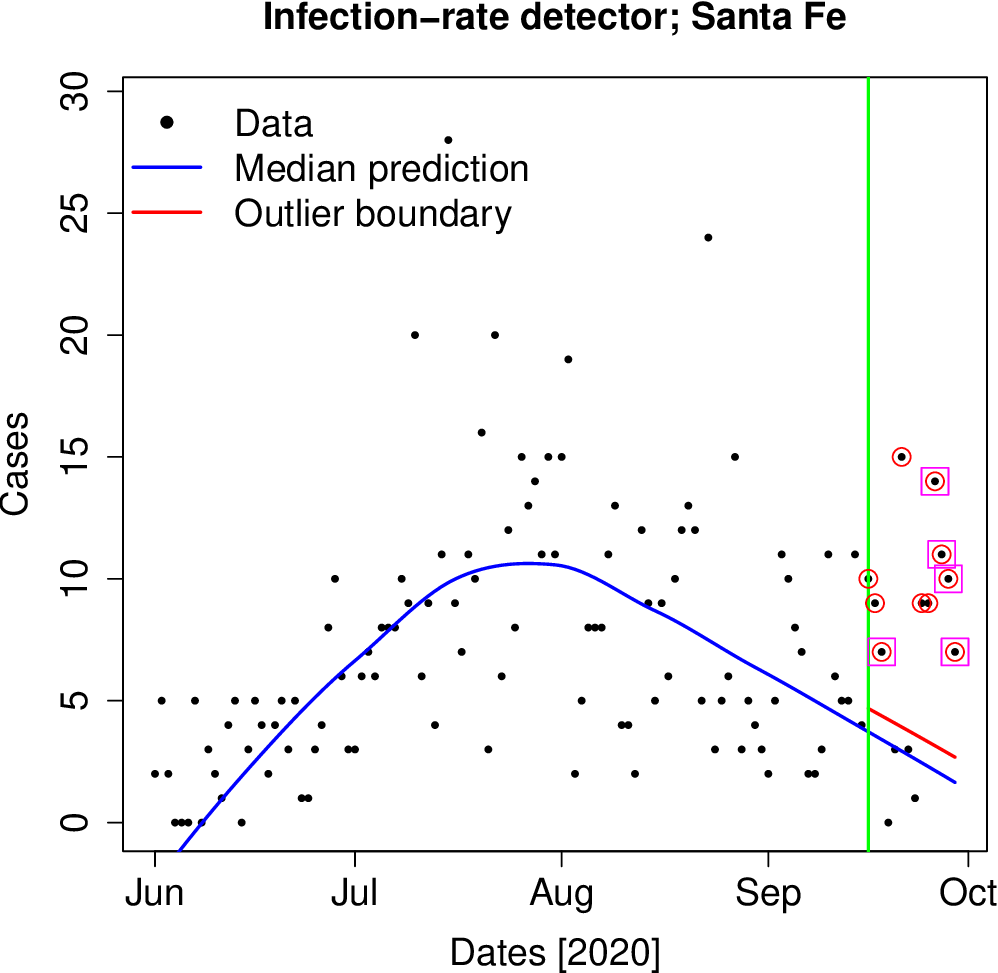}
\includegraphics[height=0.3\textheight]{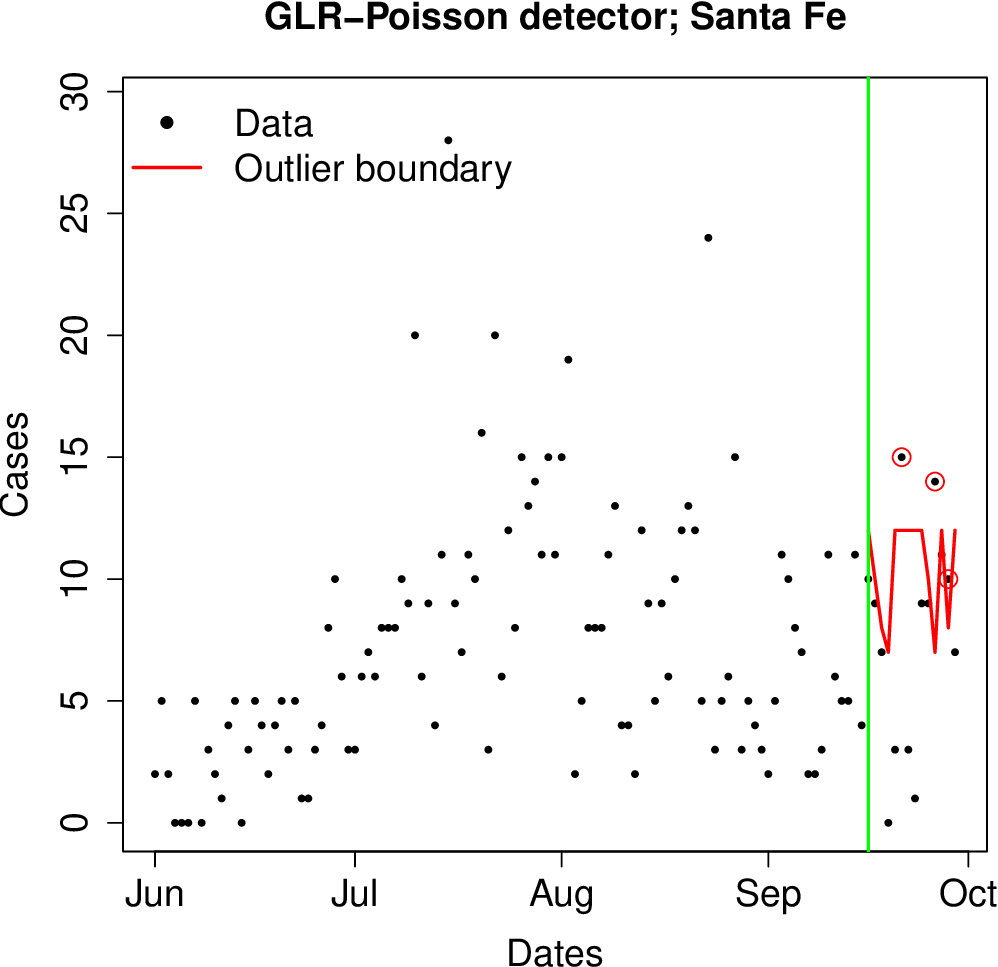}
}
\centerline{
\includegraphics[height=0.3\textheight]{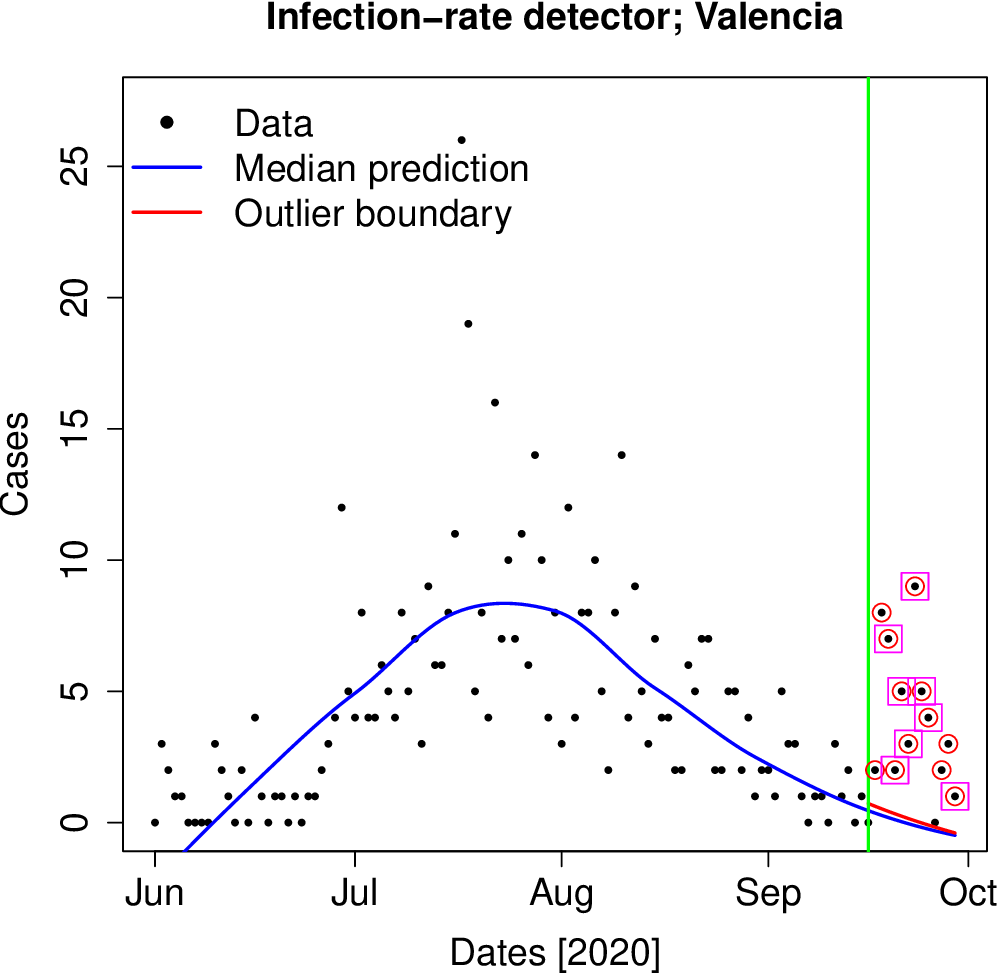}
\includegraphics[height=0.3\textheight]{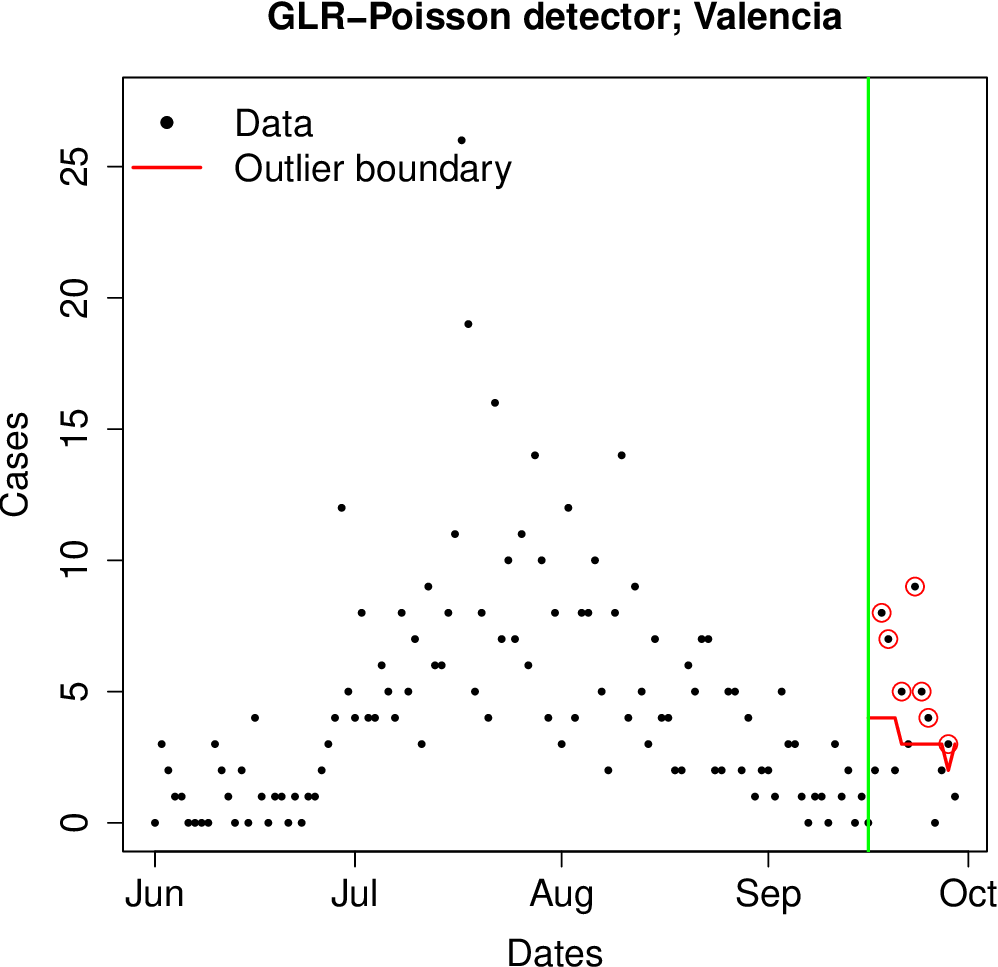}
}
\caption{Comparison of the infection-rate detector (left) compared with the GLR-Poisson detector (right), using
data from 2020-06-01 to 2020-09-15. The symbols are the observed case-counts. The Fall 2020 wave is believed to have started around September 15. The red line beyond September 15 is the outlier boundary; a day with a case-count above the dashed line is an ``outlier'' and is circled. A data point with a square box around it denotes the the last of a sequence of three consecutive alarmed days. Top row: Performance for Bernalillo county. Middle row: Results for Santa Fe county. Bottom row: Results for Valencia county. In all cases we see that the GLRNB detector misses the Fall 2020 wave.}
\label{fig:detectSept}
\end{figure}

We sample the posterior distribution for $\bp$ (plotted in Fig.~\ref{fig:BSFV_marg1D}) and produce a fantail of predictions of the evolution of the outbreak; the $99^{th}$ percentile prediction is treated as the ``outlier boundary'' (similar to SCPO\cite{11cl2a}) and any day with a case-count above the boundary is deemed an ``outlier". We treat three consecutive days of outliers as an ``alarm" indicating an anomaly in the behavior of the data with respect to the infection-rate estimated before. This is plotted for Bernalillo, Santa Fe and Valencia counties in Fig.~\ref{fig:detectSept} (left column). The green line denotes September ${\rm 15^{th}}$. Beyond this date, we see a number of days where the case-counts lie above the red ``outlier boundary''; these are circled in red. Some days also have their case-count data encased inside a box; these are the third of a 3-day sequence of outlier days (and thus an ``alarm'' day). We see that in all three counties, we could detect the arrival of the Fall 2020 wave successfully. We repeated the infection-rate estimation using data from June ${\rm 1^{st}}$ to August ${\rm 15^{th}}$ and performed a similar check for ``alarm" days between August ${\rm 15^{th}}$ and ${\rm 31^{st}}$; these are plotted in Fig.~\ref{fig:detectAug} (in the Appendix). While we do detect many ``outlier days", we do not see any ``alarm days". Thus monitoring the infection-rate allows us to detect the Fall 2020 wave when it is present; further, it does not lead to a false positive in the absence of a new wave of infection.

Next we compare the performance of the detection method using the infection-rate against a conventional detector~\cite{Hohle:2008}, which we call ``GLR-Poisson'' (for Generalized Likelihood Ratio - Poisson). This detector uses the raw case-counts to fit a time-series model (complete with prediction uncertainty bounds) and thus detect ``outlier days''. The detector has two formulations, one based on the negative binomial (NB) distribution and another based on Poisson. We use the implementation in the R Statistical Software\cite{Manual:R} (R version 4.3.2 (2023-10-31)) package \texttt{surveillance}\cite{07hm1a}. The case-count on any day is modeled as $y_t \sim {\textit{NB}} (\mu_t, \alpha)$, (or $y_t \sim {\textit{Pois}} (\mu_t)$) where $\mu_t$ is the mean and $\alpha$ is the dispersion of a NB distribution. The mean is modeled  as $\log(\mu_t) = \beta_0 + \beta_1 t + \sum_{s = 1}^{S} \beta_{2s}  \sin(\omega s t) + \beta_{2s+1} \cos(\omega s t)$, where $\omega = 2 \pi / 365$; in essence, this is a seasonal log-linear model with parameters $\mathbf{\beta}$. We set $S = 1$, since there is clearly only one mode in Fig.~\ref{fig:2020wave}. We  fit a model $\log(\mu^{0})$ using data from June ${\rm 1^{st}}$ to September ${\rm 15^{th}}$ (corresponding to $\bbeta^0$), before the arrival of the Fall 2020 wave, and test whether a new model (for $\log(\mu^{1})$) (corresponding to $\bbeta^1$), fitted solely to a moving window in the post-September ${\rm 15^{th}}$ data, explains it appreciably better than the original $\log(\mu^{0})$ model. Indexing the days after September ${\rm 15^{th}}$ as $l  = 1 \ldots L = 15$, we compute the set of days $l^{\ast}$ where
\begin{equation}
\max_{1 \le l \le L}\sup_{\mathbf{\beta}} \left[ \sum_{t = l}^{L} \log \left( \frac{f_{\bbeta^1}(y_t)}{f_{\bbeta^0}(y_t)} \right) \right] > c_{\gamma},
\end{equation}
where $f_{\bbeta}(y_t)$ is the negative binomial distribution and $c_{\gamma}= 3$. In essence, in the 15-day period between September ${\rm 16^{th}}$ and ${\rm 30^{th}}$, we search for a window where the original $\log(\mu^{0})$ model explains the data poorly.  Note that this model \emph{does} require much historical data to calibrate $\bbeta^0$ (for example, to determine the seasonal nature of the outbreaks), something that is rarely available for novel diseases such as COVID-19. Using the distribution (negative binomial or Poisson), it is also possible to predict the case-count that would have caused an ``outlier day''. Per Kim~\etal\cite{23kl6a}, the NB tends to give better fits whereas Poisson is preferable for small datasets, and so we test both formulations. The results with the NB distribution are clearly inferior and are in our technical report\cite{23rs5a}. The results with the Poisson distribution are plotted in Fig.~\ref{fig:detectSept} (right column), with the ``outlier boundary'' in red. For Bernalillo, in the post-September ${\rm 15^{th}}$ period, we see many outliers and a few alarm days, implying that the Fall 2020 wave was detected. The detector does not show any alarms for Valencia or Santa Fe, thus completely missing the Fall 2020 wave.  We repeat this analysis for data between June ${\rm 1^{st}}$ and August ${\rm 15^{th}}$ (see Fig.~\ref{fig:detectAug} in the Appendix). Here the detector identifies outliers and alarms in the data for Bernalillo and Santa Fe, thus ``detecting'' the Fall 2020 wave a full month before its arrival; clearly, this is a false positive. In contrast, the detector behaves correctly for Valencia. The reason for the poor performance of the GLR-Poisson detector  is likely due to the peculiarities of our COVID-19 data (no long historical record and low case-counts from sparsely populated areal units), which runs afoul of many assumptions embedded in conventional disease detectors.

Note that the ability to detect the Fall 2020 wave correctly does not imply that we have fashioned an infection-rate-based disease detector (e.g., we have not attempted to compute a Receiver Operating Characteristic curve); rather, it shows that the infection-rate of an outbreak of a novel disease has the information content that could be exploited within a disease detector. The smoothing effect of our estimation process (which reduces the effect of noise in the observations) and  the use of epidemiological information i.e., the incubation period distribution, compensates for the lack of long time-series data that conventional detectors rely on for information content, thus making our method particularly suited for novel outbreaks. For endemic diseases with long time-series and high-quality data, our method would possibly be unnecessarily complex.

\section{Conclusion}
\label{sec:concl}
In this paper, we explore whether it is possible to use the (latent) infection-rate of a disease as a monitoring variable in  disease surveillance. This is because the infection-rate, which is governed by mixing patterns and spreading characteristics of the pathogen in question, does not vary erratically from day-to-day; in contrast, observed case-counts, the monitoring variable for all conventional disease surveillance algorithms is contaminated by reporting errors. The difficulty, of course, lies in being able to estimate the infection-rate from the case-counts, which can have high variance if they are small numbers.

To this end, we developed a method to estimate an infection-rate (spatiotemporal) field defined over multiple areal units, conditional on case-count time-series, of various fidelities, gathered from the areal units. The aim of estimating a field, rather than a time-varying infection-rate inside an areal unit, was driven by our desire to encode spatial patterns of epidemiological dynamics into the infection-rate field, allowing us to ``borrow'' information from neighboring areal units and compensate for poor quality observations. The method was demonstrated on COVID-19 data from 3 counties of New Mexico - Bernalillo, Santa Fe and Valencia. Our method uses COVID-19 data and exogenous covariates to uncover the spatial patterns in epidemiological dynamics and encode them as a Gaussian Markov Random Field (GMRF) model. We extend our original method for estimating the infection-rate in one areal unit\cite{Safta:2021} to multiple units, and use the GMRF to impose a degree of smoothing. Joint inversions for disease parameters showed that the PDFs and posterior predictive simulations for Santa Fe (which had low case-count data) were sharper compared to inversions performed for one areal unit.

The estimated infection-rate field, estimated using data from June 1, 2020 to September 15, 2020, was used to forecast the evolution of the outbreak for two weeks ahead. The Fall 2020 wave of COVID-19 arrived on September 15 and the forecasts are expected to be erroneous i.e., our forecast acts as a detector of the new wave of infection. Our model's performance was compared with that of a conventional surveillance algorithm that, like all other surveillance algorithms, relies on a long historical training database  and which was not not available for COVID-19 because of its novelty. Our method successfully detected the arrival within  the two-week period whereas the conventional detector failed. In addition, we tested the method with data till August 15th, 2020, one month before the arrival of the Fall 2020 wave. Our method failed to detect a wave; the conventional detector detected a non-existent one for Bernalillo. The aberrant behavior of the conventional detector is easily explained by the insufficiency of training data, but this is likely to be the case for any novel disease. Thus our premise that the infection-rate could be used as a monitoring variable in surveillance algorithms seems to be a promising one and does not suffer from the need for a lot of data to function well. This robustness makes it particularly well-suited for novel diseases.

Our method suffers from two shortcomings. Our first shortcoming is that while our formulation is generalizable to many areal units, it has been demonstrated on just three areal units. This is due to the lack of scalability of MCMC. We have adapted our method to use approximate, but scalable, mean-field Variational Inference and scaled it to all 33 counties in NM; this is documented in a technical report\cite{23rs5a} and is the subject of our next paper. The second shortcoming is the use of Gaussian models throughout this paper, even though the low case-count data for some counties, e.g., Santa Fe, would have suggested a negative binomial distribution. This, however, would have requires us to develop a random field model using negative binomials, and is left to future work.

\clearpage
\section*{\bf Author contributions}

Cosmin Safta formulated the problem, wrote the software to solve it, generated the figures and wrote the paper. Jaideep Ray formulated the spatial inverse problem, wrote the software to perform the detection of the Fall 2020 wave and wrote the sections of paper describing it. Wyatt Bridgman helped with problem formulation and writing of the paper.

\section*{Acknowledgments}
This paper (SAND2024-07653O) describes objective technical results and analysis. Any subjective views or opinions that might be expressed in the paper do not necessarily represent the views of the U.S. Department of Energy or the United States Government. This article has been authored by an employee of National Technology \& Engineering Solutions of Sandia, LLC under Contract No. DE-NA0003525 with the U.S. Department of Energy (DOE). The employee owns all right, title and interest in and to the article and is solely responsible for its contents. The United States Government retains and the publisher, by accepting the article for publication, acknowledges that the United States Government retains a non-exclusive, paid-up, irrevocable, world-wide license to publish or reproduce the published form of this article or allow others to do so, for United States Government purposes. The DOE will provide public access to these results of federally sponsored research in accordance with the DOE Public Access Plan https://www.energy.gov/downloads/doe-public-access-plan.

\section*{\bf Financial disclosure}
None reported.

\section*{\bf Conflict of interest}
The authors declare no potential conflict of interests.

\clearpage
\appendix
\begin{figure}[!h]
\centering
\includegraphics[width = 0.5\textwidth]{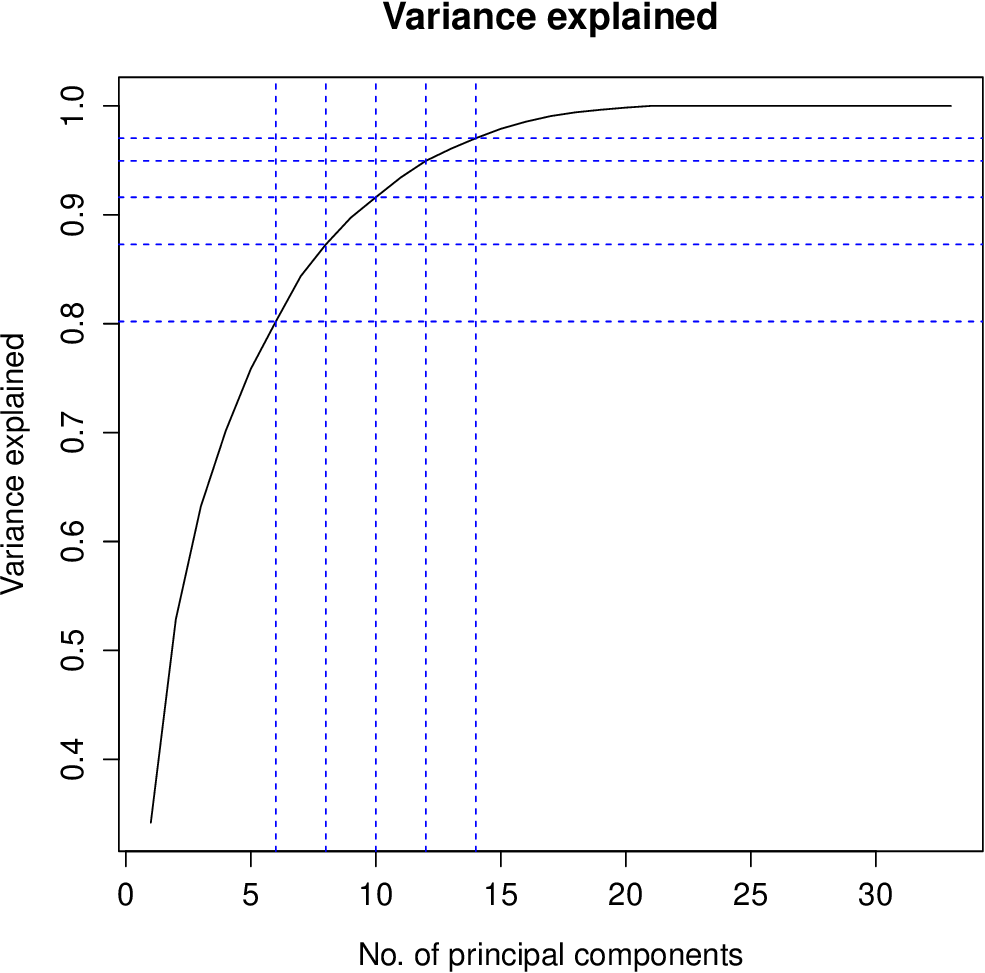}
\caption{Variation explained by the principal components obtained via sparse PCA of the 79 risk factors used to model population-normalized case-counts in the counties of New Mexico. We see that 12 principal components can cover 95\% of the variations observed in the risk-factors.}
\label{fig:scree}
\end{figure}

\begin{figure}
\centerline{
\includegraphics[height=0.3\textheight]{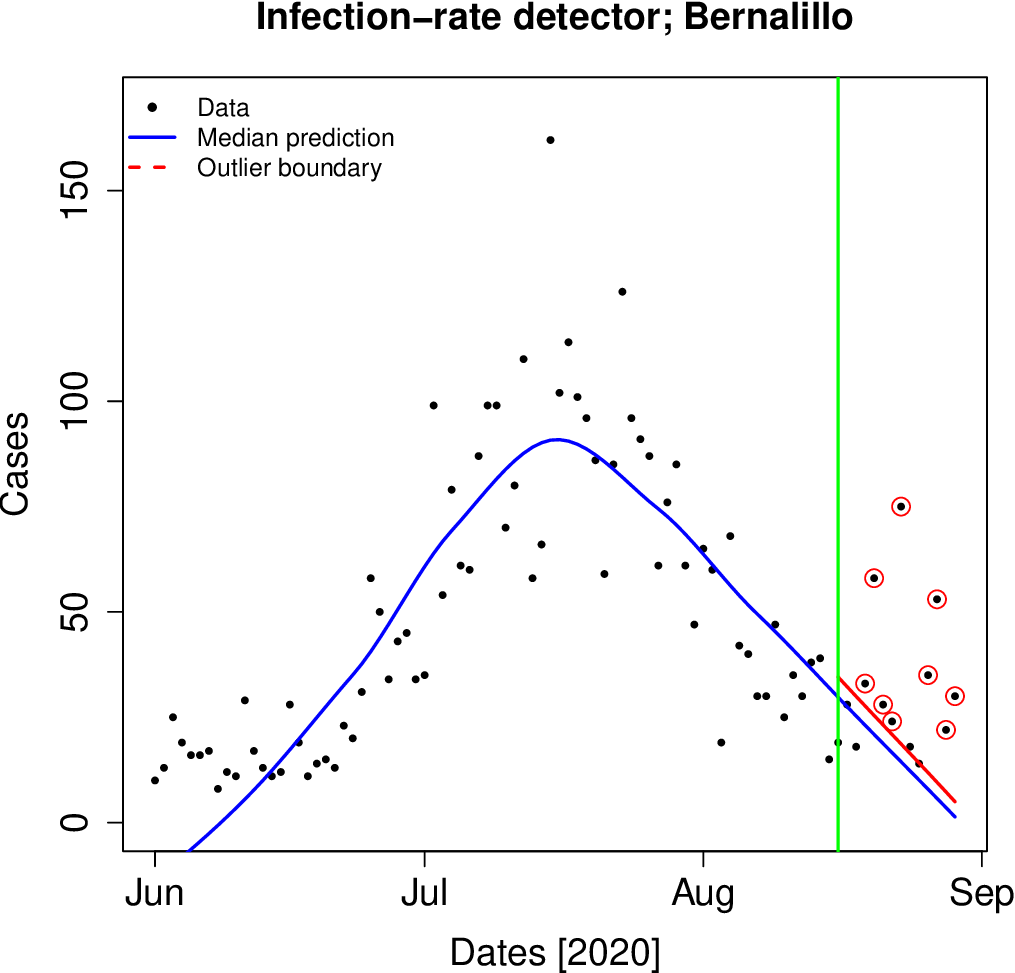}
\includegraphics[height=0.3\textheight]{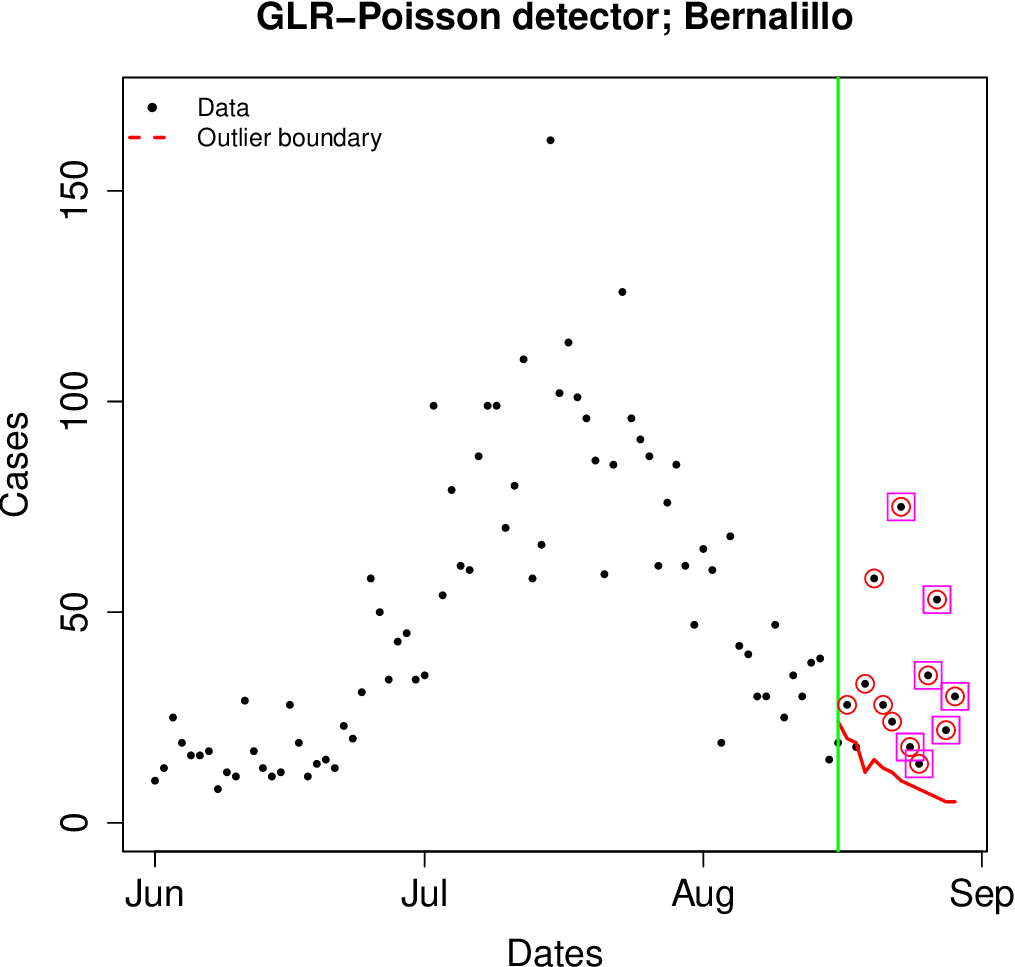}
}
\centerline{
\includegraphics[height=0.3\textheight]{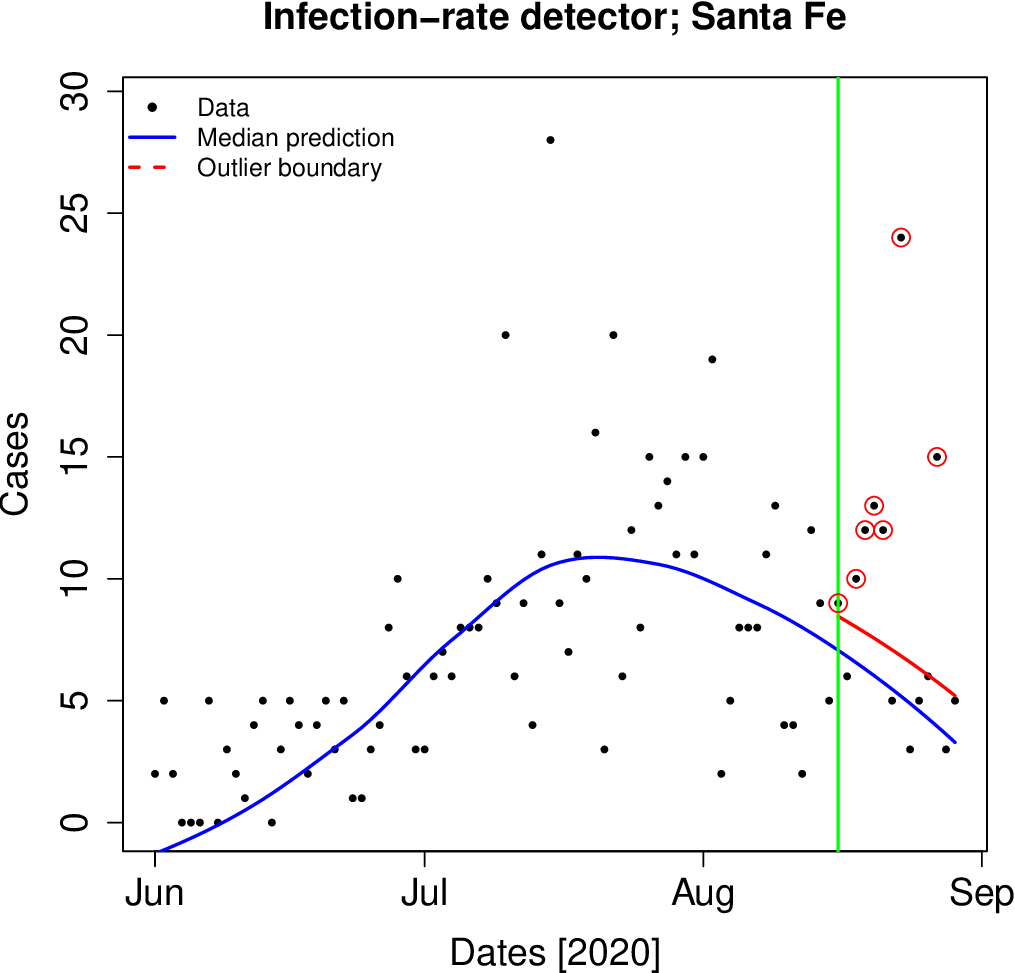}
\includegraphics[height=0.3\textheight]{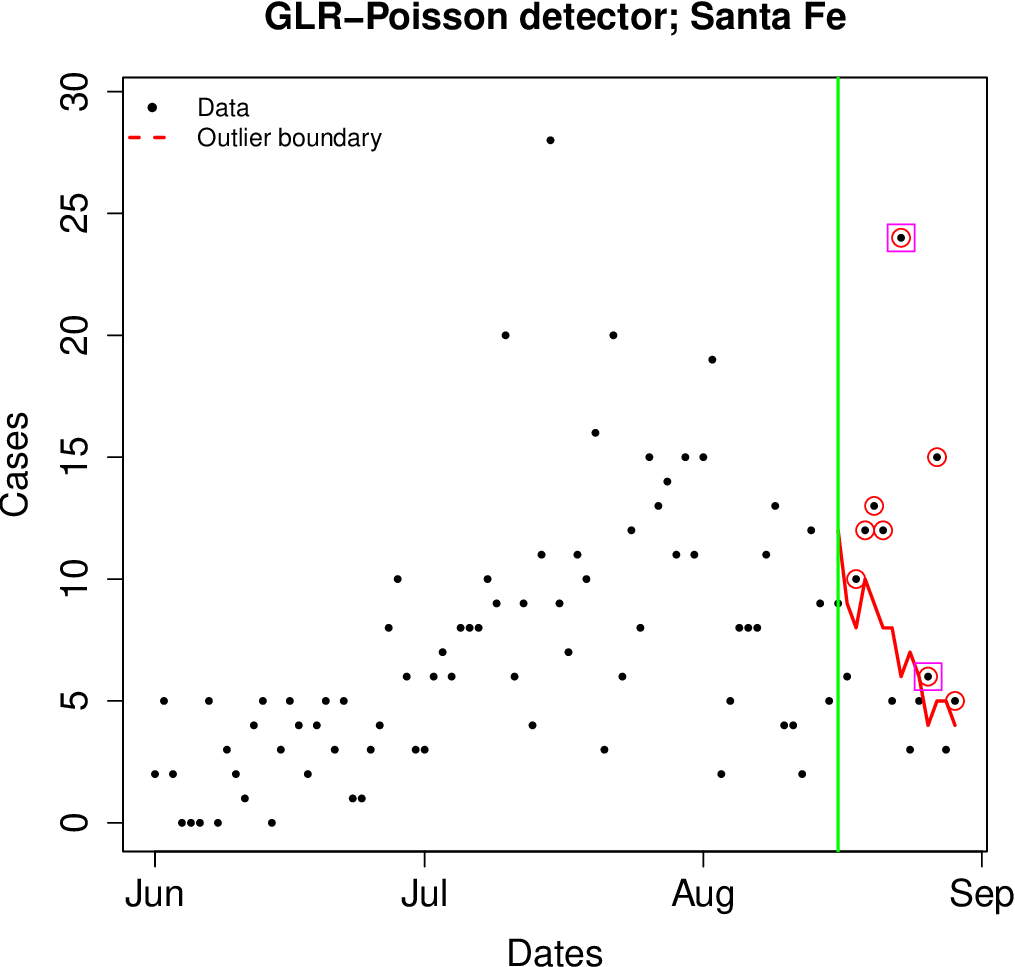}
}
\centerline{
\includegraphics[height=0.3\textheight]{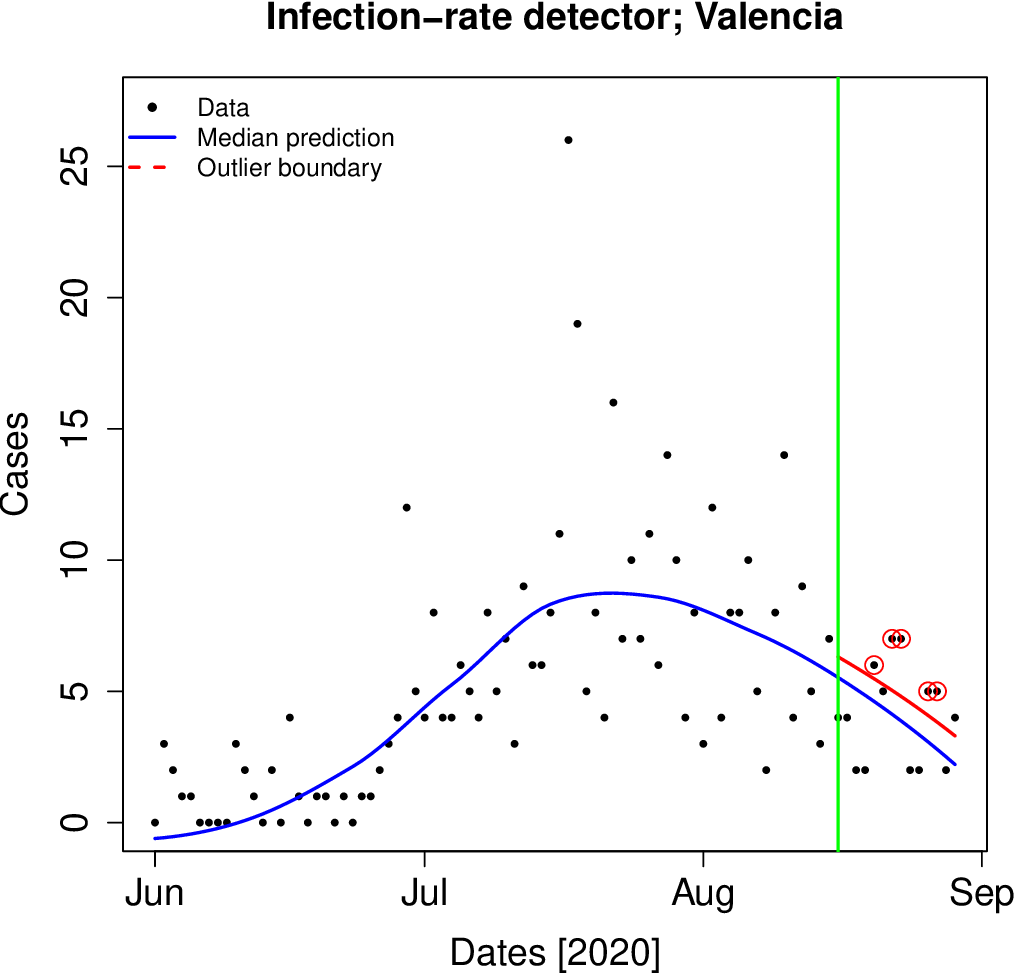}
\includegraphics[height=0.3\textheight]{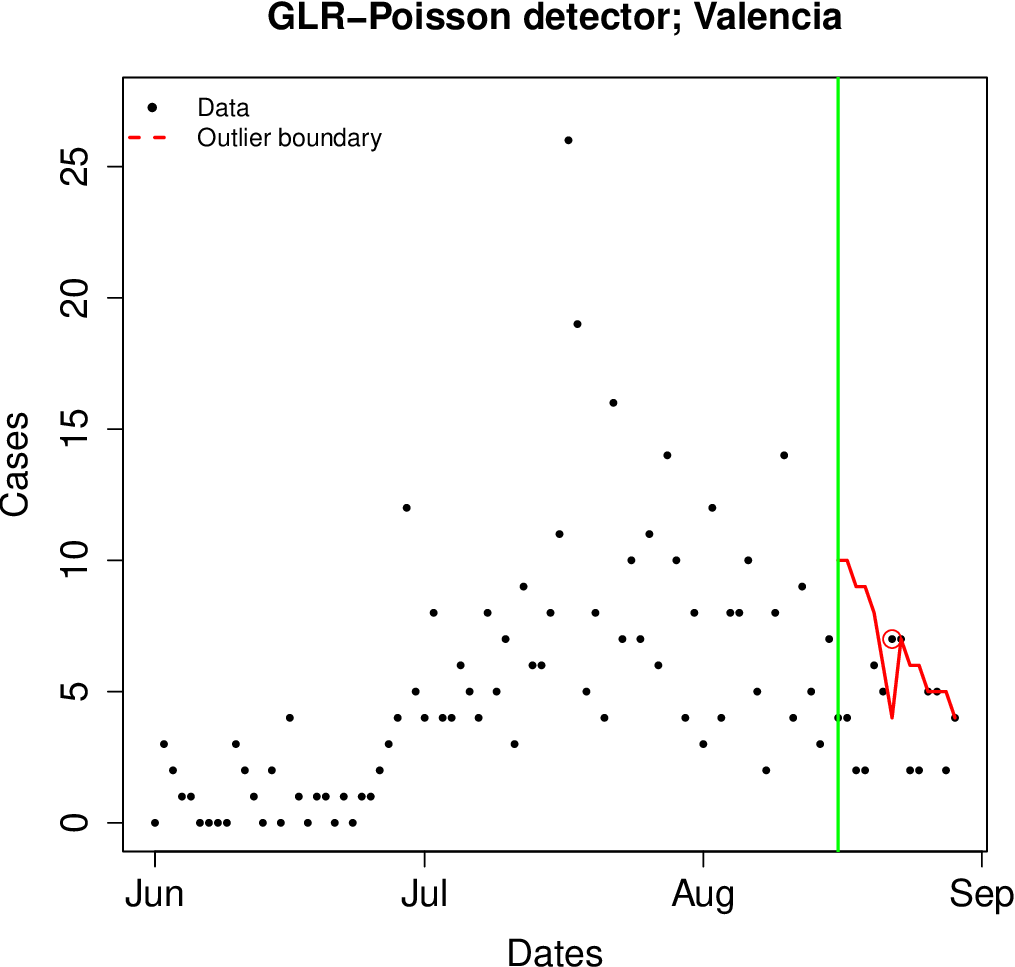}
}
\caption{Comparison of the infection-rate detector (left) compared with the GLR-Poisson detector (right), using data from 2020-06-01 to 2020-08-15. August 15 is a month before the arrival of the Fall 2020 wave. The symbols are the observed case-counts. The red line beyond August 15 is the outlier boundary; a day with a case-count above the dashed line is an ``outlier'' and is circled. Top row: Performance for Bernalillo county. Middle row: Results for Santa Fe county. Bottom row: Results for Valencia county. The absence of any days with a box around it implies that no alarms were raised, which is the correct behavior.}
\label{fig:detectAug}
\end{figure}

\begin{figure}
\centerline{
\includegraphics[width=0.4\textwidth]{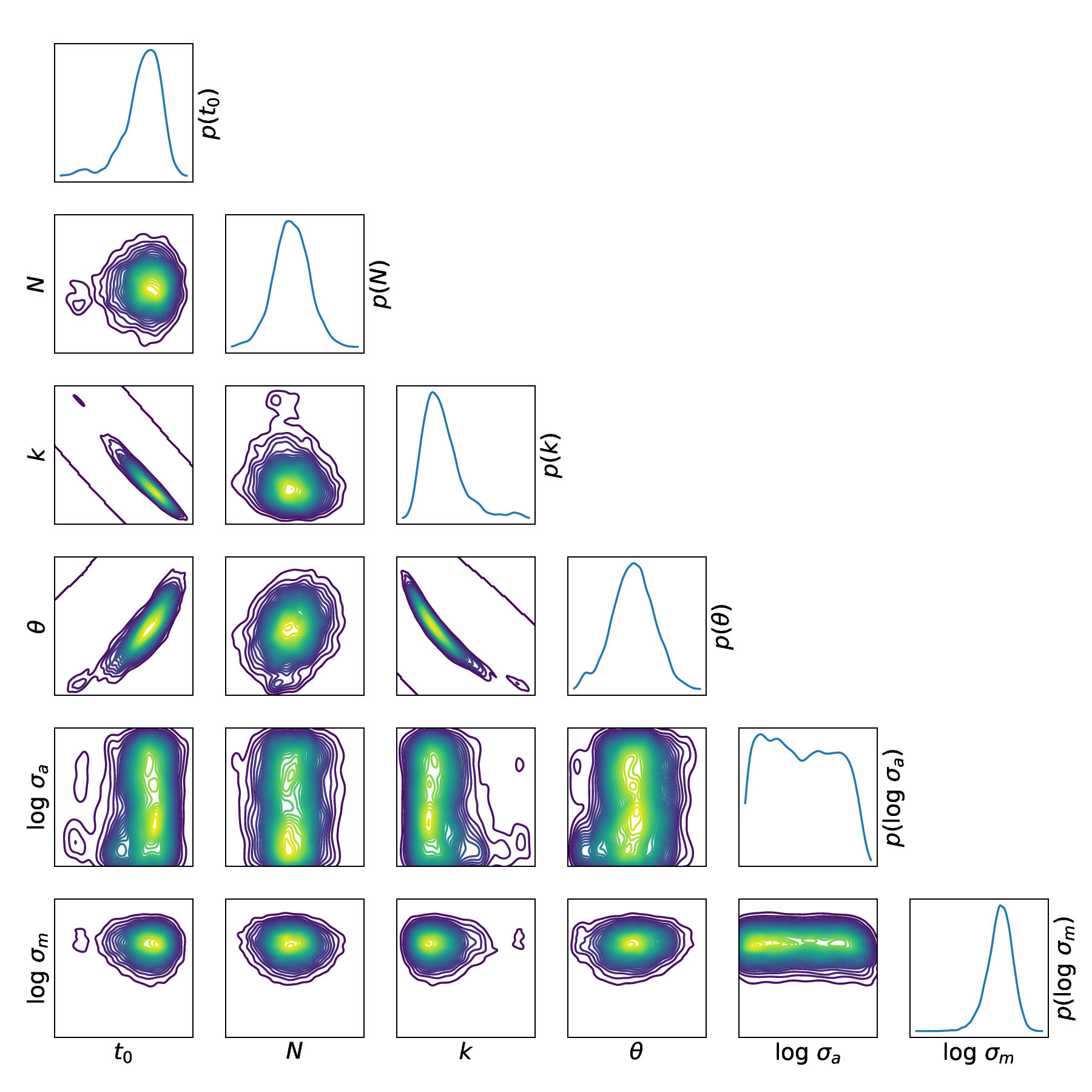}
\includegraphics[width=0.4\textwidth]{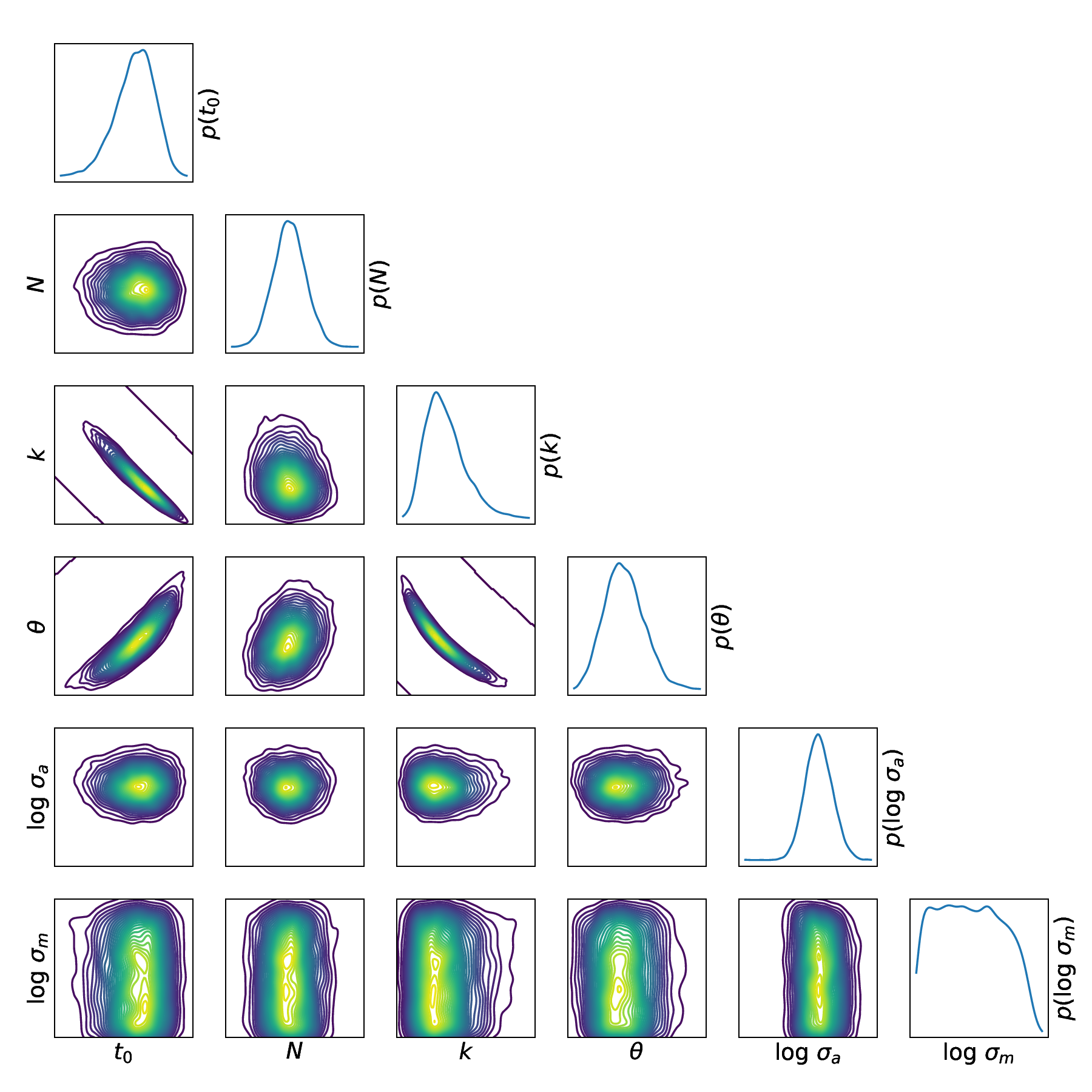}
}
\caption{One and two-dimensional marginal posterior distributions for the Bernalillo county model parameters; left: 3 region joint inversion, right: Bernalillo county only.}
\label{fig:B_marg}
\end{figure}

\begin{figure}
\centerline{
\includegraphics[width=0.4\textwidth]{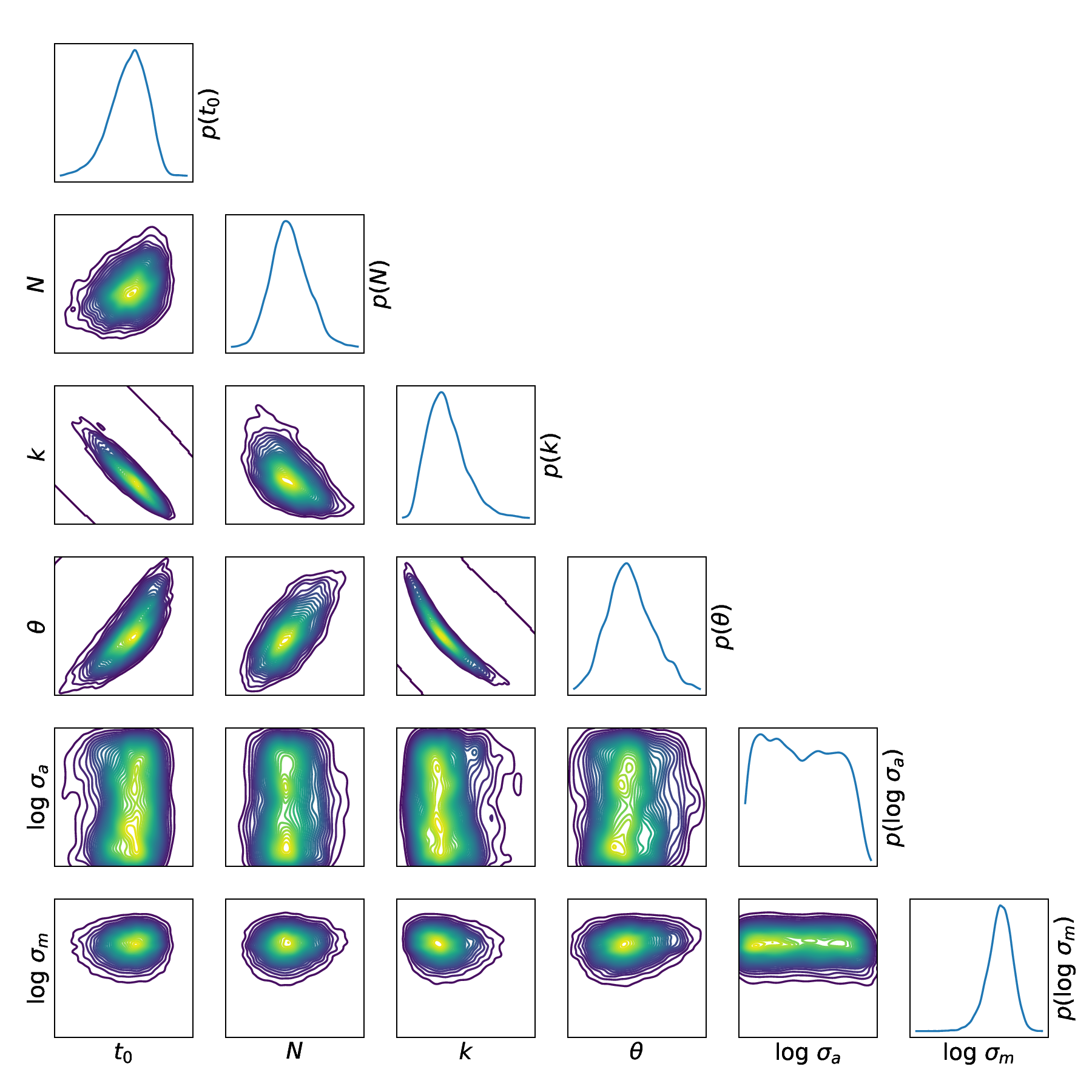}
\includegraphics[width=0.4\textwidth]{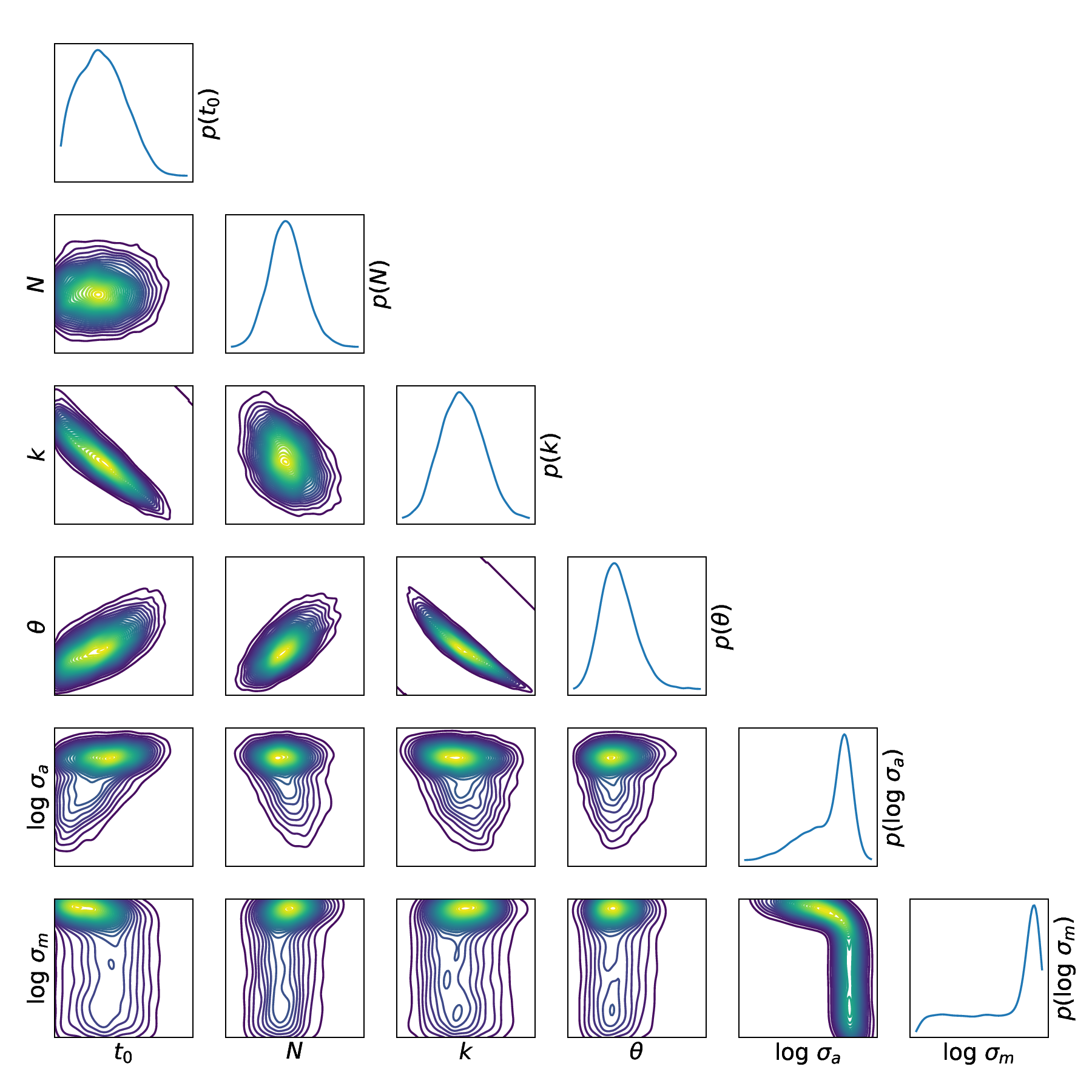}
}
\caption{One and two-dimensional marginal posterior distributions for the Santa Fe county model parameters; left: 3 region joint inversion, right: Santa Fe county only.}
\label{fig:SF_marg}
\end{figure}

\begin{figure}
\centerline{
\includegraphics[width=0.4\textwidth]{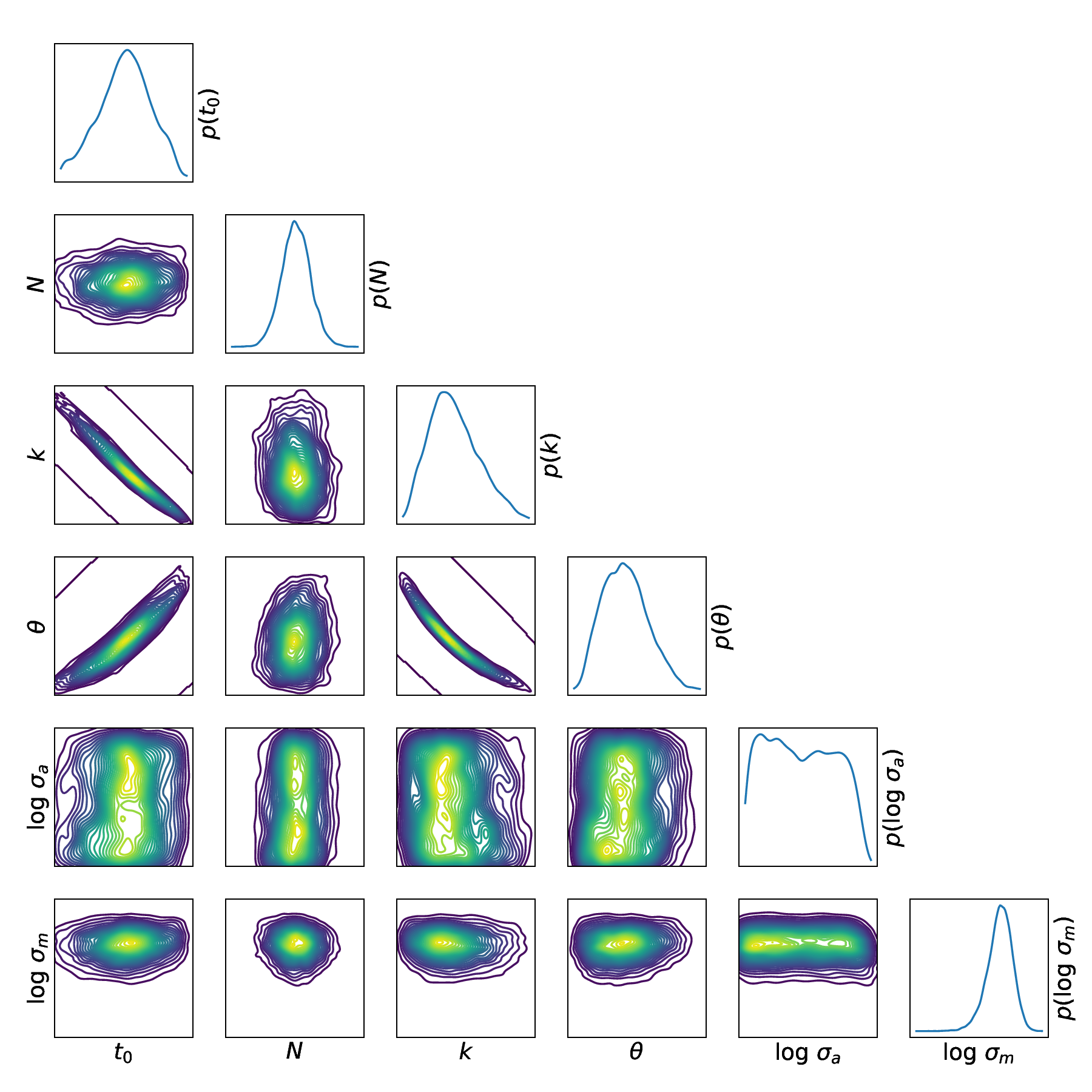}
\includegraphics[width=0.4\textwidth]{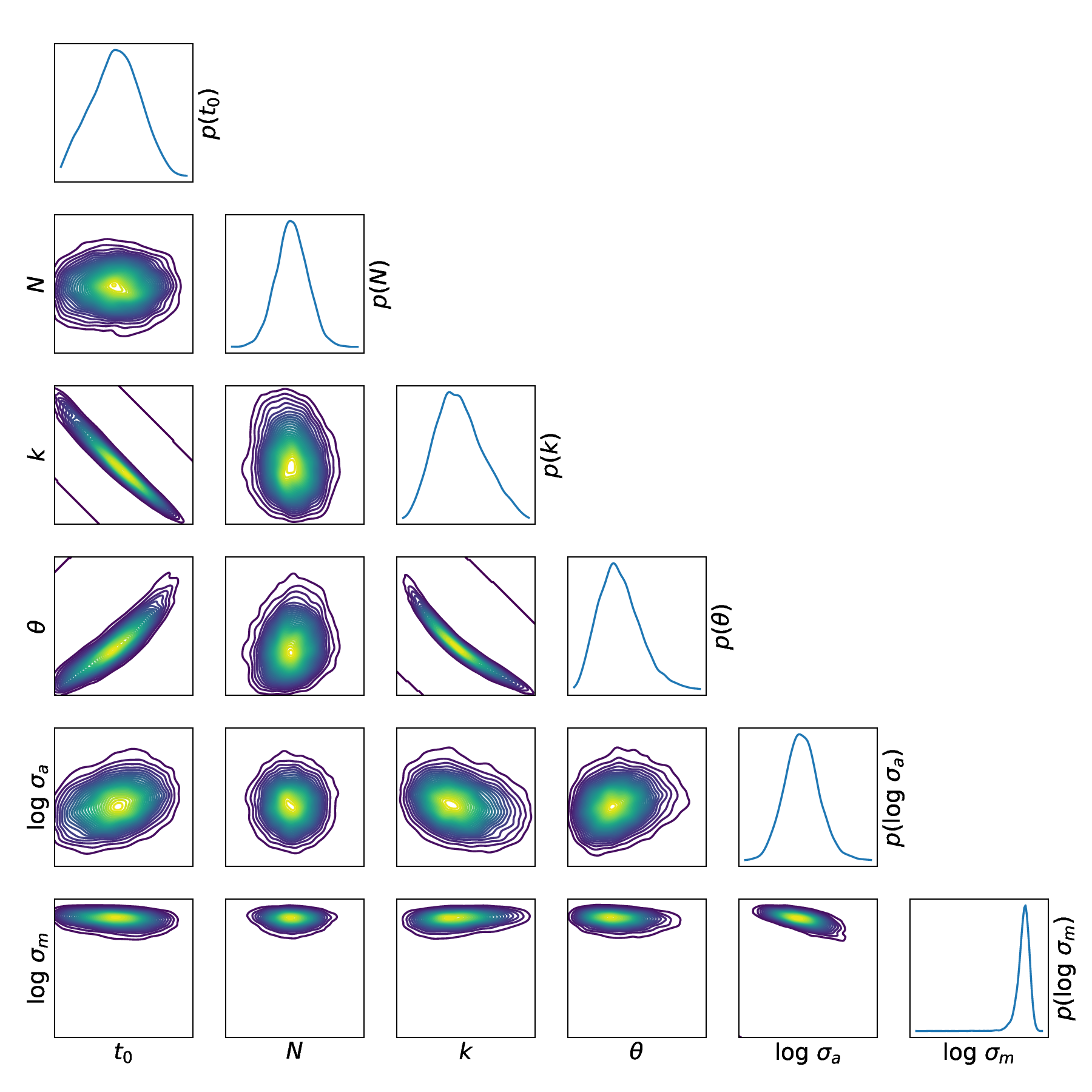}
}
\caption{One and two-dimensional marginal posterior distributions for the Valencia county model parameters; left: 3 region joint inversion, right: Valencia county only.}
\label{fig:V_marg}
\end{figure}

Figs.~\ref{fig:B_marg}-\ref{fig:V_marg} show 1D and 2D marginal posterior distributions for the three counties tackled in this study. These results indicate a strong correlation between the inferred start of the epidemic, $t_0$, and the parameters of the infection model $k$ and $\theta$ for each of these regions. When calibrating model for individual regions, the discrepancy between the model and the available observations results in an error model with both the additive $\sigma_a$ and multiplicative $\sigma_m$ components informed by the data for the counties with smaller populations, Santa Fe and Valencia. For Bernalillo only just the additive error component is sufficient to model the discrepancy. When performing the statistical inference with all three counties, the multiplicative component takes over as that error model component is less sensitive to phase shifts of the epidemic waves compared to the additive component.

\clearpage %% Force all figures to be plotted before starting bibliography

\bibliographystyle{ieeetr}

\begin{thebibliography}{10}

\bibitem{DazaTorres:2022}
M.~L. Daza-Torres, M.~A. Capistr{\'a}n, A.~Capella, and J.~A. Christen,
``Bayesian sequential data assimilation for covid-19 forecasting,'' {\em
Epidemics}, vol.~39, p.~100564, 2022.

\bibitem{WangZ:2020}
Z.~Wang, X.~Zhang, G.~Teichert, M.~Carrasco-Teja, and K.~Garikipati, ``System
inference for the spatio-temporal evolution of infectious diseases: Michigan
in the time of covid-19,'' {\em Computational Mechanics}, vol.~66,
pp.~1153--1176, 2020.

\bibitem{ChenP:2021}
P.~Chen, K.~Wu, and O.~Ghattas, ``Bayesian inference of heterogeneous epidemic
models: Application to covid-19 spread accounting for long-term care
facilities,'' {\em Computer Methods in Applied Mechanics and Engineering},
vol.~385, p.~114020, 2021.

\bibitem{Blonigan:2021}
P.~Blonigan, J.~Ray, and C.~Safta, ``Forecasting multi-wave epidemics through
bayesian inference,'' {\em Archives of Computational Methods in Engineering},
vol.~28, no.~6, pp.~4169--4183, 2021.

\bibitem{Lin:2021}
Y.~Lin, J.~Neumann, E.~Miller, R.~Posner, A.~Mallela, C.~Safta, J.~Ray,
G.~Thakur, S.~Chinthavali, and W.~Hlavacek, ``Daily forecasting of regional
epidemics of coronavirus disease with bayesian uncertainty quantification,
united states.,'' {\em Emerging Infectious Diseases}, vol.~27, no.~3,
pp.~767--778, 2021.

\bibitem{Safta:2021}
C.~Safta, J.~Ray, and K.~Sargsyan, ``Characterization of partially observed
epidemics through bayesian inference: Application to covid-19,'' {\em
Computational Mechanics}, vol.~66, no.~5, pp.~1109--1129, 2020.

\bibitem{Hohle:2008}
M.~H{\"o}hle and M.~Paul, ``Count data regression charts for the monitoring of
surveillance time series,'' {\em Computational Statistics and Data Analysis},
vol.~52, no.~9, pp.~4357--4368, 2008.

\bibitem{Haario:2001}
H.~Haario, E.~Saksman, and J.~Tamminen, ``An adaptive metropolis algorithm,''
{\em Bernoulli}, pp.~223--242, 2001.

\bibitem{23rs5a}
J.~Ray, C.~Safta, W.~Bridgman, M.~Horii, and A.~Gould, ``A spatially
regularized detector for emergent/re-emergent disease outbreaks,'' Tech. Rep.
SAND2023-09749R, Sandia National Laboratories, PO Box 5800, Albuquerque, NM
87185, September 2023.
\newblock
\url{https://www.sandia.gov/app/uploads/sites/203/2023/09/SAND2023-09749R.pdf}.

\bibitem{Best:2005}
N.~Best, S.~Richardson, and A.~Thomson, ``A comparison of bayesian spatial
models for disease mapping,'' {\em Statistical methods in medical research},
vol.~14, no.~1, pp.~35--59, 2005.

\bibitem{Waller:2010}
L.~Waller and B.~Carlin, ``Disease mapping,'' in {\em Handbook of Spatial
Statistics} (A.~E. Gelfand, P.~J. Diggle, M.~Fuentes, and P.~Guttorp, eds.),
ch.~14, Chapman \& Hall / CRC Press, 2010.

\bibitem{Huang:2021}
X.~Huang, H.~Zhou, X.~Yang, W.~Zhou, J.~Huang, and Y.~Yuan, ``Spatial
characteristics of coronavirus disease 2019 and their possible relationship
with environmental and meteorological factors in hubei province, china,''
{\em GeoHealth}, vol.~5, no.~6, p.~e2020GH000358, 2021.

\bibitem{Geng:2021}
X.~Geng, G.~G. Katul, F.~Gerges, E.~Bou-Zeid, H.~Nassif, and M.~C. Boufadel,
``A kernel-modulated sir model for covid-19 contagiousspread from county to
continent,'' {\em Proceedings of the National Academy of Sciences}, vol.~118,
no.~21, p.~e2023321118, 2021.

\bibitem{Schuler:2021}
L.~Schuler, J.~M. Calabrese, and S.~Attinger, ``Data driven high resolution
modeling and spatial analyses of the covid-19 pandemic in germany,'' {\em
PLOS ONE}, vol.~16, pp.~1--14, 08 2021.

\bibitem{McMahon:2022}
T.~McMahon, A.~Chan, S.~Havlin, and L.~K. Gallos, ``{Spatial correlations in
geographical spreading of COVID-19 in the United States},'' {\em Scientific
Reports}, vol.~12, no.~1, p.~699, 2022.

\bibitem{Rendana:2021}
M.~Rendana, W.~M.~R. Idris, and S.~{Abdul Rahim}, ``Spatial distribution of
covid-19 cases, epidemic spread rate, spatial pattern, and its correlation
with meteorological factors during the first to the second waves,'' {\em
Journal of Infection and Public Health}, vol.~14, no.~10, pp.~1340--1348,
2021.
\newblock Special Issue on COVID-19 – Vaccine, Variants and New Waves.

\bibitem{Sathish:2023}
S.~H.~S. Indika, N.~Diawara, H.~A. Jeng, B.~D. Giles, and D.~S.~K. Gamage,
``Modeling the spread of covid-19 in spatio-temporal context,'' {\em
Mathematical Biosciences and Engineering}, vol.~20, no.~6, pp.~10552--10569,
2023.

\bibitem{Lawson:2017}
A.~Lawson and D.~Lee, ``Chapter 16 - bayesian disease mapping for public
health,'' in {\em Disease Modelling and Public Health, Part A} (A.~S.
{Srinivasa Rao}, S.~Pyne, and C.~Rao, eds.), vol.~36 of {\em Handbook of
Statistics}, pp.~443--481, Elsevier, 2017.

\bibitem{MacNab:2022}
Y.~C. MacNab, ``Bayesian disease mapping: Past, present, and future,'' {\em
Spatial Statistics}, vol.~50, pp.~100593, 28, 2022.

\bibitem{Besag:1991}
J.~Besag, J.~York, and A.~Molli{\'e}, ``Bayesian image restoration, with two
applications in spatial statistics,'' {\em Annals of the institute of
statistical mathematics}, vol.~43, pp.~1--20, 1991.

\bibitem{Stern:1999}
H.~Stern and N.~A. Cressie, ``Inference for extremes in disease mapping,'' in
{\em Disease Mapping and Risk Assessment for Public Health} (A.~Lawson, ed.),
pp.~63--84, Chichester: Wiley, 1999.

\bibitem{Cressie:2015}
N.~Cressie, {\em Statistics for spatial data}.
\newblock John Wiley \& Sons, 2015.
\newblock Revised Edition.

\bibitem{Baptista:2016}
H.~Baptista, J.~M. Mendes, Y.~C. MacNab, M.~Xavier, and J.~C. de~Almeida, ``A
gaussian random field model for similarity-based smoothing in bayesian
disease mapping,'' {\em Statistical Methods in medical Research}, vol.~25,
no.~4, pp.~1166--1184, 2016.

\bibitem{Best:1999}
N.~G. Best, R.~A. Arnold, A.~Thomas, L.~A. Waller, and E.~M. Conlon, ``Bayesian
models for spatially correlated disease and exposure data,'' {\em Bayesian
statistics}, vol.~6, pp.~131--156, 1999.

\bibitem{12uf5a}
S.~Unkel, C.~P. Farrington, P.~H. Garthwaite, C.~Robertson, and N.~Andrews,
``Statistical methods for the prospective detection of infectious disease
outbreaks: a review,'' {\em Journal of the Royal Statistical Society Series
A: Statistics in Society}, vol.~175, no.~1, pp.~49--82, 2012.

\bibitem{Shewhart:1930}
W.~A. Shewhart, ``Economic quality control of manufactured product 1,'' {\em
Bell System Technical Journal}, vol.~9, no.~2, pp.~364--389, 1930.

\bibitem{Serfling:1963}
R.~E. Serfling, ``Methods for current statistical analysis of excess
pneumonia-influenza deaths,'' {\em Public health reports}, vol.~78, no.~6,
p.~494, 1963.

\bibitem{Pelat:2007}
C.~Pelat, P.-Y. Bo{\"e}lle, B.~J. Cowling, F.~Carrat, A.~Flahault, S.~Ansart,
and A.-J. Valleron, ``Online detection and quantification of epidemics,''
{\em BMC Medical Informatics and Decision Making}, vol.~7, no.~1, pp.~1--9,
2007.

\bibitem{Parker:1989}
R.~Parker, ``Analysis of surveillance data with poisson regression: a case
study,'' {\em Statistics in Medicine}, vol.~8, no.~3, pp.~285--294, 1989.

\bibitem{Jackson:2007}
M.~L. Jackson, A.~Baer, I.~Painter, and J.~Duchin, ``A simulation study
comparing aberration detection algorithms for syndromic surveillance,'' {\em
BMC medical informatics and decision making}, vol.~7, no.~1, pp.~1--11, 2007.

\bibitem{Farrington:1996}
C.~Farrington, N.~J. Andrews, A.~Beale, and M.~Catchpole, ``A statistical
algorithm for the early detection of outbreaks of infectious disease,'' {\em
Journal of the Royal Statistical Society: Series A (Statistics in Society)},
vol.~159, no.~3, pp.~547--563, 1996.

\bibitem{Reis:2003}
B.~Y. Reis and K.~D. Mandl, ``Time series modeling for syndromic
surveillance,'' {\em BMC medical informatics and decision making}, vol.~3,
no.~1, pp.~1--11, 2003.

\bibitem{Williamson:1999}
G.~D. Williamson and G.~Weatherby~Hudson, ``A monitoring system for detecting
aberrations in public health surveillance reports,'' {\em Statistics in
medicine}, vol.~18, no.~23, pp.~3283--3298, 1999.

\bibitem{LeStrat:1999}
Y.~Le~Strat and F.~Carrat, ``Monitoring epidemiologic surveillance data using
hidden markov models,'' {\em Statistics in medicine}, vol.~18, no.~24,
pp.~3463--3478, 1999.

\bibitem{MartinezBeneito:2008}
M.~A. Mart{\'\i}nez-Beneito, D.~Conesa, A.~L{\'o}pez-Qu{\'\i}lez, and
A.~L{\'o}pez-Maside, ``Bayesian markov switching models for the early
detection of influenza epidemics,'' {\em Statistics in medicine}, vol.~27,
no.~22, pp.~4455--4468, 2008.

\bibitem{Conesa:2010}
D.~Conesa, R.~Amor{\'o}s, A.~L{\'o}pez-Qu{\i}lez, and M.~A.
Mart{\i}nez-Beneito, ``Mean-variability hidden markov models for the
detection of influenza outbreaks,'' in {\em 25th International Workshop on
Statistical Modelling}, (Amsterdam, The Netherlands), Statistical Modelling
Society, 2010.

\bibitem{LuHM:2011}
H.-M. Lu, D.~Zeng, and H.~Chen, ``Markov switching models for outbreak
detection,'' in {\em Infectious Disease Informatics and Biosurveillance:
Research, Systems and Case Studies} (C.~Castillo-Chavez, H.~Chen, W.~B.
Lober, M.~Thurmond, and D.~Zeng, eds.), pp.~111--144, Springer US, 2011.

\bibitem{06hh4a}
L.~Held, M.~Hofmann, M.~H{\"o}hle, and V.~Schmid, ``A two-component model for
counts of infectious diseases,'' {\em Biostatistics}, vol.~7, no.~3,
pp.~422--437, 2006.

\bibitem{10ls2a}
A.~B. Lawson and H.-R. Song, ``Bayesian hierarchical modeling of the dynamics
of spatio-temporal influenza season outbreaks,'' {\em Spatial and
spatio-temporal epidemiology}, vol.~1, no.~2-3, pp.~187--195, 2010.

\bibitem{23la1a}
A.~B. Lawson, ``Evaluation of predictive capability of bayesian spatio-temporal
models for covid-19 spread,'' {\em BMC Medical Research Methodology},
vol.~23, no.~1, p.~182, 2023.

\bibitem{23kl6a}
J.~Kim, A.~B. Lawson, B.~Neelon, J.~E. Korte, J.~M. Eberth, and G.~Chowell,
``Evaluation of bayesian spatiotemporal infectious disease models for
prospective surveillance analysis,'' {\em BMC Medical Research Methodology},
vol.~23, no.~1, p.~171, 2023.

\bibitem{21lk2a}
A.~B. Lawson and J.~Kim, ``Space-time covid-19 bayesian sir modeling in south
carolina,'' {\em PLoS One}, vol.~16, no.~3, p.~e0242777, 2021.

\bibitem{21sl3a}
B.~Sartorius, A.~Lawson, and R.~Pullan, ``Modelling and predicting the
spatio-temporal spread of covid-19, associated deaths and impact of key risk
factors in england,'' {\em Scientific reports}, vol.~11, no.~1, p.~5378,
2021.

\bibitem{22lk2a}
A.~B. Lawson and J.~Kim, ``Issues in bayesian prospective surveillance of
spatial health data,'' {\em Spatial and Spatio-temporal Epidemiology},
vol.~41, p.~100431, 2022.

\bibitem{16rl2a}
C.~Rotejanaprasert, A.~Lawson, S.~Bolick-Aldrich, and D.~Hurley, ``Spatial
bayesian surveillance for small area case event data,'' {\em Statistical
methods in medical research}, vol.~25, no.~4, pp.~1101--1117, 2016.

\bibitem{11cl2a}
A.~Corber{\'a}n-Vallet and A.~B. Lawson, ``Conditional predictive inference for
online surveillance of spatial disease incidence,'' {\em Statistics in
medicine}, vol.~30, no.~26, pp.~3095--3116, 2011.

\bibitem{nytcovid}
``{Coronavirus (Covid-19) Data in the United States}.''
\url{https://github.com/nytimes/covid-19-data}.
\newblock Accessed: 2023-01-01.

\bibitem{jhucovid}
``{COVID-19 Data Repository by the Center for Systems Science and Engineering
(CSSE) at Johns Hopkins University}.''
\url{https://github.com/CSSEGISandData/COVID-19}.
\newblock Accessed: 2023-01-01.

\bibitem{20sf5a}
L.~Shand, A.~Foss, A.~Zhang, J.~D. Tucker, and G.~Huerta, ``A statistical model
for the spread of sars-cov-2 in new mexico,'' Tech. Rep. SAND2020-10080,
Sandia National Laboratories, PO Box 5800, Albuquerque, NM 87185, September
2020.

\bibitem{QF:NM}
``{United States Census Bureau}, quickfacts new mexico.''
\url{https://www.census.gov/quickfacts/NM}, January 2024.
\newblock Accessed January 2024.

\bibitem{Infra:NM}
``{New Mexico Primary Care Association}, find a health center.''
\url{https://www.nmpca.org/find-a-health-center}, 2024.
\newblock Accessed January 2024.

\bibitem{Infra:Screen}
``{New Mexico Department of Health}, covid-19 screening and test sites.''
\url{https://cvprovider.nmhealth.org/directory.html}.
\newblock Accessed September 2020.

\bibitem{Infra:RGIS}
``{Earth Data Analysis Center}, university of new mexico, resource geographic
information system.'' \url{https://rgis.unm.edu}.
\newblock Accessed September 2020.

\bibitem{20ez6a}
N.~B. Erichson, P.~Zheng, K.~Manohar, S.~L. Brunton, J.~N. Kutz, and A.~Y.
Aravkin, ``Sparse principal component analysis via variable projection,''
{\em SIAM Journal on Applied Mathematics}, vol.~80, no.~2, pp.~977--1002,
2020.

\bibitem{18bw2a}
R.~S. Bivand and D.~W. Wong, ``Comparing implementations of global and local
indicators of spatial association,'' {\em Test}, vol.~27, no.~3,
pp.~716--748, 2018.

\bibitem{Lauer:2020}
S.~A. Lauer, K.~H. Grantz, Q.~Bi, F.~K. Jones, Q.~Zheng, H.~R. Meredith, A.~S.
Azman, N.~G. Reich, and J.~Lessler, ``{The Incubation Period of Coronavirus
Disease 2019 (COVID-19) From Publicly Reported Confirmed Cases: Estimation
and Application},'' {\em Annals of Internal Medicine}, 2020.

\bibitem{Cressie:2008}
N.~Cressie and G.~Johannesson, ``Fixed rank kriging for very large spatial data
sets,'' {\em Journal of the Royal Statistical Society: Series B (Statistical
Methodology)}, vol.~70, no.~1, pp.~209--226, 2008.

\bibitem{Andrieu:2009}
C.~Andrieu and G.~O. Roberts, ``{The pseudo-marginal approach for efficient
Monte Carlo computations},'' {\em Ann. Statist.}, vol.~37, no.~2,
pp.~697--725, 2009.

\bibitem{Kass:1998}
R.~Kass, B.~Carlin, A.~Gelman, and R.~Neal, ``Markov chain monte carlo in
practice: A roundtable discussion,'' {\em The American Statistician},
vol.~52, no.~2, pp.~93--100, 1998.

\bibitem{14gk2a}
T.~Gneiting and M.~Katzfuss, ``Probabilistic forecasting,'' {\em Annual Review
of Statistics and Its Application}, vol.~1, pp.~125--151, 2014.

\bibitem{Szekely:2007}
G.~Sz\'{e}kely, M.~Rizzo, and N.~Bakirov, ``Measuring and testing dependence by
correlation of distances,'' {\em Annals of Statistics}, vol.~35,
pp.~2769--2794, 2007.

\bibitem{Manual:R}
{R Core Team}, {\em R: A Language and Environment for Statistical Computing}.
\newblock R Foundation for Statistical Computing, Vienna, Austria, 2023.

\bibitem{07hm1a}
M.~Höhle, ``surveillance: An {R} package for the monitoring of infectious
diseases,'' {\em Computational Statistics}, vol.~22, no.~4, pp.~571--582,
2007.

\end{thebibliography}

\end{document}